\documentclass[11pt]{article} 
\usepackage{jheppub}

\usepackage{amsmath,amsfonts,amsbsy,amssymb,array,accents,dsfont}
\usepackage{enumerate}
\usepackage[shortlabels]{enumitem}
\usepackage{graphicx} 
\usepackage{subcaption} 
\usepackage{upgreek}
\usepackage{empheq}
\usepackage[many]{tcolorbox}
\usepackage{bm}
\usepackage{cancel}
\usepackage{slashed}
\usepackage{nccmath}
\usepackage{tikz}

\newcommand*{\Scale}[2][4]{\scalebox{#1}{\ensuremath{#2}}}%

\newtcolorbox{empheqboxed}{colback=gray!30, 
 colframe=white,
 width=\textwidth,
 sharpish corners,
 top=-2mm, 
 bottom=0pt
}

\let\includefigures=\iftrue
\let\useblackboard==\iftrue
\newfam\black

\PassOptionsToPackage{ 
     colorlinks=true,
     linkcolor=myBlue,
     urlcolor=blue,
     filecolor=black,
     citecolor=myRed,
}{hyperref}

\usepackage{xparse}
\usepackage{xspace}

\NewDocumentCommand\eqn{om}{%
  \IfNoValueTF{#1}
     {\[ #2 \]}
     {\begin{equation}\label{#1} #2  \end{equation} \expandafter\newcommand\csname #1\endcsname{\eqref{#1}\xspace}\ignorespaces}
}
\NewDocumentCommand\eqna{om}{%
  \IfNoValueTF{#1}
    {\begin{align*} #2 \end{align*}}
    {\begin{equation}\label{#1}\begin{split} #2  \end{split}\end{equation} \expandafter\def\csname #1\endcsname{\eqref{#1}\xspace}\ignorespaces}
}

\newcommand{\rcite}{\cite}

\def\sl{\text{sl}}

\def\sltwo{\ensuremath{SL(2,\mathds R)}}

\def\sutwo{{SU(2)}}

\def\uone{U(1)}

\def\({\left(}
\def\){\right)}
\def\[{\left[}
\def\]{\right]}

\def\ie{{i.e.}}
\def\eg{{e.g.}}
\def\cf{{c.f.}}
\def\etc{{etc}}

\def\ttt{{\sf t}}
\def\nin{{\notin}}
\def\mfg{{\mathfrak g}}
\def\mfh{{\mathfrak h}}

\def\nfive{{n_5}}

\def\none{{n_1}}

\def\sfk{{\mathsf k}}
\def\sfm{{\mathsf m}}
\def\sfp{{\mathsf p}}

\def\mfg{{\mathfrak g}}

\DeclareMathSymbol{\medhatsym}{\mathord}{largesymbols}{"62} 

\DeclareMathSymbol{\medtildesym}{\mathord}{largesymbols}{"65}

\makeatletter
\newcommand*\rel@kern[1]{\kern#1\dimexpr\macc@kerna}
\newcommand*\widebar[1]{%
  \begingroup
  \def\mathaccent##1##2{%
    \rel@kern{0.8}%
    \overline{\rel@kern{-0.8}\macc@nucleus\rel@kern{0.2}}%
    \rel@kern{-0.2}%
  }%
  \macc@depth\@ne
  \let\math@bgroup\@empty \let\math@egroup\macc@set@skewchar
  \mathsurround\z@ \frozen@everymath{\mathgroup\macc@group\relax}%
  \macc@set@skewchar\relax
  \let\mathaccentV\macc@nested@a
  \macc@nested@a\relax111{#1}%
  \endgroup
}
\makeatother

\def\half{\frac12}

\def\One{{\hbox{1\kern-1mm l}}}

\def\barray{\begin{array}}
\def\earray{\end{array}}
\def\be{\begin{equation}}
\def\ee{\end{equation}}
\def\bea{\begin{eqnarray}}
\def\eea{\end{eqnarray}}
\def\bal{\begin{align}}
\def\eal{\end{align}}
\def\nn{\nonumber}

\newcommand{\bC}{{\mathbb C}}

\newcommand{\bR}{{\mathbb R}}
\newcommand{\bS}{{\mathbb S}}
\newcommand{\bT}{{\mathbb T}}

\newcommand{\bZ}{{\mathbb Z}}

				\def\sfk{{\mathsf k}}		
\def\sfm{{\mathsf m}}		\def\sfn{{\mathsf n}}				\def\sfp{{\mathsf p}}

\definecolor{cardinal}{rgb}{0.6,0,0}
\definecolor{darkgreen}{rgb}{0,0.4,0}
\definecolor{green}{rgb}{0,0.4,0}
\definecolor{golden}{rgb}{0.92, 0.7, 0}
\definecolor{midnight}{rgb}{0, 0, 0.5}
\definecolor{darkblue}{rgb}{0, 0, 0.7}

\numberwithin{equation}{section}

\mathchardef\mhyphen="2D

\def\cA{\mathcal {A}}  \def\cC{\mathcal {C}}

\def\cJ{\mathcal {J}}  
\def\cM{\mathcal {M}} \def\cN{\mathcal {N}} 
  \def\cR{\mathcal {R}}

\def\one{{\hbox{\kern+.5mm 1\kern-.8mm l}}}
\def\zero{{\hbox{0\kern-1.5mm 0}}}

\def\id{\textrm{id}}

\newcommand{\WZW}{\text{WZW}}

\def\id{{1 \kern-.28em {\rm l}}}

\def\journal#1&#2(#3){\unskip, \sl #1\ \bf #2 \rm(19#3) }
\def\andjournal#1&#2(#3){\sl #1~\bf #2 \rm (19#3) }

\def\ie{{\it i.e.}}
\def\eg{{\it e.g.}}
\def\cf{{\it c.f.}}

\def\etc{{\it etc}}

\def\half{\frac12}

\def\One{{1\hskip -3pt {\rm l}}}

\catcode`\@=11
\def\slash#1{\mathord{\mathpalette\c@ncel{#1}}}
\def\underrel#1\over#2{\mathrel{\mathop{\kern\z@#1}\limits_{#2}}}

\catcode`\@=12

\def\exp{{\rm exp}}

\def\ie{{\it i.e.}}
\def\eg{{\it e.g.}}

\title{
{
NS5-brane backgrounds and coset CFT partition functions
}}

\author{
Andrea Dei {\it and} Emil J. Martinec
}

\affiliation{
\vskip 0.01cm
Kadanoff Center for Theoretical Physics, Enrico Fermi Institute, and Department of Physics\\ 
University of Chicago,
5640 S. Ellis Ave.,
Chicago IL 60637\\ 
}

\emailAdd{%
adei@uchicago.edu,
e-martinec@uchicago.edu}

\abstract{%
Worldsheet string theory is solvable for a variety of backgrounds involving Neveu-Schwarz fivebranes, in terms of gauged nonlinear sigma models on group manifolds. We compute the worldsheet torus partition function of these models, and propose gauging of null isometries as a unifying principle and conceptual framework for this large family of string backgrounds. In the process, we explain how partition functions of asymmetrically gauged Wess-Zumino-Witten models can be computed from the path integral, and organize and systematize various results scattered throughout the literature.
}

\begin{document}
\hypersetup{pageanchor=false}
\maketitle
\hypersetup{pageanchor=true}
\pagenumbering{arabic}


\vskip 2cm







\section{Introduction} 
\label{sec:intro}

There are few curved backgrounds in string theory for which an exact worldsheet description is available.  Those that are known involve Wess-Zumino-Witten (WZW) models, and the coset models that arise when they are gauged (GWZW models).  
The WZW model is the non-linear sigma model on a group manifold $G$, with a covariantly constant Neveu-Schwarz 3-form flux $H_3$ given by the group structure constants.  The action is invariant under left and right group transformations of the sigma model field $g\mapsto h_L\, g\, h_R$.  One can then gauge any non-anomalous subgroup $ H\subset G_L\times  G_R$.  From these constructions (including flat spacetime and tori), a further set of models can be obtained via quotienting by discrete symmetries, \ie\ the orbifold procedure~\rcite{Dixon:1985jw}.

\smallskip

A particularly rich set of examples involves backgrounds sourced by $n_5$ Neveu-Schwarz (NS) fivebranes:  
\begin{itemize}
\item
The Euclidean black NS5-brane geometry in the decoupling limit $g_s\to 0$ is the coset model~\rcite{Witten:1991yr, Dijkgraaf:1991ba}
\be
\label{EucBH}
\frac{\sltwo_\nfive}{\uone}\times \sutwo_\nfive\times ds^2_{\parallel} \ , 
\ee
where $ds^2_\parallel$ is the spatial geometry along the fivebranes.
\item
The geometry of $\nfive$ separated BPS fivebranes is described in the same decoupling limit by the coset orbifold~\rcite{Giveon:1999px, Giveon:1999tq}
\be
\label{DSLST}
\bigg(\frac{\sltwo_\nfive}\uone\times\frac{\sutwo_\nfive}\uone\bigg)\Big/\bZ_\nfive\times ds^2_{\parallel} \ , 
\ee
\item
The decoupling limit of $\nfive$ NS5-branes and $\none$ fundamental strings is the WZW model
\be
\label{AdS3}
\sltwo_\nfive\times \sutwo_\nfive\times \cM \ , 
\ee
where $\cM=\bT^4$ or $K3$ \rcite{Giveon:1998ns, Kutasov:1999xu, Maldacena:2000hw}.
\item
The Euclidean BTZ black hole is the orbifold \rcite{Banados:1992wn, Banados:1992gq, Natsuume:1996ij}
\be
\label{EucBTZ}
\bigg(\frac{SL(2,\bC)_\nfive}{\sutwo_\nfive}\bigg)\Big/\bZ \times\sutwo_\nfive\times \cM \ . 
\ee
\end{itemize}

\smallskip 

Recently, a large class of 1/2-BPS ``supertube'' geometries \rcite{Lunin:2001fv} have been shown to be described by the cosets \rcite{Martinec:2017ztd, Martinec:2018nco, Martinec:2019wzw, Martinec:2020gkv, Bufalini:2021ndn, Bufalini:2022wyp, Bufalini:2022wzu, Martinec:2022okx}
\be
\label{MMgeoms}
\frac{\sltwo_\nfive\times\sutwo_\nfive\times\bR_t\times \uone_y}{\uone_L\times\uone_R} ~,
\ee
where $t$ ($y$) parametrize timelike (spacelike) factors in the target space.
These are highly coherent states far from the vacuum, describing fivebranes carrying momentum and/or string winding charge, as well as $\bS^3$ angular momentum, which nevertheless admit an exact description in terms of worldsheet string theory. This construction extends to a larger class of backgrounds by varying the embedding of the gauge group to involve asymmetrically gauged cosets, where $\uone_L$ and $\uone_R$ are distinct subgroups of $G$.  Examples include the 1/4-BPS ``GLMT''~\rcite{Giusto:2004id,Giusto:2012yz} and non-supersymmetric ``JMaRT''~\rcite{Jejjala:2005yu,Chakrabarty:2015foa} geometries obtained by spacetime spectral flow of the 1/2-BPS backgrounds.
In the $AdS_3$ limit $R_y\to\infty$, there is also a worldsheet construction of the backgrounds~\eqref{MMgeoms} as global orbifolds $(\sltwo\times\sutwo)/\bZ_\sfk$~\rcite{Martinec:2001cf,Martinec:2023zha}.  

Holographically, the construction~\eqref{MMgeoms} realizes particular heavy states having conformal dimension of order $N=\none\nfive$ in the dual spacetime CFT of central charge $c=6N$~\rcite{Lunin:2001jy,Giusto:2012yz,Chakrabarty:2015foa}.%
\footnote{The coset models~\eqref{MMgeoms} build states in the Ramond sector of the spacetime CFT, while the $(AdS_3\times\bS^3)/\bZ_\sfk$ orbifold construction builds the corresponding states in the NS sector.}  Perturbative fluctuations around these particular heavy states, including stringy excitations, are captured by the worldsheet theory.
The construction~\eqref{DSLST} has been used to calculate various properties of little string theory, for instance the perturbative string S-matrix~\rcite{Aharony:2004xn}; also, the elliptic genus~\rcite{Eguchi:2004yi,Ashok:2012qy,Giveon:2015raa}, which describes a large class of 1/4-BPS states.
The worldsheet partition function on the Euclidean BTZ background~\eqref{EucBTZ} has been computed in~\rcite{Maldacena:2000kv, Ashok:2020dnc, Ashok:2022vdz}, and related to the spectrum of (stringy) quasinormal modes.

\medskip

This wide variety of examples of exactly solvable string backgrounds with NS flux motivates a study of the worldsheet partition function of these models, as the generating function of the perturbative string spectrum.  Most of the above examples involve gauging of the current algebra symmetries of the worldsheet sigma-model.  The most elaborate of these gauged WZW models, eq.~\eqref{MMgeoms}, involve the gauging of null isometries of the background; but we will see that {\it all} of the above coset models can be recast as null-gauged WZW models.  In this way, null gauging serves as a sort of unifying principle.

Computing the worldsheet torus partition function of the GWZW models \eqref{MMgeoms} presents two challenges: The group $G$ is non-compact, and in general asymmetrically gauged. Since $G$ is not compact, the associated partition function cannot be computed simply by following standard CFT$_2$ methods, such as tracing over the Hilbert space (see~\rcite{DiFrancesco:1997nk} for a review). The only known approach is the path integral \rcite{Gawedzki:1991yu}, where the subtleties of the correct measure for integrating over a continuum of group representations can be sorted out. In fact, partition functions for non-compact GWZW models have been computed for all the examples~\eqref{EucBH}-\eqref{EucBTZ}, \rcite{Gawedzki:1991yu, Kutasov:2000jp,Maldacena:2000kv, Hanany:2002ev, Israel:2003ry, Israel:2004ir, Ashok:2020dnc}. While in all these cases the group $G$ was non-compact, the gauging was symmetric, \ie~it did not involve a different treatment of holomorphic and anti-holomorphic currents.  The consideration of non-simple groups opens up a much larger choice of embeddings of the gauge group~\rcite{Martinec:2017ztd,Bufalini:2021ndn}, including asymmetric embeddings. 
In this manuscript we deal with both these issues, and extend previous treatments in the literature, in which only one of these ingredients is present.%
\footnote{A notable exception is~\rcite{DAppollonio:2007ldj, DAppollonio:2008xxq}, which analyzes asymmetric cosets of $H_4$, the plane-wave limit of $SL(2,\mathds R) \times U(1)$, and computes the partition function.}

\medskip

We mentioned that the isometries gauged in the models \eqref{MMgeoms} are \emph{null}. In order to convey the intuition behind many of the computations we carry out in this manuscript, let us briefly sketch the origin of some special feautures arising when the gauged isometries are null. When gauging $H\subset G_L\times G_R$, usually one of the two components of the $\mathfrak h$-valued gauge field $A_\mu$ is locally pure gauge, while the other is a physical degree of freedom.  One can then locally rewrite the physical component of the gauge field in terms of $H$-valued fields $h$, together with the zero modes of $A_\mu$.
One then employs the Polyakov-Wiegmann identity~\rcite{Polyakov:1984et}
\begin{equation}
\kappa S^\WZW(g) - \frac{\kappa}{\pi} \int \text{d}^2 z \, \text{Tr}(g^{-1} \partial_{\bar z} g \,  \partial_{z} h h^{-1}) = \kappa S^\WZW(gh) - \kappa S^\WZW(h)  
~,
\label{PW-identity0}
\end{equation}
to decouple $A_\mu$ (parametrized in terms of $h$) from the path integral over $g\in G$.
Schematically, the path integral over $G$ with the subgroup $H\subset G_L\times G_R$ gauged yields a generalization of the LHS involving both chiralities, which after a systematic use of the Polyakov-Wiegmann identity yields a corresponding generalization of the RHS. One then absorbs the appearance of $h$ in the first term via a field redefinition to arrive at a decoupled path integral over $G\times H$.

On the other hand, when the action of $H$ on $G$ is such that the gauge orbits are null isometries (as is the case for the models \eqref{MMgeoms} of~\rcite{Martinec:2017ztd}, see also~\rcite{Klimcik:1994wp,Israel:2004ir}), {\it both} components of $A_\mu$ are pure gauge~-- there is a {\it second} gauge invariance under a group $\widetilde H$ with a closely related embedding into $ G_L\times G_R$, and the single gauge field $A_\mu$ is sufficient to gauge both $ H$ and $\widetilde H$.
In this situation, {\it both} components of $A_\mu$ are pure gauge, and can be eliminated apart from zero modes by a choice of gauge.  One is then left with an integral over $G$, together with the zero modes of $A_\mu$.

This fact suggests that ordinary GWZW models can be recast as null-gauged models, simply by gauging $ G\times H$ instead of $ G$, with a ``wrong-sign'' kinetic term for $ H$ (as on the RHS of~\eqref{PW-identity0} above).  With the diagonal embedding of $H$ into $G_L\times H_L$ and into $G_R\times H_R$, the gauge orbits are null isometries.  Using the gauge symmetry to eliminate the gauge field apart from zero modes, one is left with a path integral over $G\times H$, together with the gauge field zero modes, and thus formally the standard GWZW construction is equivalent to null gauging of $G\times H$.
However, the manipulations of the path integral involved are rather delicate; in particular, we have been rather cavalier in the treatment of the gauge field zero modes in the above discussion. 

\bigskip

After reviewing the WZW model and its gauging at the start of Section~\ref{sec:GWZW}, we go on to show how null gauging renders all but the Wilson line degrees of freedom of the gauge field to be gauge-redundant.  As warmup exercises in the computation of more general null-gauged WZW partition functions, we show in Sections~\ref{sec:su2/u1} and~\ref{sec:sl2/u1} how to recover the standard $\frac\sutwo\uone$ and $\frac\sltwo\uone$ GWZW partition functions from null gauging of $\sutwo\times\mathds R$ and $\sltwo\times\uone$, respectively.

Null-gauging can of course be applied to more general examples than those that arise from the conventional coset construction.
A more nontrivial example of null gauging arises in the example of $G=\sltwo\times\sutwo$, which we turn to in Section~\ref{sec:cosetorb}.  We show there how it reproduces the partition function of the model \eqref{DSLST} of~\rcite{Giveon:1999px}, where a background of NS5-branes on their Coulomb branch is described by the coset orbifold $\big(\frac\sltwo\uone\times\frac\sutwo\uone\big)/\bZ_{n_5}$.
In this case, the gauge group $H$ is not a factor in $G$, and the equivalence to the standard coset construction is less straightforward (as the additional orbifold in the relation between the two suggests).

Finally, we come in Section~\ref{sec:supertube} to the models of~\rcite{Martinec:2017ztd} describing NS5-F1 ``supertubes'' and related backgrounds. Here the group $G$ is $\sltwo\times\sutwo\times \mathds R\times\uone$, see eq.~\eqref{MMgeoms}, with $H=\uone$ embedded non-trivially across all the factors in $G$.  The null gauging procedure is at present the only known realization of these models, and the construction of their partition functions is our main result. See in particular eq.~\eqref{ST-pf-final}. The decomposition of this result into characters (see eqs.~\eqref{Z-ST-trace} and \eqref{Z-ST-decomp-long}) gives us confidence in the result, and confirms the analysis of the spectrum carried out in \eg~\rcite{Martinec:2018nco,Martinec:2022okx}. 

We conclude with a discussion in Section~\ref{sec:discussion}, where we list a number of possible applications of our results. Several appendices address some technical points and provide details of the various component theta series, affine characters, \etc. In particular, Appendix~\ref{app:GWZW in superspace} provides a detailed treatment of the gauging of fermions in the supersymmetric asymmetric GWZW model, which to our knowledge has not appeared previously in the literature.


\section{Review of Gauged WZW models} 
\label{sec:GWZW}

In view of the discussion on more general cosets in the next section, let us review the best known examples of GWZW: vector and axial gauging. We will refer to these two cosets as ``symmetric'' GWZW, as opposed to ``asymmetric'' gauging, for which the currents to be gauged on the left and the right are different. 

The WZW action for a simple group $G$ reads \cite{Witten:1983ar,DiFrancesco:1997nk}\footnote{For a semi-simple group, the action is simply given by the sum of terms of the form \eqref{WZW-action} for each simple factor.}
\begin{align}
\kappa S^{\WZW}(g) & = -\frac{\kappa}{8 \pi} \int_{\Sigma} \text{d}^2x \, \text{Tr}(g^{-1} \partial^\mu g g^{-1} \partial_\mu g)  - \frac{i \kappa}{12 \pi} \int_{B} \text{d}^3x\, \epsilon^{\mu \nu \rho}  \,\text{Tr}(g^{-1} \partial_\mu g g^{-1} \partial_\nu g g^{-1} \partial_\rho g) \ , 
\label{WZW-action}
\end{align}
where $g \in G$, $\kappa \in \mathds N$ and $B$ is a three-dimensional manifold having as boundary the compactification of the base space $\Sigma$.  It is easy to show that the action \eqref{WZW-action} obeys the Polyakov-Wiegmann identity written above, 
\begin{equation}
\kappa S^\WZW(gh) = \kappa S^\WZW(g) + \kappa S^\WZW(h) - \frac{\kappa}{\pi} \int \text{d}^2 v \, \text{Tr}(g^{-1} \partial_{\bar v} g \,  \partial_v h h^{-1}) \ , 
\label{PW-identity}
\end{equation}
where we introduced holomorphic coordinates\footnote{For the integral measure we adopt the following conventions, $\text{d}^2v = \text{d Re}(v) \,  \text {d Im}(v) = \text d x^0 \text d x^1 = \text d^2 x$.}
\begin{equation}
v = x^0 + i x^1 \ , \qquad \bar v = x^0 - i x^1 \ . 
\label{v-x}
\end{equation}
The Polyakov-Wiegmann identity makes manifest the gauge invariance of the action \eqref{WZW-action} under\footnote{Different conventions are frequently adopted in the math literature, where the gauge invariance is often written as $g \to h_R g h_L$. We prefer to adopt the conventions of \cite{Quella:2002fk}. }
\begin{equation}
g(v, \bar v) \to h_L(v) g(v, \bar v) h_R(\bar v) \ , 
\label{GxG-symmetry}
\end{equation}
where $h_L(v)$ and $h_R(\bar v)$ are arbitrary (anti-) holomorphic functions taking values in a subgroup $H \subset G$. The coset theory $G/H$ can be obtained by gauging the global symmetry \eqref{GxG-symmetry} and promoting it to 
\begin{equation}
g(v, \bar v) \to h_L(v, \bar v) g(v, \bar v) h_R(v, \bar v) \ , 
\label{g->hLghR}
\end{equation}
where $h_L$ and $h_R$ are again valued in some subgroup $H$ of $G$ but are no longer (anti-)holomorphic. The best known non-anomalous gauging for a simple group $G$ and a semi-simple subgroup $H$ is
\begin{equation}
 g(v,\bar v) \to h(v, \bar v) g(v, \bar v) h^{-1}(v,\bar v) \ , 
\end{equation}
known as vector gauging. When instead $H$ contains $U(1)$ factors, for each $U(1)$ factor, in addition to vector gauging also axial gauging is non-anomalous. The latter corresponds to
\begin{equation}
 g(v,\bar v) \to h^{-1}(v, \bar v) g(v, \bar v) h^{-1}(v,\bar v)  \ . 
\end{equation}
The vector gauged action \cite{Gawedzki:1988hq, Karabali:1988au, Park:1989ae, Karabali:1989dk, Bardakci:1990lbc, Kiritsis:1991zt}
\begin{equation}
    \kappa S^V(g,\mathcal A) = \kappa S^\WZW(g)+\frac{\kappa}{\pi} \int \text d^2 v \, \text{Tr}\left( \mathcal A_{\bar v} \partial_v g g^{-1} - \mathcal
 A_v g^{-1} \partial_{\bar v}g  + g^{-1} \mathcal  A_{\bar v} g \mathcal  A_{v} - \mathcal  A_v \mathcal  A_{\bar v} \right)
    \label{vector-action}
\end{equation}
is gauge-invariant under
\begin{equation}
    g \mapsto hgh^{-1} \ , \qquad \mathcal  A_v \mapsto h \mathcal A_v h^{-1} + \partial_v h \, h^{-1} \ ,  \qquad  \mathcal  A_{\bar v} \mapsto h \mathcal  A_{\bar v} h^{-1} + \partial_{\bar v} h \, h^{-1} \ , 
\end{equation}
while for the axial gauged action 
\begin{equation}
   \kappa S^A(g, \mathcal A) = \kappa S^\WZW(g)-\frac{\kappa}{\pi} \int \text d^2 v \, \text{Tr}\left( \mathcal  A_{\bar v} \partial_v g g^{-1} + \mathcal
 A_v g^{-1} \partial_{\bar v}g  + g^{-1} \mathcal  A_{\bar v} g \mathcal A_{v} + \mathcal A_v \mathcal A_{\bar v}\right) \ , 
    \label{axial-action}
\end{equation}
gauge transformations read
\begin{equation}
    g \mapsto h^{-1}gh^{-1} \ , \qquad  \mathcal  A_v \mapsto \mathcal  A_v + \partial_v h \, h^{-1} \ ,  \qquad \mathcal  A_{\bar v} \mapsto \mathcal 
 A_{\bar v} +  \partial_{\bar v} h \, h^{-1} \ . 
\end{equation}

The WZW action \eqref{WZW-action} can be promoted to an $\mathcal N=1$ supersymmetric action \cite{Abdalla:1984ef, DiVecchia:1984nyg, Figueroa-OFarrill:1996xek}
\begin{equation}
S^{\WZW}_{\mathcal N=1}(g) = (\kappa-h^{\vee}) S^\WZW(g) + \frac{1}{2 \pi} \int \text d^2 v \, \text{Tr}\left( \psi \partial_{\bar v} \psi + \tilde \psi \partial_{v} \tilde \psi \right) \ ,
\end{equation}
where $h^\vee$ is the dual Coxeter number of $\mathfrak g$, \ie~the quadratic Casimir of the adjoint representation divided by 2 and the free fermions $\psi$ and $\tilde \psi$ take values in the adjoint representation of $\mathfrak g$. The supersymmetric vector gauged action reads \cite{Schnitzer:1988qj, Witten:1991mk, Nakatsu:1991pu, Tseytlin:1993my, Figueroa-OFarrill:1996xek}
\begin{equation}
    S^V_{\mathcal N=1}(g, \mathcal A) = (\kappa-h^{\vee})S^V(g, \mathcal A) + \int \frac{\text d^2 v}{2 \pi}  \, \text{Tr}\left( \psi \partial_{\bar v} \psi - \psi \, [\mathcal A_{\bar v}, \psi]+ \tilde \psi \partial_{v} \tilde \psi - \tilde \psi [\mathcal A_v, \tilde \psi] \right) \ , 
\end{equation}
while for axial gauging we have
\begin{equation}
    S^A_{\mathcal N=1}(g, \mathcal A) = (\kappa-h^{\vee})S^A(g, \mathcal A) +  \int \frac{\text d^2 v}{2 \pi} \, \text{Tr}\left( \psi \partial_{\bar v} \psi + \psi \, [\mathcal A_{\bar v}, \psi]+ \tilde \psi \partial_v \tilde \psi - \tilde \psi [\mathcal A_v, \tilde \psi]\right) \ , 
\end{equation}
where the worldsheet fermions $\psi$ and $\tilde \psi$ take value in $\mathfrak h^\perp$, the orthogonal complement of $\mathfrak h$.

\subsection{Asymmetric gauging} 
\label{sec:asym}

A richer class of gauged WZW models arises when gauging different subgroups $H_L$ on the left and $H_R$ on the right of a (not necessarily simple) group $G$ by promoting eq.~\eqref{g->hLghR} to 
\begin{equation}
    g \to \epsilon_L(h) \, g \, \epsilon_R(h^{-1}) \ , 
\end{equation}
with $\epsilon_L$ and $\epsilon_R$ (a priori different) Lie group homomorphisms from $H$ to $G$, satisfying
\begin{equation}
    \epsilon_L(g_1 \, g_2) = \epsilon_L(g_1) \, \epsilon_L(g_2) \ , \qquad      \epsilon_R(g_1 \, g_2) = \epsilon_R(g_1) \, \epsilon_R(g_2) \ , \qquad \forall \, g_1, g_2 \in G \ . 
\end{equation}
With some slight abuse of notation, we will denote by $\epsilon_L$ and $\epsilon_R$ also the corresponding embeddings of the Lie algebras, which are Lie algebra homomorphisms satisfying 
\begin{equation}
\begin{aligned}
\epsilon_{L}([B,C]) &= [\epsilon_{L}(B), \epsilon_{L}(C)] \ , \qquad \forall \, B, C \in \mathfrak h_{L} \ , \\
\epsilon_{R}([B,C]) &= [\epsilon_{R}(B), \epsilon_{R}(C)] \ , \qquad \forall \, B, C \in \mathfrak h_{R} \ . 
\end{aligned}
\end{equation}
Let us assume that both $G$ and $H$ are reductive, \ie~they can be decomposed into a product of simple and $U(1)$ factors. When gauging a subgroup $ H = H_1 \times \dots \times H_r$ of a group $G= G_1 \times \dots \times G_n$, the asymmetric gauged action takes the form \rcite{Guadagnini:1987ty, Guadagnini:1987qc, Witten:1991mm, Bars:1991pt, Quella:2002fk}
\begin{equation}
    S^B = \sum_{i = 1}^n \kappa_i^B S^B_i \ , 
    \label{asym-act}
\end{equation}
where the superscript $B$ denotes the bosonic part of the action
\begin{multline}
S^B_{i}  = S^\WZW(g_i) + \sum_{s=1}^r \int \frac{\text d^2v}{\pi} \text{Tr}_i \Bigl[ \epsilon_L^{i,s}(\mathcal A_{\bar v}^s) \, \partial_v g_i g_i^{-1} - \epsilon_R^{i,s}(\mathcal  A_v^s) \, g_i^{-1} \partial_{\bar v} g_i + \epsilon_L^{i,s}(\mathcal A_{\bar v}^s) \,  g_i \,  \epsilon_R^{i,s}(\mathcal  A_v^s) \, g_i^{-1} \\
-\tfrac{1}{2} \epsilon_L^{i,s}(\mathcal A_{\bar v}^s) \, \epsilon_L^{i,s}(\mathcal A_v^s) -\tfrac{1}{2} \epsilon_R^{i,s}(\mathcal A_{\bar v}^s) \, \epsilon_R^{i,s}(\mathcal A_v^s)  \Bigr] 
\label{asym-act-A}
\end{multline}
and we split the embeddings $\epsilon_L$ and $\epsilon_R$ respectively into
\begin{equation}
    \epsilon_L^{i,s} : H_s \hookrightarrow G_i \ , \qquad     \epsilon_R^{i,s} : H_s \hookrightarrow G_i \ , \qquad i = 1, \dots, n \ , \qquad s = 1, \dots, r \ . 
\end{equation}
In eq.~\eqref{asym-act}, $\kappa_i^B$ denotes the bosonic level of the $i$-th factor in the decomposition $G= G_1 \times \dots \times G_n$. For the gauging to be non-anomalous one should require that~\rcite{Bars:1991pt, Witten:1991mm, Figueroa-OFarrill:1994vwl, Quella:2002fk}
\begin{equation}
    \sum_{i=1}^n \kappa_i \, \text{Tr}_i\{ \epsilon_L^{s,i}(B) \, \epsilon_L^{s,i}(C) \} = \sum_{i=1}^n \kappa_i \, \text{Tr}_i\{ \epsilon_R^{s,i}(B) \, \epsilon_R^{s,i}(C) \} \ , \qquad \forall \, B, C \in \mathfrak{h}_s  \ . 
    \label{no-anomal}
\end{equation}
When considering bosonic GWZW models, $\kappa_i$ in \eqref{no-anomal} reads $\kappa_i = \kappa^B_i$, which is the level entering the bosonic action \eqref{asym-act}. We will introduce in a moment $\mathcal N = 1$ asymmetric GWZW and see that in this case, $\kappa_i = \kappa^B_i + h^\vee_i$, where $h^\vee_i$ is the dual Coxeter number of $\mathfrak{g}_i$.%
\footnote{More precisely, $\kappa_i = \kappa_i^B + h^\vee_i$ for each factor of $\mathfrak g$ containing charged fermions. This distinction will not play a role for us.} 
Using eq.~\eqref{no-anomal} one can show that the action \eqref{asym-act} is invariant under the infinitesimal gauge transformation
\begin{align}
\delta \mathcal A^s_v = i \partial_v \omega^s + i [\omega^s, \mathcal A^s_v] \ , \quad \delta \mathcal A^s_{\bar v} = i \partial_{\bar v} \omega^s + i [\omega^s, \mathcal A^s_{\bar v}] \ , \quad \delta g_i = i \epsilon_L(\omega_s) \, g_i - i g_i \, \epsilon_R(\omega_s)  \ .
\label{gauge-trsf}
\end{align}
In fact, the variation of the action reads \rcite{Quella:2002fk}
\begin{multline}
    \delta S^B = i \sum_{i=1}^n  \sum_{s=1}^r \frac{\kappa_i^B}{2 \pi} \int \text d^2 v \, \text{Tr}_i \Bigl[ \epsilon_L^{i,s}(\mathcal A^s_{\bar v}) \partial_v \epsilon_L^{i,s}(\omega_s)  - \epsilon_R^{i,s}(\mathcal A^s_{\bar v}) \partial_v \epsilon_R^{i,s}(\omega_s) \\
    - \epsilon_L^{i,s}(\mathcal A^s_v) \partial_{\bar v} \epsilon_L^{i,s}(\omega_s) + \epsilon_R^{i,s}(\mathcal A^s_v) \partial_{\bar v} \epsilon_R^{i,s}(\omega_s)   \Bigr] \ , 
    \label{variation-SB}
\end{multline}
which vanishes upon using eq.~\eqref{no-anomal} with $\kappa_i = \kappa_i^B$. Notice that the vector and axial gauged actions \eqref{vector-action} and \eqref{axial-action} can be recovered from eq.~\eqref{asym-act-A} by dropping the index $i$ and setting 
\begin{equation}
    \epsilon^{\rm vec}_L(\mathcal A_v) =  \epsilon^{\rm vec}_R(\mathcal A_v) = \mathcal A_v ~~ , \qquad \epsilon^{\rm vec}_L(\mathcal A_{\bar v}) = \epsilon^{\rm vec}_R(\mathcal A_{\bar v}) = \mathcal A_{\bar v} \ , 
\end{equation}
and 
\begin{equation}
\epsilon^{\rm ax}_L(\mathcal A_v) = - \epsilon^{\rm ax}_R(\mathcal A_v) = -\mathcal A_v ~~ , \qquad \epsilon^{\rm ax}_L(\mathcal A_{\bar v}) = -\epsilon^{\rm ax}_R(\mathcal A_{\bar v}) = -\mathcal A_{\bar v} \ , 
\end{equation}
respectively. 

In the following we will mainly consider gauged WZW models with $H =U(1)$ and hence $r=1$. We will thus often drop the $s$ label. Moreover, we will be interested in supersymmetric GWZW model, for which the asymmetrically gauged action~\eqref{asym-act} will be accompanied by a corresponding asymmetrically gauged action of left- and right-moving fermions transforming in the adjoint of $\mathfrak g$. The action of supersymmetric GWZW takes the schematic form
\begin{equation}
    S = S^B + S^F \ , \qquad \text{with} \qquad  S^B = \sum_{i=1}^n \kappa_i^B S_i^B \ , 
    \label{S=SB+SF}
\end{equation}
and 
\begin{equation}
    S^F = \frac{1}{2 \pi} \int \text d^2 v \, \text{Tr}\Bigl( \psi \, \partial_{\bar v} \psi - \medmath{\sum_s} \psi \, [\epsilon_L( \mathcal A_{\bar v}^s), \psi] + \tilde \psi \, \partial_v \tilde \psi -\medmath{\sum_s} \tilde \psi \, [\epsilon_R( \mathcal A_v^s), \tilde \psi] \Bigr) \ , 
    \label{psi-action-general}
\end{equation}
where we denoted by $\psi$ and $\tilde \psi$ respectively holomorphic and anti-holomorphic free fermions, taking values in the Lie algebra of $G$.  As we review in Appendix~\ref{app:GWZW in superspace}, the fermions in $\mathfrak h_L$ and $\mathfrak h_R$ decouple, resulting in an action of the same form, but with the fermions taking value respectively in $\mathfrak h_L^\perp$ and $\mathfrak h_R^\perp$. One can easily check that the fermionic action \eqref{psi-action-general} is classically invariant under the gauge transformations
\begin{equation}
\begin{aligned}
   \delta \mathcal A_v^s & = i \partial_v \omega^s + i [\omega^s, \mathcal A_v^s] \ , &  \delta \mathcal A_{\bar v}^s & = i \partial_{\bar v} \omega^s + i [\omega^s, \mathcal A_{\bar v}^s] \ , \\
   \delta \psi & = i [\epsilon_L(\omega^s), \psi] \ , &   \delta \tilde \psi & = i [\epsilon_R(\omega^s), \tilde \psi ]    \ .
\end{aligned}
 \label{psi-gauge-trsf-general}%
\end{equation}
While the action \eqref{psi-action-general} is gauge invariant, the path integral measure is not, 
\begin{equation}
\prod_A \prod_{\tilde A} \mathcal{D} \psi_A \, \mathcal{D} \tilde \psi_{\tilde A}  \to \prod_A \prod_{\tilde A} \mathcal{D} \psi_A \, \mathcal{D} \tilde \psi_{\tilde A}  \, e^{\mathfrak A} \ , 
\label{anomaly-measure-gauge}
\end{equation}
where we decomposed $\psi$ and $\tilde \psi$ over a basis of generators of $\mathfrak{h}_L^\perp$ and $\mathfrak{h}_R^\perp$ respectively and $A, \tilde A$ run over $A \in \{ 1 , \dots , \text{dim Adj} (\mathfrak{h}^\perp_L) \}$ and $\tilde A \in \{ 1 , \dots , \text{dim Adj}(\mathfrak{h}^\perp_R) \} $. 
In all the examples we are going to encounter, the anomaly $\mathfrak A$ in \eqref{anomaly-measure-gauge} cancels against the variation \eqref{variation-SB} of the bosonic action, 
\begin{equation}
    - \delta S^B + \mathfrak{A} = 0 \ , 
\end{equation}
provided eq.~\eqref{no-anomal} is obeyed with $\kappa_i = \kappa_i^B + h^\vee_i$, see also \rcite{Tseytlin:1992ri, Johnson:1994kv}.


\subsection{Null gauging} 
\label{sec:null}

Let us now consider a special instance of asymmetric GWZW models: null GWZW models~\rcite{Klimcik:1994wp,Israel:2004ir,Martinec:2017ztd}. The gauging is said to be null when each side of eq.~\eqref{no-anomal} independently vanishes, 
\begin{equation}
    \sum_{i=1}^n \kappa_i \, \text{Tr}_i\{ \epsilon_L^{s,i}(B) \, \epsilon_L^{s,i}(C) \} = \sum_{i=1}^n \kappa_i \, \text{Tr}_i\{ \epsilon_R^{s,i}(B) \, \epsilon_R^{s,i}(C) \}  = 0 \ , \qquad \forall \, B, C \in \mathfrak{h}_s \ . 
    \label{null-condition-embeddings}
\end{equation}
In this case, the action \eqref{S=SB+SF} is invariant under the additional symmetry
\begin{equation}
\begin{aligned}
 \delta g_i & = -i \epsilon_L(\check\omega_s) \, g_i - i g_i \, \epsilon_R(\check\omega_s)  \ , & &\\
\delta \mathcal A^s_v & = i \partial_v \check\omega_s + i [\check\omega_s, \mathcal A^s_v] \ ,  & \delta \mathcal A^s_{\bar v} & = -i \partial_{\bar v} \check\omega_s - i [\check\omega_s, \mathcal A^s_{\bar v}] \ , \\ 
\delta \psi & = -i [\epsilon_L(\check \omega_s), \psi ] \ ,  &   \delta \tilde \psi & =i [\epsilon_R(\check \omega_s), \tilde \psi ]    \ , 
\end{aligned}
 \label{second-gauge-sym}%
\end{equation}
which slightly differs from the gauge transformations~\eqref{gauge-trsf} and \eqref{psi-gauge-trsf-general} by reversing the direction of the null gauge orbits on the right relative to the null gauge orbits on the left. Similarly to the gauge invariance we discussed in the previous section, the variation of the bosonic action under \eqref{second-gauge-sym} reads
\begin{multline}
    \check \delta S^B = -i \sum_{i=1}^n  \sum_{s=1}^r \frac{\kappa_i^B}{\pi} \int \text d^2 v \, \text{Tr}_i \Bigl[ \epsilon_L^{i,s}(\mathcal A^s_{\bar v}) \partial_v \epsilon_L^{i,s}(\check\omega_s)  + \epsilon_R^{i,s}(\mathcal A^s_{\bar v}) \partial_v \epsilon_R^{i,s}(\check\omega_s) \\
    - \partial_{\bar v} \epsilon_L^{i,s}(\check\omega_s) \epsilon_L^{i,s}(\mathcal A^s_v) - \epsilon_R^{i,s}(\mathcal A^s_v) \partial_{\bar v} \epsilon_R^{i,s}(\check\omega_s)   \Bigr] \ , 
    \label{dS-null}
\end{multline}
while the fermionic action gives rise to the anomaly factor 
\begin{equation}
\prod_A \prod_{\tilde A} \mathcal{D} \psi_A \, \mathcal{D} \tilde \psi_{\tilde A}  \to \prod_A \prod_{\tilde A} \mathcal{D} \psi_A \, \mathcal{D} \tilde \psi_{\tilde A}  \, e^{\check{\mathfrak A}} \ . 
\label{anomaly-null}
\end{equation}
In all the examples of null gauging we consider, the anomaly \eqref{anomaly-null} cancels against the variation \eqref{dS-null},
\begin{equation}
    - \check \delta S^B + \check{\mathfrak A} = 0 \ , 
\end{equation}
provided eq.~\eqref{null-condition-embeddings} is obeyed with $\kappa_i = \kappa_i^B + h^\vee_i$. 

When the gauging is null, the fermions entering the action \eqref{psi-action-general} take values in the ``transverse space'' ${\mathfrak g}_L^{\ttt}$ (respectively ${\mathfrak g}_R^{\ttt}$), which is a subspace of $\mathfrak h_L^\perp$ (respectively $\mathfrak h_R^\perp$), see eqs.~\eqref{g-null-decomposition} and \eqref{compos-ss-wzw-action-4-null}. This is simply a consequence of the fact that when the gauging is null, $\mathfrak h_L \subset \mathfrak h_L^\perp$ and $\mathfrak h_R \subset \mathfrak h_R^\perp$. 
We show in Appendix~\ref{app:GWZW in superspace} that one can also remove the fermions along $\mfh_L$ and $\mfh_R$ by a fermionic gauge transformation.
It follows that the labels $A$ and $\tilde A$ in eq.~\eqref{anomaly-null} run over the remaining directions $A \in \{ 1 , \dots , \text{dim Adj} ({\mathfrak g}_L^{\ttt})\}$ and $\tilde A \in \{ 1 , \dots , \text{dim Adj}({\mathfrak g}_R^{\ttt}) \} $. We refer to Appendix~\ref{app:GWZW in superspace} for a detailed derivation of this statement and for a discussion on its relation with the usual BRST decoupling of fermions in terms of $\beta \gamma$ superghosts. 


\subsection{Parametrization of the gauge field} 
\label{sec:parametrization-gauge-field}

In the following sections we are going to compute partition functions of coset CFT's by evaluating the path integral of the GWZW model. Let us review how gauge fields are usually parameterized in the literature \cite{Karabali:1988au, Gawedzki:1988hq, Gawedzki:1988nj, Karabali:1989dk, Gawedzki:1991yu, Hanany:2002ev, Eguchi:2010cb} and spell out our conventions. We start by defining the worldsheet torus $\Sigma$ with modular parameter $\tau$ by 
\begin{equation}
    v = \sigma^1 + \tau \sigma^2 \ ,   \qquad     \bar v = \sigma^1 + \bar \tau \sigma^2 \ ,
    \label{torus-coordinates}
\end{equation}
where $\sigma^1$ and $\sigma^2$ parameterize $\alpha$ and $\beta$ cycles and range over 
\begin{equation}
    \sigma^1 \in [0, 2 \pi) \ , \qquad     \sigma^2 \in [0, 2 \pi) \ . 
\end{equation}
The embeddings $\epsilon_{L}^i(\mathcal A_v) \in \mathfrak{h}_L$ and $\epsilon_{R}^i(\mathcal A_v) \in \mathfrak{h}_R$ can be decomposed in terms of the generators $T^i_{L,a} \,, T^i_{R,a}$ as  
\begin{equation}
\begin{aligned}
    \epsilon_L^i(\mathcal A_v) &= \sum_a T^i_{L,a} \, A^{a}_v ~~ , &    \epsilon_R^i(\mathcal A_v) &= \sum_a T^i_{R,a} \, A^{a}_v \ , \\     \epsilon_L^i(\mathcal A_{\bar v}) &= \sum_a T^i_{L,a} \, A^{a}_{\bar v} ~~ , &     \epsilon_R^i(\mathcal A_{\bar v}) &= \sum_a T^i_{R,a} \, A^{a}_{\bar v} \ , 
\end{aligned}%
    \label{embeddings-decomp}%
\end{equation}
where $A^{a}_v$ and $A^{a}_{\bar v}$ are fields in the worldsheet coordinates $v, \bar v$. Since our interest here is $H=U(1)$, in the following we will simplify the notation and write the embedding in terms of a single field $A_v$ and constant matrices $\varepsilon_{L}^i$, $\varepsilon_{R}^i$ as 
\begin{equation}
    \epsilon_L^i(\mathcal A_v) = \varepsilon_L^i A_v \ , \qquad     \epsilon_R^i(\mathcal A_v) = \varepsilon_R^i A_v \ , \qquad  \epsilon_L^i(\mathcal A_{\bar v}) = \varepsilon_L^i A_{\bar v} \ , \qquad     \epsilon_R^i(\mathcal A_{\bar v}) = \varepsilon_R^i A_{\bar v} \ .
    \label{embeddings-special}
\end{equation}
Using the Helmholtz-Hodge decomposition, the real gauge fields $A_0$ and $A_1$ can be parameterized as 
\begin{equation}
\begin{aligned}
    A_0 & = \partial_0 X - \partial_1 Y - \frac{u_1}{\tau_2} \ , \\
    A_1 & = \partial_1 X + \partial_0 Y - \frac{u_2}{\tau_2} \ , 
\end{aligned}
\label{Hodge-decomp-01}%
\end{equation}
where $X$ is a real non-compact boson, and $Y$ is a real compact boson with unit radius, 
\begin{equation}
    Y(\sigma_1 + 2 \pi, \sigma_2) = -2 \pi m_1 \ , \qquad      Y(\sigma_1, \sigma_2+ 2 \pi) = 2 \pi m_2 \ , \qquad \text{with} \qquad m_1, m_2 \in \mathds{Z} \ . 
\end{equation}
The Jacobian in the functional measure for the change of variables~\eqref{Hodge-decomp-01} leads as usual to a functional determinant that is represented by a set of Faddeev-Popov ghosts $b,c$ and $\tilde b, \tilde c$.

In the following, it will prove useful to introduce the notation 
\begin{equation}
        u = u_1 + i u_2  = s_1 \tau + s_2 \ , 
        \label{u-def}
\end{equation}
in terms of which the holonomies of the gauge field read 
\begin{equation}
    \int_0^{2 \pi} \text d \sigma^1 \, A_{\sigma^1} = - \frac{2 \pi \, \tau_1}{\tau_2}(m_1+s_1)  \ , \qquad \quad 
        \int_0^{2 \pi} \text d \sigma^2 \, A_{\sigma^2}  = - \frac{2 \pi \, \tau_1}{\tau_2}(m_2+s_2)  ~.
\end{equation}
In order to allow arbitrary real values for the gauge holonomies, since $m_1, m_2 \in \mathds Z$, we should require $s_1$ and $s_2$  to range over
\begin{equation}
   0 \leq s_1 \leq 1  \ ,  \qquad 0 \leq s_2 \leq 1 \ . 
\end{equation}
In terms of the holomorphic coordinates \eqref{torus-coordinates}, eq.~\eqref{Hodge-decomp-01} reads  
\begin{equation}    
\begin{aligned}
    A_v & = \partial_v X - i \partial_v Y - \frac{\bar u}{2 \tau_2} = \partial_v X - i \partial_v Y - i \partial_v \Phi[u]  \ , \\
    A_{\bar v} & = \partial_{\bar v} X + i \partial_{\bar v} Y - \frac{u}{2 \tau_2} = \partial_{\bar v} X + i \partial_{\bar v} Y + i \partial_{\bar v} \Phi[u] \ , 
\end{aligned}
\label{Hodge-decomp-v}%
\end{equation}
where we introduced the notation
\begin{equation} 
    \Phi[u] = -\sigma_1 s_1 + \sigma_2 s_2 = \frac{i}{2 \tau_2}(\bar v u - v \bar u) \ .
    \label{Phi[u]}
\end{equation}
In the literature, eq.~\eqref{Hodge-decomp-v} is often rewritten as 
\begin{equation}
    A_v  = \partial_v \text{a}^{\dagger} \,  \text a^{\dagger -1} \ , \qquad A_{\bar v} = \partial_{\bar v} \text a \, \text a^{-1}  \ , 
\end{equation}
where
\begin{equation}
    \text a = \Omega \, a \ , \qquad \text{with} \qquad \Omega = e^{X+iY} \ , \qquad  a = e^{i \Phi[u]} \ . 
\end{equation}
The advantage of this parametrization of gauge fields is that using eq.~\eqref{embeddings-special}, the asymmetric gauged action \eqref{asym-act-A} can be written as 
\begin{multline}
S^B_i(g_i, \text a, \text a^\dagger) = S^\WZW(g_i) - \frac{1}{\pi} \int \text d^2v \, \text{Tr}\Biggl[ -\varepsilon_L^i \, \partial_{\bar v} \text a \,  \text a^{-1} \, \partial_v g_i g_i^{-1} + \varepsilon_R^i \, \partial_v \text a^\dagger \, \text a^{\dagger-1} \, g_i^{-1} \partial_{\bar v} g_i  \\
- \varepsilon_L^i \, \partial_{\bar v} \text a \, \text a^{-1} \,  g_i \, \varepsilon_R^i \, \partial_{v} \text a^\dagger \, \text a^{\dagger-1} \, g_i^{-1} +\frac{\varepsilon_L^i}{2} \, \partial_{\bar v} \text a \, \text a^{-1} \, \varepsilon_L^i \,  \partial_v \text a^\dagger \, \text a^{\dagger-1} +\frac{\varepsilon_R^i}{2} \, \partial_{\bar v} \text a \, \text a^{-1} \, \varepsilon_R^i \, \partial_v \text a^\dagger \, \text a^{\dagger -1}  \Biggr] \ . 
\label{asym-act-a-long}
\end{multline}
One can check that using the Polyakov-Wiegmann identity, eq.~\eqref{asym-act-a-long} takes the concise form 
\begin{multline}
S^B_i(g_i, \text a, \text a^\dagger)  = S^\WZW(\text a^{-\varepsilon^i_L} g_i (\text a^\dagger)^{\varepsilon^i_R}) - S^\WZW(\text{a}^{-\varepsilon^i_L} (\text a^{\dagger-1})^{\varepsilon^i_R}) \\
- \frac{1}{2 \pi} \int \text d^2 v \, \text{Tr} [(\varepsilon^i_L+\varepsilon^i_R)^2] \, \text a^{-1} \partial_{\bar v} \text a \, \partial_v \text a^{\dagger} \text a^{\dagger -1} \ , 
\label{asym-act-a}
\end{multline}
where we adopted the notation
\begin{equation}
    \text a^{\varepsilon^i_L} = e^{(X + i Y + i\Phi[u])\varepsilon^i_L} \ ,  \qquad     (\text a^\dagger)^{\varepsilon^i_R} = e^{(X - i Y - i\Phi[u])\varepsilon^i_R} \ , \qquad (\text a^{\dagger-1})^{\varepsilon^i_R} = e^{(-X + i Y + i\Phi[u])\varepsilon^i_R} \ ,
\end{equation}
and so on. We will soon see that writing the asymmetric gauged action as in \eqref{asym-act-a} makes it easier to decouple the group elements $g_i$ from the free bosons $X$ and $Y$, which enter the parametrization of the gauge field~\eqref{Hodge-decomp-v}.

We already mentioned that the free fermions entering the action \eqref{psi-action-general} take value in the ``transverse space'' ${\mathfrak g}_L^{\ttt}$ (respectively ${\mathfrak g}_R^{\ttt}$). The gauge field only couples to charged fermions, so we can split $\mfg^\ttt$ into the action for fermions in the charged subspace, which will involve the gauge field, and the remaining uncharged fermions, whose action is that of a collection of free fermions and whose partition function is trivial.  The action of the fermions is
\begin{multline}
     S^F = \int  \frac{\text d^2 v}{4 \pi} \Bigl( \psi^A \bigl(\delta^{AB}\partial_{\bar v} - 2\text{Tr}(T^A [\varepsilon_L, T^B]) \, A_{\bar v} \bigr) \psi^B  +  \tilde \psi^A \bigl( \delta^{AB}\partial_v - 2\text{Tr}(\tilde T^A [\varepsilon_R, \tilde T^B]) \, A_{\bar v}\bigr) \tilde \psi^B \Bigr) \ , 
     \label{psi-action-ab}
\end{multline}
where $A,B$ run over the directions in $\mfg^\ttt$.  Only charged fermions couple to the gauge field.  We will suppress explicit mention of the remaining uncharged fermions in the following, while remembering to include their partition function in the final result.

\paragraph{Parametrization of the gauge field for null gauging} When the gauging is null, an alternative approach~\rcite{Martinec:2020gkv} is to gauge the null currents of $U(1)_L$ and $U(1)_R$ using separate gauge fields; one finds that half the components of the two gauge fields simply drop out of the action.  The action formally looks like that of the single gauge field formalism we employ here, however the treatment of the zero modes is different~-- in the Euclidean formalism, the zero modes of $\mathcal A_v$ and $\mathcal A_{\bar v}$ here are complex conjugates of one another, whereas in the treatment of~\rcite{Martinec:2020gkv}, $A$ and $\bar A$ come from different gauge fields and so their zero modes a priori have nothing to do with one another.  As we will see, it is the approach we use here using null gauging with a single gauge field that reproduces known results in the literature. In fact, as we are going to see in the following sections, each gauge field zero mode gives rise to \emph{two} coset constraints of the schematic form $\mathcal J_0= \bar{\mathcal J}_0=0$, 
where $\mathcal J_0$ and $\bar{\mathcal J}_0$ are the zero modes of respectively the holomorphic and anti-holomorphic currents to be gauged. Treating the gauge fields as in~\rcite{Martinec:2020gkv} one would then end up with twice as many constraints as the decomposition of the partition function into characters would suggest.


\subsection{Decoupling}
\label{sec:decoupling}

In the following, we will be interested in evaluating path integrals of the form 
\begin{equation}
    Z^{\frac{\mathfrak g}{\mathfrak h}} = \int \mathcal D A_v \, \mathcal D A_{\bar v} \, \prod_i \int \mathcal D g_i \prod_{A, \tilde A} \int \mathcal D \psi_A \, \mathcal D \tilde \psi_{\tilde A} \, e^{-S} \ . 
    \label{Z-general}
\end{equation}
In terms of the parametrization of the gauge field introduced in the previous section we can rewrite eq.~\eqref{Z-general} as
\begin{equation}
    Z^{\frac{\mathfrak g}{\mathfrak h}}  = \int_0^1 \text ds_1 \int_0^1 \text ds_2 \int \mathcal D X \, \mathcal D Y \int \mathcal D b \, \mathcal D \tilde b \, \mathcal D c \, \mathcal D \tilde c \, \prod_i \int \mathcal D g_i \prod_{A, \tilde A} \int \mathcal D \psi_A \, \mathcal D \tilde \psi_{\tilde A}  \, e^{-S^B-S^F-S_{gh}} \ ,  
    \label{rough-path-int}
\end{equation}
where we introduced the usual $b \, c$ ghost action $S_{gh}$, arising from the path integral Jacobian of the transformation $\mathcal{\mathcal D} A_v \mathcal D A_{\bar v} \to \mathcal D X \mathcal D Y$. Assuming invariance of the path integral measure~\rcite{Gawedzki:1988nj},
\begin{equation}
    \mathcal D g_i = \mathcal D (\Omega^{\varepsilon_L^i} \, g_i \, \Omega^{\dagger-\varepsilon_R^i} ) \ , \qquad \text{with} \qquad \Omega^{\varepsilon_L^i} \equiv e^{\varepsilon_L^i(X + i Y)} \ , \qquad \Omega^{\dagger-\varepsilon_R^i} \equiv e^{-\varepsilon_R^i(X - i Y)} \ , 
\end{equation}
and using eq.~\eqref{asym-act-a} we can replace the action $S^B$ in the path integral \eqref{rough-path-int} with
\begin{equation}
   S^{B,d} = \sum_{i = 1}^n \kappa_i^B S^B_i(\Omega^{\varepsilon_L^i}g_i \Omega^{\dagger -\varepsilon_R^i}, \text a, \text a^\dagger) = \sum_{i=1}^n (\kappa_i^B S_{g_i} + \kappa_i^B S^{X,Y}_i )  \ , 
   \label{S-decoupled}
\end{equation}
where
\begin{subequations}
\label{decoupling}%
\begin{align}
S_{g_i} &= S^\WZW(a^{-\varepsilon_L^i} \, g_i \, a^{\dagger \varepsilon_R^i}) \ , \\
S^{X,Y}_i & = - S^\WZW(a^{-\varepsilon_L^i} \, \Omega^{-\varepsilon_L^i} \, \Omega^{\dagger -\varepsilon_R^i} \, a^{\dagger -\varepsilon_R^i})  - \int \frac{\text d^2v}{2\pi} \text{Tr}[(\varepsilon_L^i+\varepsilon_R^i)^2] \, | a^{-1} \partial_{\bar v} a + \Omega^{-1} \partial_{\bar v} \Omega|^2 \ .
\label{decoupling-b}
\end{align}
\end{subequations}
Notice that in the action \eqref{S-decoupled} the gauge degrees of freedom $X$ and $Y$ decoupled from the group valued fields $g_i$. This will greatly simplify the computation of partition functions in the following sections. 

In a similar spirit, let us see how also the fermions can be decoupled from the gauge degrees of freedom $X$ and $Y$. In all the examples we are going to consider, the embeddings take values in the Cartan subalgebra of $\mathfrak{g}$ and we can introduce the charges $q_a$ as
\begin{equation}
    [\varepsilon_L^m, T^a_m] = q_a \, \ell_m \,  T^a_m \ , \qquad      [\varepsilon_R^m, \tilde T^a_m] = \tilde q_a \, r_m \, \tilde T^a_m \ , 
\end{equation}
where $m$ runs over the factors of $\mfg$ containing charged fermions, which we denote by $\psi_m^a$; $\ell_m,r_m$ prescribe the embedding of $\mathfrak h$ into each factor, and $q_a,\tilde q_a$ are the weights of the generators $T^a_m,\tilde T^a_m$ under those Cartan generators, respectively.
Eq.~\eqref{psi-action-ab} thus reduces to
\begin{equation}
     S^F = \sum_m S^F_m = \sum_m   \int  \frac{\text d^2 v}{4 \pi} \Bigl( \delta^{ab} \psi_m^a ( \partial_{\bar v} - q_b \, \ell_m \, A_{\bar v} ) \psi_m^b + \delta^{ab} \tilde \psi_m^a ( \partial_v - \tilde q_b \, r_m \, A_v ) \tilde \psi_m^b \Bigr) \ , 
     \label{psi-action-qa}
\end{equation}
and is invariant under the gauge transformations
\begin{equation}
\begin{aligned}
\psi^a_m & \mapsto e^{q_a \, \ell_m \,  \gamma} \, \psi^a_m \ , & \qquad  \tilde \psi^a_m & \mapsto e^{\tilde q_a \, r_m \, \gamma} \, \tilde \psi^a_m \ , \label{psi-gauge-trsf-lr}\\
A_v & \mapsto A_v + \partial_v\gamma  \ , & \qquad A_{\bar v} & \mapsto A_{\bar v} + \partial_{\bar v} \gamma  \ . 
\end{aligned}
\end{equation}

Recall that the gauge fields $A_{v}$, $A_{\bar v}$ are parametrized as in eq.~\eqref{Hodge-decomp-v}. Performing a gauge transformation with parameter $\gamma= - X$, we can eliminate $X$ from the action \eqref{psi-action-qa} without producing any anomaly factor. We then introduce the fermions
\begin{equation}
\eta^a_m = e^{-i q_a \ell_m (Y+\Phi)} \psi^a_m  \ , \qquad  \tilde \eta^a_m = e^{i q_a r_m (Y+\Phi)} \tilde \psi^a_m \ . 
\label{twisted-fermions-lr}
\end{equation}
The fermion measure is not invariant and \eqref{twisted-fermions-lr} gives rise to an anomaly of the form \eqref{anomaly-null}, depending on $Y$, $\Phi$, $\ell_m$ and $r_m$, 
\begin{equation}
\prod_m \prod_{a \in \mfg^\ttt_m} \mathcal{D} \psi^a_m \, \mathcal{D} \tilde \psi^a_m  = \prod_m \prod_{a \in \mfg^\ttt_m} \mathcal{D} \eta^a_m \, \mathcal{D} \tilde \eta^a_m  \, e^{\check{\mathfrak A}} \ . 
\label{psi-eta-anomaly}
\end{equation}
The path integral over the fermions can then be rewritten as 
\begin{align}
Z^F & = \prod_m \prod_{a \in \mfg^\ttt_m} \int \mathcal D \psi_m^a \mathcal D \tilde \psi_m^a \, e^{-S^F} \\
& = e^{\check{\mathfrak A}} \prod_m \prod_{a \in \mfg^\ttt_m} \int \mathcal D \eta_m^a \, \mathcal D \tilde \eta_m^a \, \exp\left( \int  \frac{\text d^2 v}{4 \pi} \delta^{ab}\Bigl( \eta_m^a \partial_{\bar v}  \eta_m^b + \tilde \eta_m^a  \partial_v \tilde \eta_m^b \Bigr) \right) \ .  
\label{decoupled-fermion-path-integral}
\end{align}
Notice that the fermionic degrees of freedom $\eta_m^a$ are now decoupled from the gauge field components $X$ and $Y$. 

The path-integral \eqref{rough-path-int} thus factorizes as 
\begin{equation}
Z^{\frac{\mathfrak g}{\mathfrak h}}  = \int \mathcal D X \, \mathcal D Y \, \int_0^1 \text d s_1 \int_0^1 \text ds_2 \,  Z_{gh} \, \prod_i  \, Z^{B}_i \prod_ m \, Z^{F}_m  \ , 
\label{Z-tot-general} 
\end{equation}
where the product over $i$ runs over the various group factors $G_i$ of $G$, while the product over $m$ runs over the various factors of $\mfg$ with charged fermions, and we introduced the notation
\begin{equation}
    Z^{B}_i =  e^{-S^{X,Y}_i} \int \mathcal D g_i \,  e^{-S_{g_i}} \ , \qquad   Z^{F}_m  = \prod_{a \in \mfg^\ttt_m} \int \mathcal D \psi_m^a \, \mathcal D \tilde \psi_m^a  \, e^{-S_m^F} \ . 
    \label{ZB-ZF-i}
\end{equation}
Finally, 
\begin{equation}
    Z_{gh} = \int \mathcal D b \, \mathcal D \tilde b \, \mathcal D c \, \mathcal D \tilde c \, e^{-S_{gh}} = \tau_2 |\eta(\tau)|^4 \ . \label{ghost-pf}
\end{equation}

\subsection{Simple gauged WZW models as null gauging} 
\label{sec:standard=null}

In the following sections we will consider various examples of WZW models $G = G_L \times G_R$ and gauge a null subgroup $H = H_L \times H_R$ with $H_L = U(1)$ and $H_R = U(1)$. The corresponding coset theories are usually denoted as 
\begin{equation}
   \frac{G}{H} = \frac{G}{U(1)_L \times U(1)_R} \ . 
    \label{G/U(1)LR}
\end{equation}
In all the examples we are going to encounter, the numerator group $G$ contains a non-compact direction, directly involved in the gauging. As a consequence, while the notation \eqref{G/U(1)LR} is standard in the literature, strictly speaking the group to be gauged is not $H = U(1)_L \times U(1)_R$, but rather $H = \mathds R \times U(1)$. See \rcite{Martinec:2017ztd, Martinec:2018nco, Martinec:2019wzw, Martinec:2020gkv} for a discussion on this point.  

We are going to investigate whether standard GWZW models can be equivalently recast in terms of associated \emph{null} GWZW models. 
As anticipated in the Introduction, when the gauging of the WZW model is null, the presence of the additional symmetry \eqref{second-gauge-sym} suggests that the standard GWZW $G/H$ can schematically be recast as
\begin{equation}
    \frac{G}{H} \simeq \frac{G \times \tilde H}{U(1)_L \times U(1)_R} \ , 
    \label{standard-null-GWZW}
\end{equation}
where $U(1)_L$ and $U(1)_R$ are null isometries involving $\tilde H$. One of our goals is to make eq.~\eqref{standard-null-GWZW} precise in various examples and check whether the partition functions of the models in the LHS and RHS of \eqref{standard-null-GWZW} agree. This equivalence is not obvious since the naive argument discussed above and leading to eq.~\eqref{standard-null-GWZW} does not take into account gauge field zero modes. We begin in Section \ref{sec:su2/u1} by showing that the partition function of the $\mathcal N=2$ supersymmetric coset
\begin{equation}
    \frac{SU(2)}{U(1)_{V,A}}
    \label{su2/u1-va}
\end{equation}
agrees with the partition function of the null gauged model 
\begin{equation}
    \frac{\sutwo\times \mathds R}{\uone_L \times \uone_R} \ . 
    \label{su2xR-null}
\end{equation}
In \eqref{su2/u1-va} the subscript $V,A$ stands for ``vector'' or ``axial''. The distinction between these two options here is not important as the partition functions of these two models agree \rcite{Kac:1984mq, Fateev:1985mm, Gepner:1986hr, Huitu:1990xv}. Beyond \eqref{su2xR-null} having an extra factor in the numerator group, the main a priori difference between the GWZW models \eqref{su2/u1-va} and \eqref{su2xR-null} is the following. While in eq.~\eqref{su2/u1-va} the non-gauge part of $A_v$ and $A_{\bar v}$ is a physical field $Y$, see eq.~\eqref{Hodge-decomp-v}, in eq.~\eqref{su2xR-null} all of the gauge field is pure gauge, apart from zero modes. Although it is somewhat non-trivial that the two are the same, to some extent we are just relabelling the $Y$ field from parametrizing the gauge field to parametrizing an extra component of the upstairs group. Similar statements may be formulated for the putative equivalence 
\begin{equation}
    \frac{SL(2, \mathds R)}{U(1)_{A}} \simeq \frac{SL(2, \mathds R) \times U(1)}{\uone_L \times \uone_R} \ , 
\end{equation}
which is analyzed in Section \ref{sec:sl2/u1}. Things become more nontrivial when we get to product groups like $G = \sltwo\times\sutwo$ and beyond. In fact, in \cite{Israel:2004ir} equivalence of the supersymmetric GWZW models
\begin{equation}
    \left( \frac{SL(2, \mathds R)_{n_5}}{U(1)_A} \times \frac{SU(2)_{n_5}}{U(1)_V} \right) \!\bigg/ \mathds Z_{n_5} \simeq \frac{SL(2, \mathds R)_{n_5} \times SU(2)_{n_5}}{U(1)_L \times U(1)_R}
    \label{cosetorb-equiv}
\end{equation}
was argued.  
This equivalence has been confirmed by a more detailed study of the spectrum in \rcite{Martinec:2018nco}. In Section~\ref{sec:cosetorb}, we compute the partition function of the null gauged model in the RHS of eq.~\eqref{cosetorb-equiv} and show it indeed reproduces the partition function of the coset orbifold on the~LHS.


\section{Warmup: \texorpdfstring{$\boldsymbol{\frac{\sutwo}{\uone}}$}{} } 
\label{sec:su2/u1}

In preparation for Sections \ref{sec:cosetorb} and \ref{sec:supertube}, let us warm up with one of the best known coset CFT's, the $\mathcal N=2$ supersymmetric GWZW model
\begin{equation}
    \frac{SU(2)_{\kappa_{su}}}{U(1)} \ . 
    \label{su2/u1-coset-intro}
\end{equation}
The distinction between axial and vector gauging is not important here, since the partition function does not depend on this choice. The partition function is well-known \rcite{Kac:1984mq, Fateev:1985mm, Gepner:1986hr, Huitu:1990xv} and for the reader's convenience it is reviewed in Appendix \ref{app:su(2)}. As is typical of supersymmetric theories, the partition function depends on whether we are working in the NS or R sector and on whether or not we consider insertions of $(-1)^{F_L+ F_R}$ operators, where $F_L$ and $F_R$ count the number of worldsheet fermions in the holomorphic and anti-holomorphic sector respectively. To be concrete, we will consider the partition function 
\begin{equation}
\mathcal Z^{\frac{\mathfrak{su}(2)}{\mathfrak{u}(1)}}(\tau, z) = \text{Tr}_\text{R}[(-1)^{F_L + F_R} q^{L_0-\frac{c}{24}}\,  \bar q^{\bar L_0-\frac{c}{24}} \, e^{2 \pi i z (J^{\cR}_{su})_0} \,  e^{-2 \pi i \bar z \bar (J^{\cR}_{su})_0}] \ , 
\label{trace-partfn-su2}
\end{equation}
with the trace taken over the worldsheet R-R sector. The computation we are going to present is completely analogous for the other sectors. The current $J^{\cR}_{su} + \bar J^{\cR}_{su}$ entering eq.~\eqref{trace-partfn-su2} is the $\cR$-current of the $\mathcal N=2$ model \eqref{su2/u1-coset-intro}. The $\cR$-current $J^{\cR}_{su}$ and the $\mathcal N=1$ Cartan $J^3_{su}$ of the $\mathfrak{su}(2)_{\kappa_{su}}$ WZW model can be written as the sum of the bosonic and fermionic Cartans $j^3_{su}$ and $j^3_{f,su}$, 
\begin{equation}
    J^3_{su} = j^3_{su} + j^3_{f,su} \ , \qquad J^{\cR}_{su} = -\frac{2}{\kappa_{su}}j^3_{su} + \frac{\kappa_{su}-2}{\kappa_{su}} j^3_{f,su}  \ , 
    \label{JBFR-rel-su}
\end{equation}
where the bosonic level $\kappa_{su}^B$ is related to the supersymmetric level $\kappa_{su}$ as
\begin{equation}
    \kappa_{su} = \kappa^B_{su} + 2 \ . 
\end{equation}
We associate chemical potentials to each current according to Table~\ref{tab:chemical-potentials} and similarly for the anti-holomorphic sector. 
\begin{table}
\centering
\begin{tabular}{c|c|c|c|c}
Current & $j^3_{su}$ & $j^3_{f,su}$ & $J^3_{su}$ & $J^\cR_{su}$ \\[0.1cm]
\hline
Chemical potential & $z_B$ &  $z_F$ & $u$ & $z$ \\
\end{tabular}
\caption{We list various currents together with the associated chemical potentials.}
\label{tab:chemical-potentials}
\end{table}
Requiring that 
\begin{equation}
    z \, J^{\cR}_{su} + u \, J^3_{su} = z_B \, j^3_{su} + z_F \, j^3_{f,su} \ , 
    \label{su(2)-ch-potentials}
\end{equation}
we find that the chemical potentials are related by
\begin{equation}
z_{B} = u - \frac{2}{\kappa_{su}}z \ , \qquad z_{F} =  u + \frac{\kappa_{su}-2}{\kappa_{su}} z \ . 
\label{su2-chemical-potentials-relation}
\end{equation}

Following the discussion in Section~\ref{sec:standard=null}, we are going to show that the partition function of the coset theory~\eqref{su2/u1-coset-intro} exactly matches the one derived from the null gauged coset\footnote{A similar computation has been briefly sketched in \rcite{Gawedzki:1991yu} for the bosonic parafermion theory. }
\begin{equation}
\frac{SU(2) \times {\mathds R}_t}{U(1)_L \times U(1)_R} \ , 
\label{su2xRt-coset}
\end{equation}
where $\mathds R_t$ is a non-compact time-like free boson while $U(1)_L$ and $U(1)_R$ are parameterized by the embeddings 
\begin{equation}
\varepsilon_L^{su} = \varepsilon_R^{su} = -\frac{\sigma_3}{2} \ , \qquad \varepsilon_L^{t} = -\varepsilon_R^{t} = -\frac{\sqrt{\kappa_{su}}}{2} \ . 
\label{SU(2)-U(1)-axial-embeddings}
\end{equation}
As anticipated in Section~\ref{sec:standard=null}, the non-compact direction $\mathds R_t$ enters the gauging in eq.~\eqref{su2xRt-coset} and the denominator should be more appropriately described as $U(1) \times \mathds R$. Formally, we will model the $\mathds R_t$ time-like free boson as a WZW model at level $\kappa_t = -2$ with group element 
\begin{equation}
    g_t = e^{i t} 
\end{equation}
and OPE
\begin{equation}
    \partial_v t(v_1)  \, \partial_v t(v_2) = \frac{1}{2 \, (v_1 - v_2)^2} \ .  
\end{equation}
Notice that the gauging \eqref{SU(2)-U(1)-axial-embeddings} is null, 
\begin{equation}
\sum_{i} \kappa_i \text{Tr}[(\varepsilon_L^i)^2] = \sum_{i} \kappa_i \text{Tr}[(\varepsilon_R^i)^2] = \kappa_{su} \, \text{Tr}\left[\left(\frac{\sigma_3}{2}\right)^2\right] - 2 \left(\tfrac{\sqrt{\kappa_{su}}}{2} \right)^2 = 0 \ . 
\end{equation}

Since the chemical potentials $z_B$ and $z_F$, associated respectively to the currents $j^3_{su}$ and $j^3_{f,su}$, are independent and not related to one another, one should in principle have two gauge fields: one associated to the bosonic current $j^3_{su}$ and a second one associated to the fermionic current $j^3_{f,su}$. Having the same gauge field for \emph{both} the bosonic and the fermionic currents amounts to having a unique chemical potential $u$ associated to $J^3_{su} = j^3_{su} + j^3_{f,su}$ and no chemical potential associated to the orthogonal current $J^{\cR}_{su}$. This is implicitly what we did at the level of the general discussion of Section~\ref{sec:GWZW}. On the other hand, here we are interested in keeping track of the $\cR$-current quantum numbers, associated to the chemical potential $z$. Following \cite{Eguchi:2010cb, Eguchi:2004yi}, we will do so by evaluating the bosonic and fermionic $SU(2)$ partition functions \eqref{ZB-ZF-i} respectively at the bosonic and fermionic chemical potentials $z_B$ and $z_F$. Using eq.~\eqref{su2-chemical-potentials-relation}, it is then easy to recover the dependence of the partition function \eqref{Z-tot-general} on the chemical potentials $z$ and $u$. The partition function of the $\mathcal N=2$ coset \eqref{su2/u1-coset-intro} thus reads,\footnote{We will see momentarily that the fields $X$ and $Y$ decouple completely. We then omit the infinite volume factor arising from path integrating over these fields, \ie\ we ``factor out the gauge group". }
\begin{equation}
Z^{\frac{\mathfrak{su}(2)}{\mathfrak{u}(1)}}(\tau, z) = \int_0^1 \text d s_1 \int_0^1 \text ds_2 \, Z_{gh}(\tau) \, Z^{B}_{su}(\tau, z_B) \, Z^{B}_{t}(\tau, u) \,  Z^{F}_{su}(\tau, z_F) \ . 
\label{su(2)-B-F-gh}
\end{equation}
As discussed in Section \ref{sec:decoupling} and explained in detail in Appendix~\ref{app:GWZW in superspace}, when the gauging is null, worldsheet fermions take value in the adjoint representation of the \emph{transverse} space $\mathfrak g^\ttt$ to the gauge direction in the Lie algebra, see eq.~\eqref{compos-ss-wzw-action-4-null}. In our case, this means only the fermions $\psi^+_{su}$ and $\psi^-_{su}$ (together with the anti-holomorphic fields $\tilde \psi^+_{su}$ and $\tilde \psi^-_{su}$) enter the partition function \eqref{su(2)-B-F-gh}. Let us now derive the various terms entering eq.~\eqref{su(2)-B-F-gh}. 

\subsection{Bosonic sector}

In this section we evaluate the bosonic contributions $Z^{B}_{su}(\tau, z_B)$ and $Z^{B}_{t}(\tau, u)$ entering eq.~\eqref{su(2)-B-F-gh}. We begin with $Z^{B}_{su}(\tau, z_B)$. Using eq.~\eqref{S(hgh)-su2} we find 
\begin{multline}
\int \mathcal D g_{su} \, \exp \left( - \kappa^B_{su} \, S_{g_{su}} \right) = \int \mathcal D g_{su} \, \exp \left( - \kappa^B_{su} \, S^\WZW \bigl( a[z_B]^{-\varepsilon_L^{su}} \, g_{su} \, a[z_B]^{\dagger \varepsilon_L^{su}}\bigr) \right) \\
= \exp \left( \frac{\pi \, (\kappa_{su}-2)}{4\tau_2} \, (z_B -\bar z_B)^2 \right) \mathcal Z_{B}^{\mathfrak{su}(2)}(\tau, z_B) \ ,  
\label{path-int-su2} 
\end{multline}
where $\mathcal Z_{B}^{\mathfrak{su}(2)}$ is the trace partition function of the bosonic $\mathfrak{su}(2)$ model at level $\kappa^B_{su} = \kappa_{su}-2$, see eq.~\eqref{SU2-pf}. For $S_{su}^{X,Y}$ we find
\begin{equation}
S_{su}^{X,Y} = - \frac{\kappa_{su}-2}{\pi} \int \text d^2 v \, \big|\partial_v(Y + \Phi[z_B])\big|^2 \ , 
\end{equation}
and hence using eq.~\eqref{ZB-ZF-i},
\begin{equation}
Z^{B}_{su}(\tau, z_B) = \exp \left( \mfrac{\pi \, (\kappa_{su}-2)}{4\tau_2} \, (z_B -\bar z_B)^2 + \mfrac{\kappa_{su}-2}{\pi} \int \text d^2 v \, \big|\partial_v(Y + \Phi[z_B])\big|^2 \right) \mathcal Z_{B}^{\mathfrak{su}(2)}(\tau, z_B) \ . 
\label{Z-B-su}
\end{equation}
Let us now consider $Z^{B}_t(\tau, u)$. The actions $S_{g_t}$ and $S^{X,Y}_t$ read, see eq.~\eqref{decoupling}
\begin{align}
S_{g_t} = \kappa_t \, S^\WZW \bigl( a[u]^{-\varepsilon_L^t} \, g_t \,  a[u]^{\dagger \varepsilon_R^t} \bigr) & = - \frac{1}{\pi} \int \text d^2 v \, |\partial_v t|^2 \ , \\
S^{X,Y}_t = - \kappa_t \, S^\WZW \bigl(a[u]^{-\varepsilon_L^t} \, \Omega^{-\varepsilon_L^t} \, \Omega^{\dagger -\varepsilon_R^t} \, a[u]^{\dagger -\varepsilon_R^t} \bigr) & = \frac{\kappa_{su}}{\pi} \int \text d^2v \, \big|\partial_v (Y + \Phi[u])\big|^2 \ , 
\end{align}
and hence 
\begin{equation}
Z^{B}_t(\tau, u) = \frac{1}{\sqrt{\tau_2} |\eta(\tau)|^2} \, \exp\left(-\frac{\kappa_{su}}{\pi} \int \text d^2v \,  \big|\partial_v (Y+\Phi[u])\big|^2  \right) \ . 
\end{equation}
Collecting the various terms, we find
\begin{multline}
Z^{B}_{su}(\tau, z_B) \, Z^{B}_t(\tau, u) = \frac{\exp \left( \frac{\pi \, (\kappa_{su}-2) }{4\tau_2} (z_B-\bar z_B)^2 \right)}{\sqrt{\tau_2} |\eta(\tau)|^2} \, \sum_{j=0}^{\kappa_{su}-2}  \chi_{j,B}^{\mathfrak{su}(2)}(\tau,z_B) \, \overline{\chi_{j,B}^{\mathfrak{su}(2)}(\tau,z_B)} \\
\times \, \exp\left(\frac{\kappa_{su}-2}{\pi} \int \text d^2v \,  |\partial_v (Y+\Phi[z_B])|^2  -\frac{\kappa_{su}}{\pi} \int \text d^2v \, \big|\partial_v (Y+\Phi[u])\big|^2 \right) \ , 
\label{SU(2)-U(1)-bosons-pf}
\end{multline}
where we used eq.~\eqref{SU2-pf} to express the $\mathfrak{su}(2)_{\kappa_{su}-2}$ trace partition function in terms of characters. 

\subsection{Fermionic sector}
\label{sec:su(2)/u(1)-fermions}

Following \rcite{Henningson:1993nr, Eguchi:2004yi, Eguchi:2010cb, Ashok:2011cy} let us explain how to compute the $SU(2)$ fermionic partition function%
\footnote{The reduction of the initial path integral involving the four fermions $\psi^\pm_{su}, \psi^3_{su} ,\psi^t$ to those spanning the transverse space to the gauge deformations (namely $\psi^\pm_{su}$) is discussed in Appendix~\ref{app:GWZW in superspace}.}
\begin{equation}
Z^F_{su}(\tau, z_F) = \int \mathcal{D} \psi^+_{su} \, \mathcal{D} \psi^-_{su} \, \mathcal{D} \tilde \psi^+_{su} \, \mathcal{D} \tilde \psi^-_{su} \, e^{-S_{su}^F} \ , 
\label{fermion-path-integral}
\end{equation}
where 
\begin{equation}
S_{su}^F =  \int \frac{\text d^2v}{2 \pi} \left[  \psi_{su}^+(\partial_{\bar v} +   A_{\bar v} )\psi^-_{su} + \psi^-_{su}(\partial_{\bar v} -  A_{\bar v})\psi^+_{su} + \tilde \psi^+_{su}(\partial_v + A_v ) \tilde \psi^-_{su} + \tilde \psi^-_{su}(\partial_v - A_v ) \tilde \psi^+_{su} \right] \ . 
\label{SU(2)-fermionic-action} 
\end{equation}
The action \eqref{SU(2)-fermionic-action} is gauge invariant under 
\begin{equation}
\begin{aligned}
\psi^\pm_{su}  \mapsto e^{\pm \gamma} \psi^\pm_{su} \quad &,\qquad 
A_v  \mapsto A_v + \partial_v \gamma  ~~ , & 
\\[.2cm]
\tilde \psi^\pm_{su}  \mapsto e^{\pm \gamma} \tilde \psi^\pm_{su} \quad &,\qquad 
A_{\bar v}  \mapsto A_{\bar v} + \partial_{\bar v} \gamma ~~ . 
\end{aligned}
\end{equation}
In the parametrization \eqref{Hodge-decomp-v} of the gauge field, we can eliminate $X$ from \eqref{SU(2)-fermionic-action} by a gauge transformation with gauge parameter $\gamma = -X$. Let us now introduce the fermions
\begin{equation}
\begin{aligned}
\eta^+_{su}  = e^{-iY-i \Phi} \psi^+_{su} \quad &,\qquad 
\tilde \eta^+_{su}  = e^{+iY+i \Phi} \tilde \psi^+_{su} ~~, 
\\[.2cm]
\eta^-_{su} = e^{+iY+i \Phi} \psi^-_{su} \quad &,\qquad 
\tilde \eta^-_{su} = e^{-iY-i \Phi} \tilde \psi^-_{su} ~~,
\end{aligned}
\label{twisted-fermions}%
\end{equation}
in terms of which the action \eqref{SU(2)-fermionic-action} becomes 
\begin{equation}
S_{su}^F = \frac{1}{2 \pi} \int \text d^2v \left[ \eta^+_{su} \partial_{\bar v} \eta^-_{su} + \eta^-_{su} \partial_{\bar v} \eta^+_{su} + \tilde \eta^+_{su} \partial_v \tilde \eta^-_{su} + \tilde \eta^-_{su} \partial_v \tilde \eta^+_{su} \right] \ .  
\end{equation}
As anticipated in Section~\ref{sec:GWZW}, the ``rotation'' \eqref{twisted-fermions} is anomalous and produces the anomaly factor 
\begin{equation}
\mathcal{D} \psi^+_{su} \,  \mathcal{D} \psi^-_{su} \, \mathcal{D} \tilde \psi^+_{su} \, \mathcal{D} \tilde \psi^-_{su}  = \mathcal{D} \eta^+ \, \mathcal{D} \eta^- \, \mathcal{D} \tilde \eta^+ \, \mathcal{D} \tilde \eta^- \, e^{\check{\mathfrak A}} \ , 
\label{axial-anomaly}
\end{equation}
where for a chiral rotation of the form 
\begin{equation}
\begin{pmatrix} \delta \psi^\pm_{su} \\ \delta \tilde \psi^{\pm}_{su} \end{pmatrix} = \pm i \,  \begin{pmatrix} \alpha & 0 \\ 0 & - \alpha \end{pmatrix} \, \begin{pmatrix} \psi^\pm_{su} \\ \tilde \psi^{\pm}_{su} \end{pmatrix}
\label{su2-fermion-anomalous-rotation}
\end{equation}
the anomaly reads \rcite{Weinberg:1996kr}
\begin{equation}
\check{\mathfrak A} = -\frac{1}{2\pi} \int \text dx^0 \text dx^1 \, \varepsilon^{\mu \nu} \, \partial_\mu \alpha \, A_\nu 
= \frac{2}{\pi} \int \text d^2v \, \partial_v (Y + \Phi[z_F]) \, \partial_{\bar v}(Y + \Phi[z_F]) \ . 
\label{su(2)-chiral-anomaly}
\end{equation}
In the second equality in \eqref{su(2)-chiral-anomaly} we used that 
\begin{equation}
\alpha = - Y - \Phi[z_F] \ , \qquad A_v = -i \partial_v Y - i \partial_v \Phi[z_F] \ , \qquad  A_{\bar v} = i \partial_{\bar v} Y + i \partial_{\bar v} \Phi[z_F] \ . 
\end{equation}
The path integral \eqref{fermion-path-integral} then becomes  
\begin{equation}
Z^F_{su}(\tau, z_F) = \exp \left(\frac{2}{\pi} \int \text d^2v \, \partial_v (Y + \Phi[z_F]) \, \partial_{\bar v}(Y + \Phi[z_F]) \right) \, Z^\eta_{su} \ , 
\end{equation}
with 
\begin{equation}
Z^\eta_{su} = \int \mathcal{D} \eta^+_{su} \mathcal{D} \eta^-_{su} \mathcal{D} \tilde \eta^+_{su}  \mathcal{D} \tilde \eta^-_{su} \, \exp \left(- \int \frac{\text d^2v}{2 \pi} \left[ \eta^+_{su} \partial_{\bar v} \eta^-_{su} + \eta^-_{su} \partial_{\bar v} \eta^+_{su} + \tilde \eta^+_{su} \partial_v \tilde \eta^-_{su} + \tilde \eta^-_{su} \partial_v \tilde \eta^+_{su} \right] \right) \ . 
\label{Z-eta}
\end{equation}
It remains to compute the partition function $Z^\eta_{su}$. Notice that while the fermions $\psi^{\pm}_{su}, \tilde \psi^{\pm}_{su}$ have periodic boundary conditions (we are in the $\tilde{\text{R}}$-sector, \ie~the \text{R}-sector with additional insertions of $(-1)^{F}$), the fermions $\eta^{\pm}_{su}, \tilde \eta^{\pm}_{su}$ obey
\begin{equation}
\begin{aligned}
\eta^\pm_{su}(v+2 \pi, \bar v + 2 \pi) &= e^{\pm 2 \pi i s_1'} \eta^{\pm}_{su}(v,\bar v) \ , \\
\eta^\pm_{su}(v+2 \pi \tau, \bar v + 2 \pi \bar \tau) &= e^{\mp 2 \pi i s_2'} \eta^{\pm}_{su}(v,\bar v) \ , \\
\tilde \eta^\pm_{su}(v+2 \pi, \bar v + 2 \pi) &= e^{\mp 2 \pi i s_1'} \tilde \eta^{\pm}_{su}(v,\bar v) \ , \\
\tilde \eta^\pm_{su}(v+2 \pi \tau, \bar v + 2 \pi \bar \tau) &= e^{\pm 2 \pi i s_2'} \tilde \eta^{\pm}_{su}(v,\bar v) \ , 
\end{aligned}
\label{eta-boundary-conditions-su(2)}%
\end{equation}
where we defined $s_1'$ and $s_2'$ by
\begin{equation}
    z_F = s_1' \tau + s_2' \ . 
\end{equation}
The partition function of free fermions with arbitrary boundary conditions has been computed in~\rcite{Alvarez-Gaume:1986rcs}. Exploiting their findings and using eq.~\eqref{eta-boundary-conditions-su(2)} we obtain
\begin{equation}
Z^\eta_{su} = \Bigl| q^{-\frac{1}{24}}q^{\frac{1}{2}(s_1'+\frac{1}{2})} \prod_{n=1}^\infty (1-q^{n+s_1'} e^{2 \pi i s_2'})(1-q^{n-s_1'-1} e^{-2 \pi i s_2'}) \Bigr|^2 = e^{\frac{\pi}{2\tau_2} \, (z_F-\bar z_F)^2} \frac{| \theta_1(\tau, z_F)|^2}{|\eta(\tau)|^2} \ , 
\end{equation}
see Appendix~\ref{app:theta} for the definitions of Theta functions. All together, the partition function \eqref{fermion-path-integral} reads
\begin{equation}
Z^F_{su}(\tau, z_F) = \frac{\exp \left( \frac{\pi}{2\tau_2} \, (z_F-\bar z_F)^2 \right) }{|\eta(\tau)|^2}  \, | \theta_1(\tau, z_F)|^2 \, \exp \left(\frac{2}{\pi} \int \text d^2v \, \big|\partial_v (Y + \Phi[z_F]) \big|^2  \right) \ . 
\label{SU(2)-U(1)-fermions-pf}
\end{equation}

\subsection{The coset partition function}

The coset partition function follows directly from eqs.~\eqref{ghost-pf}, \eqref{su(2)-B-F-gh}, \eqref{SU(2)-U(1)-bosons-pf} and \eqref{SU(2)-U(1)-fermions-pf},  
\begin{equation}
 Z^{\frac{\mathfrak{su}(2)}{\mathfrak u(1)}}(\tau, z)   
= \sqrt{\tau_2} \, e^{\frac{2 \pi}{\tau_2} \frac{\kappa_{su}-2}{\kappa_{su}}z_1^2} \, \int_0^1 \text ds_1 \int_0^1 \text ds_2 \, e^{-\frac{\pi \kappa_{su}}{\tau_2}u_2^2}\, \left| \theta_1(\tau, z_F) \right|^2 \, \sum_{j=0}^{\kappa_{su}-2} \left| \chi_{j,B}^{\mathfrak{su}(2)}(\tau,z_B) \right|^2 \ . 
\end{equation}
Let us now see how evaluating the integral over $s_1$ and $s_2$ we will recover the trace partition function \eqref{super-pf-pf}. We first relate the characters of the bosonic $\mathfrak{su}(2)_{\kappa_{su}-2}$ to the characters $I^j_\ell$ of the $\mathcal N =2$ coset model \eqref{su2/u1-coset-intro} and from eq.~\eqref{chi-theta1=ITheta-identity} we obtain
\begin{multline}
Z^{\frac{\mathfrak{su}(2)}{\mathfrak u(1)}}(\tau, z) = \sqrt{\tau_2} \, \exp \left( \mfrac{2 \pi}{\tau_2} \mfrac{\kappa_{su}-2}{\kappa_{su}}z_1^2 \right) \, \sum_{j=0}^{\kappa_{su}-2} \sum_{\ell, \ell'=-\kappa_{su}+1}^{\kappa_{su}} I^j_\ell(\tau, z) \, \overline{I^j_{\ell'}(\tau, z)}  \\
\times \, \int_0^1 \text ds_1 \int_0^1 \text ds_2 \, \exp \left( -\frac{\pi \kappa_{su}}{\tau_2}u_2^2 \right) \,  \Theta_\ell^{(\kappa_{su})}(\tau, u) \,  \overline{ \Theta_{\ell'}^{(\kappa_{su})}(\tau, u)} ~. \hskip .5cm~
\label{SU(2)-U(1)-II-TT}
\end{multline} 
See Appendices~\ref{app:theta} and \ref{app:su(2)} for the precise definition of the various quantities entering eq.~\eqref{SU(2)-U(1)-II-TT}. Orthogonality of the $\Theta_\ell^{(k)}(\tau, u)$ functions, see eq.~\eqref{Theta2-integral}, implies
\begin{multline}
Z^{\frac{\mathfrak{su}(2)}{\mathfrak{u}(1)}}(\tau, z) = \sqrt{\tau_2} \, \exp \left( \mfrac{2 \pi}{\tau_2} \mfrac{\kappa_{su}-2}{\kappa_{su}}z_1^2\right) \, \sum_{j=0}^{\kappa_{su}-2} \sum_{\ell = 1}^{\kappa_{su}} |I_\ell^j(\tau,z) |^2 \\
\times \, \sum_{n \in \mathds Z} \int_0^{1} \text d s_1 \left( e^{-{\pi \kappa_{su}}{\tau_2} \left(s_1 + 2 n+\frac{\ell}{\kappa_{su}} \right)^2} + e^{-{\pi \kappa_{su}}{\tau_2} \left(s_1 + 2 n - 1+\frac{\ell}{\kappa_{su}} \right)^2 } \right) \ , 
\label{parafermion-int-1}
\end{multline}
where we have split the sum over $\ell$, used eq.~\eqref{Is-identification} and renamed the summation labels $j \mapsto \kappa_{su}-2-j$ and $\ell \mapsto \ell-\kappa_{su}$. We can then trade the sum over $n$ and the integral over $s_1$ for an integral over the real line,  
\begin{align}
Z^{\frac{\mathfrak{su}(2)}{\mathfrak{u}(1)}}(\tau, z) &= \sqrt{\tau_2} \, \exp \left( \mfrac{2 \pi}{\tau_2} \mfrac{\kappa_{su}-2}{\kappa_{su}}z_1^2 \right) \, \sum_{j=0}^{\kappa_{su}-2} \sum_{\ell = 1}^{\kappa_{su}} |I_\ell^j(\tau,z) |^2 \int_{- \infty}^{+\infty} \text d s_1 e^{-{\pi \kappa_{su}}{\tau_2} ( s_1 +\tfrac{\ell}{\kappa_{su}})^2 }  \nn\\
& = \frac{ \exp \left( \mfrac{2 \pi}{\tau_2} \mfrac{\kappa_{su}-2}{\kappa_{su}}z_1^2 \right) }{\sqrt{\kappa_{su}}} \, \sum_{j=0}^{\kappa_{su}-2} \sum_{\ell = 1}^{\kappa_{su}} |I_\ell^j(\tau,z) |^2 \ . 
\end{align}
This does not yet look like the expected result, compare with eq.~\eqref{super-pf-pf}. Let us then double the sum over $\ell$, use the character identity~\eqref{Is-identification} and rename the sum label as $\ell \mapsto \ell + \kappa_{su}$, 
\begin{align}
Z^{\frac{\mathfrak{su}(2)}{\mathfrak{u}(1)}}(\tau, z) & = \frac{ \exp \left( \frac{2 \pi}{\tau_2} \frac{\kappa_{su}-2}{\kappa_{su}}z_1^2 \right) }{2\sqrt{\kappa_{su}}} \, \sum_{j=0}^{\kappa_{su}-2} \sum_{\ell = 1}^{\kappa_{su}}  \left(|I_\ell^j(\tau,z)|^2 + |I_{\ell-\kappa}^j(\tau,z)|^2 \right) \\
& = \frac{ \exp \left( \frac{2 \pi}{\tau_2} \frac{\kappa_{su}-2}{\kappa_{su}}z_1^2 \right)}{2\sqrt{\kappa_{su}}} \sum_{j=0}^{\kappa_{su}-2}  \sum_{\ell = 1-\kappa}^{\kappa_{su}} |I_\ell^j(\tau,z) |^2 \ . 
\label{su(2)/u(1)-pf-final}
\end{align}
The path integral partition function \eqref{su(2)/u(1)-pf-final} agrees with the trace partition function~\eqref{super-pf-pf} up to the multiplicative factor
\begin{equation}
\frac{\exp \left( \frac{2 \pi}{\tau_2} \frac{\kappa_{su}-2}{\kappa_{su}}z_1^2 \right)}{\sqrt \kappa_{su}} \ . 
\end{equation} 
The factor of $\sqrt \kappa_{su}$ in the denominator can be absorbed in the overall normalisation, which still has to be fixed. The exponential factor is of the form 
\begin{equation}
\exp \left( \frac{2 \pi}{\tau_2}\hat c \, z_1^2 \right) \ , \qquad \text{with} \qquad \hat c = \frac{c}{3} = \frac{\kappa_{su}-2}{\kappa_{su}} \ . 
 \end{equation}
This is the expected relative factor relating the \emph{path integral} partition function to the \emph{trace} partition function, see Appendix \ref{app:different-pfs} where following \rcite{Kraus:2006nb} we recap its origin.


\section{Warmup: \texorpdfstring{$\boldsymbol{\frac{\sltwo}{\uone}}$}{} } 
\label{sec:sl2/u1}

Let us consider a further warmup example, the $\mathcal N=2$ ``cigar'' theory defined by the axial coset~\rcite{Witten:1991yr,Dijkgraaf:1991ba,Gawedzki:1991yu} 
\begin{equation}
    \frac{SL(2, \mathds R)_{\kappa_{sl}}}{U(1)_A} \ . 
\end{equation}
The path integral partition function 
\begin{equation}
Z^{\frac{\mathfrak{sl}(2,\mathds R)}{\mathfrak u(1)}}(\tau, z) = \text{Tr}_R\Big[(-1)^{F_L + F_R} q^{L_0-\frac{c}{24}}\,  \bar q^{\bar L_0-\frac{c}{24}} \, e^{2 \pi i z J^{\cR}_{sl}} \,  e^{-2 \pi i \bar z \bar J^{\cR}_{sl}}\Big]
\end{equation} 
of the supersymmetric cigar theory has been computed in \rcite{Eguchi:2004yi, Israel:2004ir, Eguchi:2010cb}. To strengthen the claims of Section~\ref{sec:standard=null}, we are going to rederive it in terms of the null gauged coset\footnote{Strictly speaking, in order to compute the path integral, we will consider the analytic continuation of $SL(2,\mathds R)$ to $H_3^+$, see \eg~\cite{Gawedzki:1988nj, Gawedzki:1991yu, Maldacena:2000kv, Hanany:2002ev, Eguchi:2004yi} for more details on this.}
\begin{equation}
\frac{SL(2,\mathds R)_{\kappa_{sl}} \times U(1)_y}{U(1)_L \times U(1)_R} \ ,
\end{equation}
with embeddings
\begin{equation}
\varepsilon_L^{sl} = -\varepsilon_R^{sl} = -\frac{\sigma_3}{2} \ , \qquad \varepsilon_L^{y} = \varepsilon_R^{y} = \frac{\sqrt{\kappa_{sl}}}{2} \ .  
\label{SL2-U1-embeddings}
\end{equation}
Similarly to Section \ref{sec:su2/u1}, the currents $J^3_{sl}$ and $J^{\mathcal R}_{sl}$ can be written as 
\begin{equation}
    J^3_{sl} = j^3_{sl} + j^3_{f,sl} \ , \qquad J^{\cR}_{sl} = \frac{2}{\kappa_{sl}}j^3_{sl} + \frac{\kappa_{sl}+2}{\kappa_{sl}} j^3_{f,sl}  \ , 
    \label{JBFR-rel-sl}
\end{equation}
and chemical potentials $u$ and $z$ associated respectively to $J^3_{sl}$ and $J^{\mathcal R}_{sl}$ are related to the bosonic and fermionic chemical potentials $z_B$ and $z_F$ by  
\begin{equation}
z_B = u + \frac{2}{\kappa_{sl}} z \ , \qquad z_F =  u + \frac{\kappa_{sl}+2}{\kappa_{sl}} z \ . 
\label{sl2-chemical-potentials-relation}
\end{equation}
We parameterize the gauge field entering the decoupled action \eqref{S-decoupled}, 
\begin{align}
   S^{B,d} = S_{g_{sl}} + S_{g_y} + S^{X,Y}_{sl} + S^{X,Y}_{y} \ , 
   \label{S-decoupled-sl}
\end{align}
as discussed in Section \ref{sec:parametrization-gauge-field}. The path integral factorizes as 
\begin{equation}
Z^{\frac{\mathfrak{sl}(2, \mathds R)}{\mathfrak{u}(1)}}(\tau, z) = \int_0^1 \text d s_1 \int_0^1 \text ds_2 \, Z_{gh}(\tau) \, Z^{B}_{sl}(\tau, z_B) \, Z^{B}_{y}(\tau, u) \,  Z^{F}_{sl}(\tau, z_F)  \ . 
\label{sl(2)-B-F-gh}
\end{equation} 
Also in this case, since the gauging is null, the only fermions entering the action are those taking values in the ``transverse space'', \ie~$\psi^\pm_{sl}$ and $\tilde \psi^\pm_{sl}$ (again, see Appendix~\ref{app:GWZW in superspace}). 
Formally, we model the free compact boson $y$ with radius 
\begin{equation}
R = \sqrt{\kappa_{sl}}
\end{equation} 
as a $U(1)$ WZW model at level 2, with group element
\begin{equation}
g_y = e^{i y } 
\end{equation}
and OPE
\begin{equation}
\partial_v y(z) \, \partial_v y(w) = -\frac{1}{2(z-w)^2} \ . 
\end{equation}
With these conventions, it is easy to check that the gauging is indeed null,\footnote{The WZW action for $SL(2,\mathds{R})$ has an additional minus sign relative to~\eqref{WZW-action} so that the metric has the correct signature, see \eg~\cite{Eguchi:2010cb}.
}
  \begin{equation}
\sum_{i} \kappa_i \text{Tr}[(\varepsilon_L^i)^2] = \sum_{i} \kappa_i \text{Tr}[(\varepsilon_R^i)^2] = -\kappa_{sl} \, \text{Tr}\left[\left(\frac{\sigma_3}{2}\right)^2\right] + 2 \Big(\frac{\sqrt{\kappa_{sl}}}{2} \Big)^2 = 0 \ . 
\end{equation}

\subsection{Bosonic sector}

Let us compute $Z^{B}_{sl}(\tau, z_B)$ and $Z^{B}_{y}(\tau, u)$. We begin with $Z^{B}_{sl}(\tau, z_B)$ and consider 
\begin{align}
 \int \mathcal D g_{sl} \, e^{-S_{g_{sl}}} & = \int \mathcal D g_{sl} \, \exp \left[ -S^\WZW \left( a[z_B]^{-\varepsilon_L^{sl}} \, g_{sl} \, a[z_B]^{\dagger \varepsilon_R^{sl}} \right) + S^\WZW (a[z_B]^{\varepsilon_R^{sl}} \, a[z_B]^{\dagger \varepsilon_R^{sl}} ) \right] \nonumber \\
  & = \frac{e^{- \pi \frac{(z_B - \bar z_B)^2}{2\tau_2}}}{\sqrt{\tau_2} \, |\theta_1(\tau,z_B)|^2} \ , 
  \label{SL2-partition-function}
\end{align}
where in the first equality we used that
\begin{equation}
S^\WZW \left(a[u]^{\varepsilon_R^{sl}} \, a[u]^{\dagger \varepsilon_R^{sl}} \right)=0 \ , 
\end{equation}
while in the second one we recognised the path integral partition function of the bosonic $\mathfrak{sl}(2,\mathds R)$ WZW model \rcite{Gawedzki:1988nj, Gawedzki:1991yu, Eguchi:2010cb},\footnote{Eq.~\eqref{bosonic-sl2-pf} is correct up to some normalization constant. However, this will not be relevant for us. }
\begin{equation}
Z^{GK}_{sl}(\tau, z_B) = \frac{e^{- \pi \frac{(z_B - \bar z_B)^2}{2\tau_2}}}{\sqrt{\tau_2} \, |\theta_1(\tau,z_B)|^2}  \ .     \label{bosonic-sl2-pf}
\end{equation}
For $S^{X,Y}_{sl}$ we find
\begin{align}
S^{X,Y}_{sl} & = -S^\WZW (a[z_B]^{-\varepsilon_L^{sl}} \Omega^{-\varepsilon_L^{sl}} \Omega^{\dagger -\varepsilon_R^{sl}} a[z_B]^{\dagger -\varepsilon_R^{sl}}) = \frac{\kappa_{sl}+2}{\pi} \int \text d^2 v \, \big|\partial_v(Y + \Phi[z_B])\big|^2 \ ,
 \label{S-XY-sl} 
\end{align}
The two $U(1)_y$ terms in \eqref{S-decoupled-sl} read
\begin{equation}
S_{g_y} = S^\WZW \left( a[u]^{-\varepsilon_L^y} \, g_y \, a[u]^{\dagger \varepsilon_R^y}\right) = \frac{1}{\pi} \int \text d^2 v  \, \partial_v(y + \sqrt{\kappa_{sl}} \Phi[u]) \, \partial_{\bar v}(y + \sqrt{\kappa_{sl}} \Phi[u]) \ ,  \label{S-gy}
\end{equation}
and 
\begin{align}
S^{X,Y}_y &= - S^\WZW(a[u]^{-\varepsilon_L^y} \Omega^{-\varepsilon_L^y} \Omega^{\dagger -\varepsilon_R^y} a[u]^{\dagger -\varepsilon_R^y}) -\frac{2}{2\pi} \int \text d^2v \, (\varepsilon_L^y+\varepsilon_R^y)^2 \, | a[u]^{-1} \partial_{\bar v} a[u] + \Omega^{-1} \partial_{\bar v} \Omega|^2 \nonumber \\
& = \frac{\kappa_{sl}}{\pi} \int \text d^2 v \, \partial_v X \, \partial_{\bar v} X  -\frac{\kappa_{sl}}{\pi} \int \text d^2v \, \partial(Y + \Phi[u]) \, \partial_{\bar v}(Y + \Phi[u]) -\frac{\kappa_{sl}}{\pi} \int \text d^2v \, \partial_v X \, \partial_{\bar v} X \nonumber \\
& = -\frac{\kappa_{sl}}{\pi} \int \text d^2v \, \partial(Y + \Phi[u]) \, \partial_{\bar v}(Y + \Phi[u]) \ , 
\label{S-XY-y}
\end{align}
where we used that 
\begin{equation}
\int \text d^2 v \, \Bigl( - \partial_{\bar v} X \, \partial_v (\Phi[u]+Y) + \partial_v X \, \partial_{\bar v}(\Phi[u]+Y)\Bigr) = 0 
\end{equation}
vanishes by integration by parts. Assembling eqs.~\eqref{SL2-partition-function}, \eqref{S-XY-sl}-\eqref{S-XY-y} we obtain the bosonic contribution to the partition function,
\begin{multline}
Z^B_{sl}(\tau, z_B) Z^B_y(\tau, u) =  \frac{e^{- \pi \frac{(z_B - \bar z_B)^2}{2\tau_2}}}{\sqrt{\tau_2} \, |\theta_1(\tau,z_B)|^2} \, \exp\Bigl(-\frac{1}{\pi} \int \text d^2v \, \big|\partial_v(y + \sqrt \kappa_{sl} \Phi[u])\big
|^2 \Bigr) \, \times \\
\times \, \exp \Bigl( -\frac{\kappa_{sl}+2}{\pi} \int \text d^2v \, \big|\partial_v(Y+\Phi[z_B])\big|^2 + \frac{\kappa_{sl}}{\pi} \int \text d^2v \, \big|\partial_v(Y+\Phi[u])\big|^2 \Bigr) \ .  
\label{SL-U(1)-bosons-pf}
\end{multline}

\subsection{The coset partition function}

The fermion action reads
\begin{equation}
S_{sl}^F = \int \frac{\text d^2v}{2 \pi} \left[  \psi_{sl}^+(\partial_{\bar v} +  A_{\bar v} )\psi^-_{sl} + \psi^-_{sl}(\partial_{\bar v} - A_{\bar v})\psi^+_{sl} + \tilde \psi^+_{sl}(\partial_v - A_v ) \tilde \psi^-_{sl} + \tilde \psi^-_{sl}(\partial_v + A_v ) \tilde \psi^+_{sl} \right] \ . 
\label{SL(2)-fermionic-action} 
\end{equation}
The computation of the fermion path integral again reduces to the space transverse to the gauge action
\begin{equation}
Z^F_{sl}(\tau, z_F) = \int \mathcal{D} \psi^+_{sl} \, \mathcal{D} \psi^-_{sl} \, \mathcal{D} \tilde \psi^+_{sl} \, \mathcal{D} \tilde \psi^-_{sl} \, e^{-S_{sl}^F} \ . 
\label{fermion-path-integral-sl}
\end{equation}
One proceeds as in Section~\ref{sec:su(2)/u(1)-fermions} and we have \rcite{Eguchi:2004yi, Eguchi:2010cb}
\begin{equation}
Z^F_{sl}(\tau, z_F) = e^{\frac{\pi}{2\tau_2} (z_F-\bar z_F)^2} \frac{| \theta_1(\tau, z_F)|^2}{|\eta(\tau)|^2} \, \exp \left(\frac{2}{\pi} \int \text d^2v \, \left| \partial_v (Y + \Phi[z_F]) \right|^2 \right) \ . 
\label{SL-U(1)-fermions-pf}
\end{equation}
All together, from eqs.~\eqref{sl(2)-B-F-gh}, \eqref{SL-U(1)-bosons-pf}, and \eqref{SL-U(1)-fermions-pf} we obtain 
\begin{multline}
Z^{\frac{\mathfrak{sl}(2, \mathds R)}{\mathfrak{u}(1)}}(\tau, z)  = \, \sqrt{\tau_2} \, |\eta(\tau)|^2  \, e^{-\frac{2\pi}{\tau_2} \frac{\kappa_{sl}+4}{\kappa_{sl}} z_2^2} \int_0^1 \text d s_1 \int_0^1 \text ds_2 \, \frac{| \theta_1(\tau, u+ \frac{\kappa_{sl}+2}{\kappa_{sl}}z)|^2}{|\theta_1(\tau,u+\frac{2}{\kappa_{sl}}z)|^2} \, e^{-4 \pi \frac{u_2 z_2}{\tau_2}} \\
\times \exp \Bigl(  \medint\int \text d^2v \Bigl[ -\tfrac{\kappa_{sl}+2}{\pi} \left|\partial_v(Y+\Phi[u + \tfrac{2}{\kappa_{sl}}z])\right|^2 + \tfrac{\kappa_{sl}}{\pi} \left|\partial_v(Y+\Phi[u])\right|^2  + \tfrac{2}{\pi} \left| \partial_v(Y + \Phi[u + \tfrac{\kappa_{sl}+2}{\kappa_{sl}}z]) \right|^2 \Bigr] \Bigr) \\
\times \int  \mathcal D y \, \exp\Bigl(-\tfrac{1}{\pi} \medint\int \text d^2v \, |\partial_v(y + \sqrt \kappa_{sl} \Phi[u])|^2  \Bigr)  \ . 
\label{SL-U(1)-pf-intermediate}
\end{multline}
Notice that both $y$ and $\sqrt{\kappa_{sl}} \, \Phi[u]$ have radius $\sqrt{\kappa_{sl}}$, so that $y + \sqrt{\kappa_{sl}} \, \Phi[u]$ obeys the periodicity conditions
\begin{align}
(y + \sqrt{\kappa_{sl}} \, \Phi[u])(v+ 2 \pi, \bar v + 2 \pi) &= (y + \sqrt{\kappa_{sl}} \, \Phi[u])(v, \bar v) - 2 \pi \sqrt{\kappa_{sl}} (m_1 + s_1) \ , \\
(y + \sqrt{\kappa_{sl}} \, \Phi[u])(v+ 2 \pi \tau, \bar v + 2 \pi \bar \tau) &= (y + \sqrt{\kappa_{sl}} \, \Phi[u])(v, \bar v) + 2 \pi \sqrt{\kappa_{sl}} (m_2 + s_2) \ .  
\end{align}
We can then identify the last line of eq.~\eqref{SL-U(1)-pf-intermediate} as the partition function of a twisted compact boson of radius $\sqrt{\kappa_{sl}}$, \ie 
\begin{equation}
\int \mathcal D y \, \exp\Bigl(-\medint\int  \mfrac{\text d^2v}{\pi} \, | \partial_v(y + \sqrt{\kappa_{sl}}\,  \Phi[u]) |^2 \Bigr) = \frac{\sqrt{\kappa_{sl}}}{\sqrt{\tau_2} |\eta(\tau)|^2} \medmath{\sum_{m_1, m_2 \in \mathds Z}} \exp \Bigl(-\mfrac{\pi \kappa_{sl} }{\tau_2}|m_1 \tau + m_2 + u|^2 \Bigr) \ . 
\label{SL2-U1-y-pf}
\end{equation}
Since $\partial_v\Phi[u]$ is linear in $u$, the second line in \eqref{SL-U(1)-pf-intermediate} simplifies, 
\begin{equation}
\exp \biggl[ \biggl( -\frac{4(\kappa_{sl}+2)}{\pi \kappa_{sl}} + \frac{2(\kappa_{sl}+2)^2}{\pi \kappa_{sl}^2} \biggr) |\partial_v\Phi[z]|^2 \medint\int \text d^2 v  \biggr] 
= \exp \left(\frac{2 \pi}{\tau_2} \frac{\kappa_{sl}+2}{\kappa_{sl}} |z|^2 \right) \ , 
\end{equation}
and eq.~\eqref{SL-U(1)-pf-intermediate} becomes\footnote{Our computation is not sensitive to the overall normalization. However, as was done in \rcite{Eguchi:2010cb},  this can in principle be fixed by studying the decomposition into characters. }
\begin{multline}
Z^{\frac{\mathfrak{sl}(2, \mathds R)}{\mathfrak{u}(1)}}(\tau, z)  = e^{\frac{2\pi}{\tau_2} (\frac{\kappa_{sl}+2}{\kappa_{sl}} |z|^2 -\frac{\kappa_{sl}+4}{\kappa_{sl}} z_2^2)} \int_0^1 \text d s_1 \int_0^1 \text ds_2 \, \frac{| \theta_1(\tau, u+ \frac{\kappa_{sl}+2}{\kappa_{sl}}z)|^2}{|\theta_1(\tau,u+\frac{2}{\kappa_{sl}}z)|^2} \\
\times \, e^{-\frac{4 \pi  u_2 z_2}{\tau_2}} \sum_{m_1, m_2 \in \mathds Z} e^{-\frac{\pi \kappa_{sl} }{\tau_2}|m_1 \tau + m_2 + u|^2 } \ ,  
\label{sl-u1-pf-final}
\end{multline}
which agrees with the axial gauge partition function computed in \rcite{Eguchi:2010cb}. 

\subsection{Evaluating the gauge field zero mode integrals}
\label{sec:zeromodeints-sl2}

Let us review how the partition function \eqref{sl-u1-pf-final} can be decomposed into characters. This subsection is essentially a summary of \cite{Maldacena:2000kv, Hanany:2002ev, Eguchi:2010cb} and has the purpose to warm up in view of the decomposition into characters of the supertube partition function we discuss in Section~\ref{sec:supertube}.

We begin by combining the integrals over $s_1$ and $s_2$ with the sum over $m_1, m_2$ in \eqref{sl-u1-pf-final} into an integral over the complex plane. Using eq.~\eqref{theta1-period} we find 
\begin{equation}
Z^{\frac{\mathfrak{sl}(2, \mathds R)}{\mathfrak{u}(1)}}(\tau, z)  = e^{\frac{2\pi}{\tau_2} (\frac{\kappa_{sl}+2}{\kappa_{sl}} |z|^2 -\frac{\kappa_{sl}+4}{\kappa_{sl}} z_2^2)} \int_{\mathds C} \frac{\text d^2u}{\tau_2} \, \frac{| \theta_1(\tau, u+ \frac{\kappa_{sl}+2}{\kappa_{sl}}z)|^2}{|\theta_1(\tau,u+\frac{2}{\kappa_{sl}}z)|^2} \, e^{-\frac{4 \pi  u_2 z_2}{\tau_2}} e^{-\frac{\pi \kappa_{sl} }{\tau_2}|u|^2 } \ ,  
\label{sl-u1-pf-Cplane}
\end{equation}
where $\frac{\text d^2u}{\tau_2} = \text ds_1 \text ds_2$. This allows us to shift the integration variable according to $u \mapsto u - \frac{2}{\kappa_{sl}}z$, 
\begin{align}
Z^{\frac{\mathfrak{sl}(2, \mathds R)}{\mathfrak{u}(1)}}(\tau, z)  &= e^{\frac{2\pi}{\tau_2} (\frac{\kappa_{sl}+2}{\kappa_{sl}} |z|^2 - z_2^2)} \int_{\mathds C} \frac{\text d^2u}{\tau_2} \, \frac{| \theta_1(\tau, u+ z)|^2}{|\theta_1(\tau,u)|^2} \, e^{-\frac{4 \pi  u_2 z_2}{\tau_2}} e^{-\frac{\pi \kappa_{sl} }{\tau_2}|u-\frac{2}{\kappa_{sl}}z|^2 } \\
& = e^{\frac{2\pi}{\tau_2} (\frac{\kappa_{sl}+2}{\kappa_{sl}} |z|^2 - z_2^2)} \int_0^1 \text d s_1 \int_0^1 \text ds_2 \, \frac{| \theta_1(\tau, s_1 \tau + s_2 + z)|^2}{|\theta_1(\tau,s_1 \tau + s_2 )|^2} e^{- 4 \pi  s_1 z_2} \nonumber \\
& \hspace{150pt} \times \, \sum_{w,m \in \mathds Z} e^{-\frac{\pi \kappa_{sl} }{\tau_2}|(s_1 +w) \tau + (s_2+m) -\frac{2}{\kappa_{sl}}z|^2 } \ . 
\label{sl-u1-pf-shifted}
\end{align}

\paragraph{Integral over $\boldsymbol{s_2}$:} Let us first perform the integral over $s_2$. In order to do so, we expand the Theta functions in the numerator, see eq.~\eqref{theta-1-def}
\begin{multline}
    |\theta_1(\tau,u+z)|^2 = \sum_{n_f^{sl}, \bar n_f^{sl} \in \mathds Z + \frac{1}{2}} q^{\frac{1}{2}(n_f^{sl})^2} \bar q^{\frac{1}{2}(\bar n_f^{sl})^2} \, e^{2 \pi i s_1 \tau_1 (n_f^{sl}-\bar n_f^{sl})} \, e^{-2 \pi s_1 \tau_2 (n_f^{sl}+\bar n_f^{sl})} \\[-0.4cm]
    \times \, e^{2 \pi i (z n_f^{sl}-\bar z \bar n_f^{sl})} \, \exp \left(2 \pi i s_2 (n_f^{sl}-\bar n_f^{sl}) \right) \ , 
\end{multline}
and use the identity \rcite{Pakman:2003kh, Israel:2004ir}
\begin{equation}
\frac{1}{|\theta_1(\tau,u)|^2} = \left| \frac{1}{\eta(\tau)^3} \sum_{r \in \mathds Z} e^{2 \pi i u(r+\frac{1}{2})} S_r(\tau) \right|^2 \ , \qquad \text{with} \qquad S_r(\tau) = \sum_{n=0}^\infty (-1)^n q^{\frac{n(n+2r+1)}{2}} \ , 
\label{theta1-denom}
\end{equation}
to expand the Theta functions in the denominator, 
\begin{equation}
    \frac{1}{|\theta_1(\tau,u)|^2} = \frac{1}{|\eta(\tau)|^6} \sum_{r, \bar r \in \mathds Z} S_r(\tau) \overline{S_{\bar r}(\tau)} \, e^{2 \pi i s_1 \tau_1(r-\bar r)} \, e^{-2 \pi s_1 \tau_2(r+\bar r+1)} \, \exp \left(2 \pi i s_2(r-\bar r)\right) \ . 
\end{equation}
In order to linearize the $s_2$ dependence, we Poisson resum the last line of \eqref{sl-u1-pf-shifted}, 
\begin{multline}
\hspace{-10pt} \sum_{w,m \in \mathds Z} e^{-\frac{\pi \kappa_{sl} }{\tau_2}|(s_1 +w) \tau + (s_2+m) -\frac{2}{\kappa_{sl}}z|^2 } = \frac{\sqrt{\tau_2}}{\sqrt{\kappa_{sl}}} \, e^{-\frac{\pi \kappa_{sl} u_2^2}{\tau_2}+\frac{4 \pi u_2 z_2}{\tau_2}-\frac{4 \pi z_2^2}{\kappa_{sl} \tau_2}} \, \sum_{w, p \in \mathds Z} q^{\frac{1}{4 \kappa_{sl}}(p+w \kappa_{sl})^2} \bar q^{\frac{1}{4 \kappa_{sl}}(p-w \kappa_{sl})^2} \\
 \times  e^{-2 \pi i(\frac{p}{\kappa_{sl}}+w)z-2 \pi i(\frac{p}{\kappa_{sl}}-w)\bar z} \, e^{2 \pi i s_1 \tau_1 p-2 \pi s_1 \tau_2 w \kappa_{sl}} \, \exp \left(2 \pi i s_2 p \right) \ . 
\end{multline}
The integral over $s_2$ is now easy to perform and gives rise to the Kronecker delta
\begin{equation}
    \int_0^1 \text ds_2 \, e^{2 \pi i s_2(r+n_f^{sl}-\bar r -\bar n_f^{sl}+p)} = \delta_{\mathcal J_0 - \bar{\mathcal J}_0,0} \ , 
\end{equation}
where we made the identifications 
\begin{equation}
    (J^3_{sl})_0 - (\bar J^3_{sl})_0 = r + n_f^{sl} -(\bar r + \bar n_f^{sl})\ , 
\end{equation}
and introduced the gauged current zero modes
\begin{equation}
\mathcal J_0 = (J^3_{sl})_0 + \frac{\sqrt{\kappa_{sl}} \, p_L}{2} \ , \qquad \bar{\mathcal J}_0 = (\bar J^3_{sl})_0 - \frac{\sqrt{\kappa_{sl}} \, p_R}{2} \ , 
\label{gauged-currents-sl}
\end{equation}
with
\begin{equation}
    p_L= \frac{p}{\sqrt{\kappa_{sl}}} + w \sqrt{\kappa_{sl}} \ , \qquad      p_R= \frac{p}{\sqrt{\kappa_{sl}}} - w \sqrt{\kappa_{sl}} \ . 
\end{equation}
We see that the integral over the spatial component of the gauge field zero mode enforces the zero mode of the axial null constraint, much as the integral over the Dehn twist of the torus (the parameter $\tau_1$) implements the axial Virasoro constraint in string amplitudes.
The partition function can thus be rewritten as 
\begin{multline}
    Z^{\frac{\mathfrak{sl}(2, \mathds R)}{\mathfrak{u}(1)}}(\tau, z)  = \frac{\sqrt{\tau_2}}{\sqrt{\kappa_{sl}} \, |\eta(\tau)|^6} \, e^{\frac{2\pi}{\tau_2} \frac{\kappa_{sl}+2}{\kappa_{sl}} z_1^2}  \sum_{w, p \in \mathds Z} \ \sum_{r, \bar r \in \mathds Z} \ \sum_{n_f^{sl}, \bar n_f^{sl} \in \mathds Z + \frac{1}{2}} \delta_{\mathcal J_0 - \bar{\mathcal J}_0,0} \, S_r(\tau) \overline{S_{\bar r}(\tau)}  \\
    \times q^{\frac{p_L^2}{4}} \bar q^{\frac{p_R^2}{4}} e^{-\frac{2}{\sqrt{\kappa_{sl}}} \pi i p_L z-\frac{2}{\sqrt{\kappa_{sl}}} \pi i p_R \bar z}\, q^{\frac{1}{2}(n_f^{sl})^2} \bar q^{\frac{1}{2}(\bar n_f^{sl})^2} \, e^{2 \pi i (z n_f^{sl}-\bar z \bar n_f^{sl})} \int_0^1 \text d s_1 \, \mathcal I_{s_1} \ ,
    \label{sl2-pf-before-split}
\end{multline}
where
\begin{equation}
    \mathcal I_{s_1} = e^{-\pi \kappa_{sl} \tau_2 s_1^2} \, e^{-2 \pi s_1 \tau_2 (w \kappa_{sl}+r+\bar r+1+n_f^{sl}+\bar n_f^{sl})} \ . 
\end{equation}

\paragraph{Integral over $\boldsymbol{s_1}$:} 
The integral over $s_1$ does not yield a similar delta function imposing the zero mode of the vector null constraint, since the integrand is Gaussian in $s_1$.  One might worry that this means that unphysical states (\ie\ not satisfying this constraint) can flow through the torus amplitude; however, we will see that only physical states appear when we decompose the partition sum into characters.

\begin{figure}
\begin{center}
\begin{tikzpicture}
\draw[->,gray,thick] (0,-3) -- (0,+3);
\draw[->,gray,thick] (-4.1,0) -- (4.5,0);
\draw[line width=0.75, red] (-3,0.05) -- (3,0.05);
\draw[line width=0.75, red, dashed] (-3.8,0.05) -- (-3,0.05);
\draw[line width=0.75, red, dashed] (3,0.05) -- (3.8,0.05);
\draw[-stealth,red] (1.5,0.05) -- (1.7,0.05);
\draw[-stealth,red] (-1.7,0.05) -- (-1.5,0.05);

\draw[line width=0.75, blue] (-3,-0.05) -- (3,-0.05);
\draw[line width=0.75, blue, dashed] (-3.8,-0.05) -- (-3,-0.05);
\draw[line width=0.75, blue, dashed] (3,-0.05) -- (3.8,-0.05);
\draw[-stealth,blue] (1.5,-0.05) -- (1.7,-0.05);
\draw[-stealth,blue] (-1.7,-0.05) -- (-1.5,-0.05);

\draw[line width=0.75, blue] (-3,-1.00) -- (3,-1.00);
\draw[line width=0.75, blue, dashed] (-3.8,-1) -- (-3,-1);
\draw[line width=0.75, blue, dashed] (3,-1) -- (3.8,-1);
\draw[stealth-,blue] (1.5,-1) -- (1.7,-1);
\draw[stealth-,blue] (-1.7,-1) -- (-1.5,-1);

\node at (0.5,-0.8) {$\Scale[0.7]{-i \, \kappa_{sl} \, \tau_2}$};

\node at (-2.5,0.3) {$\Scale[0.9]{\textcolor{red}{\Gamma}}$};
\node at (-2.5,-0.3) {$\Scale[0.9]{\textcolor{blue}{\Gamma_{dis}}}$};
\node at (-2.5,-1.3) {$\Scale[0.9]{\textcolor{cyan}{\Gamma_{con}}}$};
\node at (4.1,0.2) {$\Scale[0.7]{\textcolor{gray}{\text{Im}\,c}}$};
\node at (0.3,2.7) {$\Scale[0.7]{\textcolor{gray}{\text{Re}\,c}}$};

\draw[line width=0.75, cyan] (-3,-1.05) -- (3,-1.05);
\draw[line width=0.75, cyan, dashed] (-3.8,-1.05) -- (-3,-1.05);
\draw[line width=0.75, cyan, dashed] (3,-1.05) -- (3.8,-1.05);
\draw[-stealth,cyan] (1.2,-1.05) -- (1.4,-1.05);
\draw[-stealth,cyan] (-1.5,-1.05) -- (-1.3,-1.05);

\end{tikzpicture}
\end{center}
\caption{The integration contour $\Gamma$ (in red) is equivalent to the sum of the contours $\Gamma_{dis}$ (in blue) and $\Gamma_{con}$ (in cyan), $\Gamma = \Gamma_{dis} + \Gamma_{con}$. This decomposition of integration contours induces the decomposition of the spectrum into discrete and continuous representations of $\mathfrak{sl}(2, \mathds R)$.}
\label{fig:int-contours}
\end{figure}
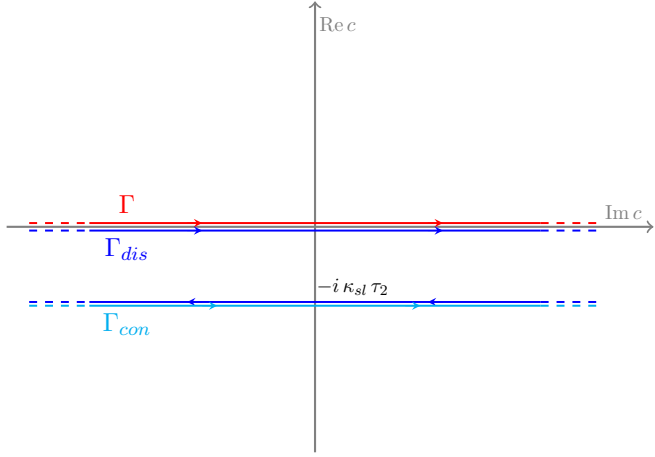

We can perform the integral over $s_1$ making use of the identity
\begin{equation}
\sqrt a \, e^{-\pi a s^2} = \int_{-\infty}^\infty \text dc \, e^{-\frac{\pi}{a}c^2-2 \pi i \, c \, s} \ ,
\label{c-linearize}
\end{equation}
which can be checked simply by completing the square. The integral over $s_1$ then gives
\begin{align}
    \int_0^1 \text d s_1 \, \mathcal I_{s_1} = & -\frac{1}{2 \pi \sqrt{\kappa_{sl}  \tau_2}} \int_{-\infty}^\infty \text dc \, \frac{e^{-\frac{\pi c^2}{\kappa_{sl} \tau_2 }} \, e^{- 2\pi (i c + \tau_2 (w \kappa_{sl} + r + \bar r + 1 + n_f^{sl} + \bar  n_f^{sl}))}}{i c  + \tau_2 (w \kappa_{sl} + r + \bar r + 1 + n_f^{sl} + \bar  n_f^{sl})} \nonumber \\
    & +\frac{1}{2 \pi \sqrt{\kappa_{sl}  \tau_2}} \int_{-\infty}^\infty \text dc \, \frac{e^{-\frac{\pi c^2}{\kappa_{sl} \tau_2 }}}{i c  + \tau_2 (w \kappa_{sl} + r + \bar r + 1 + n_f^{sl} + \bar  n_f^{sl})} \ . 
\label{Is1-to-decompose}
\end{align}
Let us now consider the first integral in the right-hand-side of \eqref{Is1-to-decompose} and shift the integration contour from $\text{Im}\, c = 0$ to $\text{Im}\, c = - \kappa_{sl} \, \tau_2$, see Figure~\ref{fig:int-contours}. 
We can rewrite eq.~\eqref{Is1-to-decompose} as
\begin{equation}
\int_0^1 \text d s_1 \, \mathcal I_{s_1} = \mathcal I_{dis} + \mathcal I_{con} \ , 
\label{split-dis-con}
\end{equation}
with 
\begin{align}
\mathcal I_{dis} & = -\frac{1}{2 \pi \sqrt{\kappa_{sl}  \tau_2}} \oint_{\Gamma_{dis}} \text dc \, \frac{e^{-\frac{\pi c^2}{\kappa_{sl} \tau_2 }} \, e^{- 2\pi (i c + \tau_2 (w \kappa_{sl} + r + \bar r + 1 + n_f^{sl} + \bar  n_f^{sl}))}}{i c  + \tau_2 (w \kappa_{sl} + r + \bar r + 1 + n_f^{sl} + \bar  n_f^{sl})} \ , 
\label{short-string-int-sl} \\[0.2cm]
\mathcal I_{con} & =  \mathcal I_{con}^+  + \mathcal I_{con}^- \label{Icon-def}\\
\mathcal I_{con}^- & = -\frac{1}{2 \pi \sqrt{\kappa_{sl} \tau_2}} \, \int_{-\infty-i \tau_2 \kappa_{sl}}^{\infty-i \tau_2 \kappa_{sl}} \text dc \, \frac{e^{-\frac{\pi c^2}{\kappa_{sl} \tau_2 }} \, e^{- 2\pi (i c + \tau_2 (w \kappa_{sl} + r + \bar r + 1 + n_f^{sl} + \bar  n_f^{sl}))}}{i c  + \tau_2 (w \kappa_{sl} + r + \bar r + 1 + n_f^{sl} + \bar  n_f^{sl})} \label{Icon-def-m} \\[0.2cm]
\mathcal I_{con}^+ & = \frac{1}{2 \pi \sqrt{\kappa_{sl} \tau_2}} \, \int_{-\infty}^\infty \text dc \, \frac{e^{-\frac{\pi c^2}{\kappa_{sl} \tau_2}}}{i c  + \tau_2 (w \kappa_{sl} + r + \bar r + 1 + n_f^{sl} + \bar  n_f^{sl})} \ . \label{Icon-def-p}
\end{align}
The split performed in eq.~\eqref{split-dis-con} induces an analogous split for the partition function. We rewrite it as
\begin{equation}
Z^{\frac{\mathfrak{sl}(2, \mathds R)}{\mathfrak{u}(1)}}(\tau, z) = Z^{\frac{\mathfrak{sl}(2, \mathds R)}{\mathfrak{u}(1)}}_{dis}(\tau, z) + Z^{\frac{\mathfrak{sl}(2, \mathds R)}{\mathfrak{u}(1)}}_{con}(\tau, z) \ . 
\end{equation}
While legitimate, this decomposition may seem arbitrary at this stage. However, we will see momentarily that $Z^{\frac{\mathfrak{sl}(2, \mathds R)}{\mathfrak{u}(1)}}_{dis}$ and $Z^{\frac{\mathfrak{sl}(2, \mathds R)}{\mathfrak{u}(1)}}_{con}$ will respectively give rise to the discrete and continuous spectrum on the worldsheet. 

\paragraph{Short string spectrum:} Let us first consider the integral in eq.~\eqref{short-string-int-sl} and perform the change of variable 
\begin{equation}
c = i \tau_2 (1-2j_{sl}) \ . 
\label{c-to-jsl}
\end{equation}
We obtain 
\begin{equation}
    \mathcal I_{dis}  = \frac{i}{ \pi  \sqrt{\kappa_{sl}  \tau_2}} \oint_{\Gamma_{dis}} \text dj_{sl} \, \frac{e^{-\frac{\pi \tau_2}{\kappa_{sl}}(2 j_{sl}-1)^2} \, e^{- 2\pi \tau_2(2 j_{sl} + w \kappa_{sl} + r + \bar r + n_f^{sl} + \bar  n_f^{sl})}}{2 j_{sl} + w \kappa_{sl} + r + \bar r + n_f^{sl} + \bar  n_f^{sl}} \ . 
    \label{Idis-jsl}
\end{equation}
This integral can be evaluated by Cauchy's theorem. As the notation suggests, $j_{sl}$ is to be identified with the $\mathfrak{sl}(2,\mathds R)$ spin. In fact, the region \eqref{c-to-jsl} in terms of $j_{sl}$ reads   
\begin{equation}
\frac{1}{2} < j_{sl} < \frac{\kappa_{sl}+1}{2} \ , 
\end{equation}
which is the correct Maldacena-Ooguri bound \rcite{Maldacena:2000hw} for the short string spectrum. Moreover, poles only appear for $j_{sl} \in \mathds R$, again in agreement with the expectation for the short string spectrum. Notice that making the identifications 
\begin{equation}
(j^3_{sl})_0  = r+j_{sl}  \ ,  \qquad (\bar j^3_{sl})_0 = \bar r+j_{sl}  \ , \qquad  (j^3_{f,sl})_0 = n_f^{sl} \ , \qquad (\bar j^3_{f,sl})_0 = \bar n_f^{sl} \ , 
\end{equation}
characteristic of the short string spectrum, the denominator of the integrand of \eqref{Idis-jsl} can be written as the sum of the gauged currents \eqref{gauged-currents-sl}
\begin{equation}
    \mathcal J_0 + \bar{\mathcal J}_0 = 2 j_{sl} + w \kappa_{sl} + r + \bar r + n_f^{sl} + \bar  n_f^{sl} \ .
\end{equation}
Cauchy's theorem then gives
\begin{equation}
    \mathcal I_{dis}  = -\frac{2}{\sqrt{\kappa_{sl}  \tau_2}} (q \bar q)^{-\frac{1}{4 \kappa_{sl}}} \int_{\frac{1}{2}}^{\frac{\kappa_{sl}+1}{2}} \text dj_{sl} \, \delta(\mathcal J_0 + \bar{\mathcal J}_0) (q \bar q)^{-\frac{j_{sl}(j_{sl}-1)}{\kappa_{sl}}}\ . 
\end{equation}
Using equations \eqref{JBFR-rel-sl} and \eqref{gauged-currents-sl} the $z$-dependent terms in the second line of \eqref{sl2-pf-before-split} can be rewritten as
\begin{equation}
    e^{-\frac{2}{\sqrt{\kappa_{sl}}} \pi i p_L z-\frac{2}{\sqrt{\kappa_{sl}}} \pi i p_R \bar z} \, e^{2 \pi i (z n_f^{sl}-\bar z \bar n_f^{sl})} = e^{2\pi i z (J^{\mathcal R}_{sl})_0} \, e^{-2\pi i \bar z (\bar J^{\mathcal R}_{sl})_0} \ , 
\end{equation}
and the short string spectrum partition function reads 
\begin{multline}
    Z^{\frac{\mathfrak{sl}(2, \mathds R)}{\mathfrak{u}(1)}}_{dis}(\tau, z)  = \frac{e^{\frac{2\pi}{\tau_2} \frac{\kappa_{sl}+2}{\kappa_{sl}} z_1^2} }{|\eta(\tau)|^6} \, (q \bar q)^{-\frac{1}{4 \kappa_{sl}}}  \,  \sum_{w, p \in \mathds Z} \ \sum_{r, \bar r \in \mathds Z} \ \sum_{n_f^{sl}, \bar n_f^{sl} \in \mathds Z + \frac{1}{2}} \int_{\frac{1}{2}}^{\frac{\kappa_{sl}+1}{2}} \text dj_{sl} \, \delta_{\mathcal J_0 - \bar{\mathcal J}_0,0} \, \delta(\mathcal J_0 + \bar{\mathcal J}_0)  \\
    \times  \, q^{\frac{p_L^2}{4}} \bar q^{\frac{p_R^2}{4}} \, q^{\frac{1}{2}(n_f^{sl})^2} \bar q^{\frac{1}{2}(\bar n_f^{sl})^2} e^{2\pi i z (J^{\mathcal R}_{sl})_0} \, e^{-2\pi i \bar z (\bar J^{\mathcal R}_{sl})_0}  \, (q \bar q)^{-\frac{j_{sl}(j_{sl}-1)}{\kappa_{sl}}} \, S_r(\tau) \overline{S_{\bar r}(\tau)}  \ ,
\end{multline}
where we reabsorbed various $\kappa_{sl}$ and numerical factors into the overall normalization. Recognising the character of $\mathcal D^+_{j}$ representations at level $\kappa_{sl}+2$ \cite{Pakman:2003kh}, 
\begin{equation}
    \chi^+_{dis,j}(\tau) \overline{\chi^+_{dis,j}(\tau)} = \frac{(q\bar q)^{-\frac{j(j-1)}{\kappa_{sl}}}}{|\eta(\tau)|^6} \sum_{r, \bar r \in \mathds Z} S_r(\tau) \overline{S_{\bar r}(\tau)} \ , 
\end{equation}
it becomes clear that the discrete representation contribution to the partition function can be written as a sum of characters over a constrained Hilbert space, 
\begin{equation}
     Z^{\frac{\mathfrak{sl}(2, \mathds R)}{\mathfrak{u}(1)}}_{dis}(\tau, z)  = e^{\frac{2\pi}{\tau_2} \frac{\kappa_{sl}+2}{\kappa_{sl}} z_1^2} \, \text{Tr}_{\mathcal H \otimes \bar{\mathcal H}} \, q^{L_0 -\frac{c}{24} } \, \bar q^{\bar L_0 -\frac{c}{24} } \, e^{2 \pi i z (J^{\mathcal R}_{sl})_0} \, e^{-2 \pi i \bar z (\bar J^{\mathcal R}_{sl})_0} \Bigl|_{\mathcal J_0 = \bar{\mathcal J}_0=0} \ ,
     \label{sl2-pf-dis}
\end{equation}
where
\begin{equation}
    c = \frac{3(\kappa_{sl} + 2)}{\kappa_{sl}} 
\end{equation}
is the central charge of the $\mathcal N=2$ cigar theory and the Hilbert space $\mathcal H$ (similarly for $\bar{\mathcal H}$) is obtained by tensoring $\mathcal D^+_{j_{sl}}$ representations with free boson representations. The exponential factor $e^{\frac{2\pi}{\tau_2} \frac{\kappa_{sl}+2}{\kappa_{sl}} z_1^2}$ in \eqref{sl2-pf-dis} is the usual prefactor relating path integral and trace partition function, see Appendix~\ref{app:different-pfs}.

\paragraph{Long string spectrum:} We are left with the integral $\mathcal I_{con}$, see eqs.~\eqref{Icon-def}-\eqref{Icon-def-p}. The basic idea now is to manipulate $\mathcal I_{con}^-$, see eq.~\eqref{Icon-def-m}, so that the numerator of the integrand resembles the one in $\mathcal I_{con}^+$, see eq.~\eqref{Icon-def-p}. Let us first shift the integration contour of $\mathcal I_{con}^-$ to the real line, 
\begin{equation}
    \mathcal I_{con}^- = -\frac{1}{2 \pi \sqrt{\kappa_{sl} \tau_2}} \, \int_{-\infty}^{\infty} \text dc \, \frac{e^{-\frac{\pi c^2}{\kappa_{sl} \tau_2 }} \, e^{-\pi \tau_2 \kappa_{sl}} \, e^{- 2\pi\tau_2 (w \kappa_{sl} + r + \bar r + 1 + n_f^{sl} + \bar  n_f^{sl})}}{i c + \tau_2 \kappa_{sl} + \tau_2 (w \kappa_{sl} + r + \bar r + 1 + n_f^{sl} + \bar  n_f^{sl})} \ . 
\end{equation}
Consider the associated partition function 
\begin{multline}
    Z^{\frac{\mathfrak{sl}(2, \mathds R)}{\mathfrak{u}(1)}, -}_{con}(\tau, z) = \frac{\sqrt{\tau_2}}{\sqrt{\kappa_{sl}} \, |\eta(\tau)|^6} \, e^{\frac{2\pi}{\tau_2} \frac{\kappa_{sl}+2}{\kappa_{sl}} z_1^2}  \sum_{w, p \in \mathds Z} \ \sum_{r, \bar r \in \mathds Z} \ \sum_{n_f^{sl}, \bar n_f^{sl} \in \mathds Z + \frac{1}{2}} \delta_{\mathcal J_0 - \bar{\mathcal J}_0,0} \, S_r(\tau) \overline{S_{\bar r}(\tau)}  \\
    \times q^{\frac{p_L^2}{4}} \bar q^{\frac{p_R^2}{4}} e^{-\frac{2}{\sqrt{\kappa_{sl}}} \pi i p_L z-\frac{2}{\sqrt{\kappa_{sl}}} \pi i p_R \bar z}\, q^{\frac{1}{2}(n_f^{sl})^2} \bar q^{\frac{1}{2}(\bar n_f^{sl})^2} \, e^{2 \pi i (z n_f^{sl}-\bar z \bar n_f^{sl})} \mathcal I_{con}^- \ . 
\end{multline}
Shifting the summation labels by 
\begin{equation}
w \mapsto w -1 \ , \qquad     n_f^{sl} \mapsto n_f^{sl} -1 \ , \qquad \bar n_f^{sl} \mapsto \bar n_f^{sl} -1 \ , 
\end{equation}
and exploiting the presence of the Kronecker delta, we obtain
\begin{multline}
    Z^{\frac{\mathfrak{sl}(2, \mathds R)}{\mathfrak{u}(1)}, -}_{con}(\tau, z) = -\frac{e^{\frac{2\pi}{\tau_2} \frac{\kappa_{sl}+2}{\kappa_{sl}} z_1^2}}{2 \pi \kappa_{sl} \, |\eta(\tau)|^6}  \sum_{w, p \in \mathds Z} \ \sum_{r, \bar r \in \mathds Z} \ \sum_{n_f^{sl}, \bar n_f^{sl} \in \mathds Z + \frac{1}{2}} \delta_{\mathcal J_0 - \bar{\mathcal J}_0,0} \, q^{r} S_r(\tau) \bar{q}^{\bar r} \, \overline{S_{\bar r}(\tau)}  \\
    \times q^{\frac{p_L^2}{4}} \bar q^{\frac{p_R^2}{4}} e^{-\frac{2}{\sqrt{\kappa_{sl}}} \pi i p_L z-\frac{2}{\sqrt{\kappa_{sl}}} \pi i p_R \bar z}\, q^{\frac{1}{2}(n_f^{sl})^2} \bar q^{\frac{1}{2}(\bar n_f^{sl})^2} \, e^{2 \pi i (z n_f^{sl}-\bar z \bar n_f^{sl})} \\
    \times \int_{-\infty}^{\infty} \text dc \, \frac{e^{-\frac{\pi c^2}{\kappa_{sl} \tau_2 }}}{i c + \tau_2 (w \kappa_{sl} + r + \bar r - 1 + n_f^{sl} + \bar  n_f^{sl})} \ . 
\end{multline}
Using the identity \cite{Israel:2004ir}
\begin{equation}
q^r S_r(\tau) = S_{-r}(\tau) \ , 
\label{rmr-id}
\end{equation}
to change the sign of $r, \bar r$ and changing integration variable as $c = 2 \tau_2 s$ we find
\begin{multline}
    Z^{\frac{\mathfrak{sl}(2, \mathds R)}{\mathfrak{u}(1)}}_{con}(\tau, z) = \frac{e^{\frac{2\pi}{\tau_2} \frac{\kappa_{sl}+2}{\kappa_{sl}} z_1^2}}{\pi \kappa_{sl} \, |\eta(\tau)|^6}  \sum_{w, p \in \mathds Z} \ \sum_{r, \bar r \in \mathds Z} \ \sum_{n_f^{sl}, \bar n_f^{sl} \in \mathds Z + \frac{1}{2}} \delta_{\mathcal J_0 - \bar{\mathcal J}_0,0} \, S_r(\tau) \, \overline{S_{\bar r}(\tau)}  \\
    \times q^{\frac{p_L^2}{4}} \bar q^{\frac{p_R^2}{4}} \, q^{\frac{1}{2}(n_f^{sl})^2} \bar q^{\frac{1}{2}(\bar n_f^{sl})^2} \, e^{2 \pi i z( n_f^{sl} -\frac{2}{\kappa_{sl}} \frac{\sqrt{\kappa_{sl}}p_L}{2})} e^{-2 \pi i \bar z (\bar n_f^{sl} + \frac{2}{\kappa_{sl}}\frac{\sqrt{\kappa_{sl}}p_R}{2})} \\
    \times \int_{-\infty}^{\infty} \text ds \, (q \bar q)^{\frac{s^2}{\kappa_{sl}}} \left( \tfrac{1}{2 i s + w \kappa_{sl} + r + \bar r + 1 + n_f^{sl} + \bar  n_f^{sl}} - \tfrac{1}{ 2i s + w \kappa_{sl} - r - \bar r - 1 + n_f^{sl} + \bar  n_f^{sl}}\right) \ . 
    \label{Z-con-sl-int-1}
\end{multline}
Making use of the identity \cite{Israel:2004ir}
\begin{equation}
S_{-r-1}(\tau) = 1 - S_r(\tau) \ , 
\label{S-r-1=1-Sr}
\end{equation}
and neglecting $\kappa_{sl}$ and numerical prefactors, we can then rewrite \eqref{Z-con-sl-int-1} as 
\begin{multline}
    Z^{\frac{\mathfrak{sl}(2, \mathds R)}{\mathfrak{u}(1)}}_{con}(\tau, z) = \frac{e^{\frac{2\pi}{\tau_2} \frac{\kappa_{sl}+2}{\kappa_{sl}} z_1^2}}{|\eta(\tau)|^6}  \sum_{w, p \in \mathds Z} \ \sum_{n_f^{sl}, \bar n_f^{sl} \in \mathds Z + \frac{1}{2}}  \\
    q^{\frac{p_L^2}{4}} \bar q^{\frac{p_R^2}{4}} \, q^{\frac{1}{2}(n_f^{sl})^2} \bar q^{\frac{1}{2}(\bar n_f^{sl})^2} \, e^{2 \pi i z( n_f^{sl} -\frac{2}{\kappa_{sl}} \frac{\sqrt{\kappa_{sl}}p_L}{2})} e^{-2 \pi i \bar z (\bar n_f^{sl} + \frac{2}{\kappa_{sl}}\frac{\sqrt{\kappa_{sl}}p_R}{2})} \, \int_{-\infty}^{\infty} \text ds \, (q \bar q)^{\frac{s^2}{\kappa_{sl}}} \\
    \sum_{r, \bar r \in \mathds Z} \Bigl\{ \Bigl( \tfrac{1}{2 i s + w \kappa_{sl} + r + \bar r + 1 + n_f^{sl} + \bar  n_f^{sl}} - \tfrac{1}{ 2i s + w \kappa_{sl} - r - \bar r - 1 + n_f^{sl} + \bar  n_f^{sl}}\Bigr)(S_r \overline{S_{\bar r}} - (1-S_r)(1-\overline{S_{\bar r}})) \\
    + \Bigl( \tfrac{1}{2 i s + w \kappa_{sl} + r - \bar r + n_f^{sl} + \bar  n_f^{sl}} - \tfrac{1}{ 2i s + w \kappa_{sl} - r + \bar r + n_f^{sl} + \bar  n_f^{sl}}\Bigr)(S_r (1-\overline{S_{\bar r}}) - (1-S_r)\overline{S_{\bar r}}) \Bigr\} \, \delta_{\mathcal J_0 - \bar{\mathcal J}_0, 0} \ . 
\end{multline}
The role of $\mathfrak{sl}(2, \mathds R)_{n_5}$ descendants in the path integral formalism is not well understood, see the comments about this in \cite{Israel:2004ir, Ashok:2020dnc}. We will then focus on the contribution of primaries, which corresponds to setting $S_r(\tau) \to 1$. Proceeding along the lines of \cite{Maldacena:2000kv, Hanany:2002ev} and introducing the density of states, 
\begin{equation}
\rho_{cig}(s) = \sum_{r_+ =0}^\infty  \Bigl( \frac{1}{2is + r_+ +1 + w \kappa_{sl} + n_f^{sl} + \bar n_f^{sl} }+ \frac{1}{2is + r_+ +1 - w \kappa_{sl} - n_f^{sl} - \bar n_f^{sl} } \Bigr) \ ,
\label{rho-cigar}
\end{equation} 
the contribution of $\mathfrak{sl}(2, \mathds R)_{n_5}$ primaries reads
\begin{multline}
    Z^{\frac{\mathfrak{sl}(2, \mathds R)}{\mathfrak{u}(1)}}_{con}(\tau, z)\Big|_{\mathfrak{sl}(2, \mathds R) \text{ prim.}} = \frac{e^{\frac{2\pi}{\tau_2} \frac{\kappa_{sl}+2}{\kappa_{sl}} z_1^2}}{|\eta(\tau)|^6}  \sum_{w, p \in \mathds Z} \ \sum_{n_f^{sl}, \bar n_f^{sl} \in \mathds Z + \frac{1}{2}}  \, q^{\frac{p_L^2}{4}} \bar q^{\frac{p_R^2}{4}} \, q^{\frac{1}{2}(n_f^{sl})^2} \bar q^{\frac{1}{2}(\bar n_f^{sl})^2}  \\
    e^{2 \pi i z( n_f^{sl} -\frac{2}{\kappa_{sl}} \frac{\sqrt{\kappa_{sl}}p_L}{2})} e^{-2 \pi i \bar z (\bar n_f^{sl} + \frac{2}{\kappa_{sl}}\frac{\sqrt{\kappa_{sl}}p_R}{2})} \, \int_{-\infty}^{\infty} \text ds \, (q \bar q)^{\frac{s^2}{\kappa_{sl}}} \rho_{cig}(s) \ ,
\end{multline}
where we used that for $r_- = r - \bar r$, 
\begin{equation}
 \sum_{r_- \in \mathds Z}  \, \delta_{\mathcal J_0 - \bar{\mathcal J}_0,0} = 1 \ .  
\end{equation}
The density of states \eqref{rho-cigar} is strictly speaking diverging and should be regularized. The correct regularization and the associated physics has been explained in detail in \cite{Maldacena:2000kv}, see also \cite{Hanany:2002ev, Israel:2004ir, Eguchi:2004yi,  Eguchi:2010cb, Troost:2010ud}.
Notice that 
\begin{align}
    e^{2 \pi i z( n_f^{sl} -\frac{2}{\kappa_{sl}} \frac{\sqrt{\kappa_{sl}}p_L}{2})} &= e^{2 \pi i z (J^{\mathcal R}_{sl})_0}\Big|_{\mathcal J_0 = 0}\\
    e^{-2 \pi i \bar z (\bar n_f^{sl} + \frac{2}{\kappa_{sl}}\frac{\sqrt{\kappa_{sl}}p_R}{2})} & =  e^{-2 \pi i \bar z (\bar J^{\mathcal R}_{sl})_0}\Big|_{\bar{\mathcal J}_0 = 0} \ , 
\end{align}
and hence the partition function can be written in a form making manifest the gauge constraints $\mathcal J_0 =  \bar{\mathcal J}_0 =  0$, 
\begin{align}
    Z^{\frac{\mathfrak{sl}(2, \mathds R)}{\mathfrak{u}(1)}}_{con}(\tau, z)\Big|_{\mathfrak{sl}(2, \mathds R) \text{ prim.}} = & \,  \frac{e^{\frac{2\pi}{\tau_2} \frac{\kappa_{sl}+2}{\kappa_{sl}} z_1^2}}{|\eta(\tau)|^6} \sum_{n_f^{sl}, \bar n_f^{sl} \in \mathds Z + \frac{1}{2}}  \, q^{\frac{1}{2}(n_f^{sl})^2} \, e^{2 \pi i z \frac{\kappa_{sl}+2}{\kappa_{sl}} n_f^{sl}} \bar q^{\frac{1}{2}(\bar n_f^{sl})^2} e^{-2 \pi i \bar z \frac{\kappa_{sl}+2}{\kappa_{sl}} \bar n_f^{sl}} \nonumber \\
   & \sum_{w, p \in \mathds Z} q^{\frac{p_L^2}{4}} \bar q^{\frac{p_R^2}{4}} \int_{-\infty}^\infty \text d((j^3_{sl})_0+(\bar j^3_{sl})_0) \sum_{(j^3_{sl})_0-(\bar j^3_{sl})_0 \in \mathds Z} \nonumber \\
   & \int_{-\infty}^{\infty} \text ds \, \rho_{cig}(s) \, \chi_c(s, (j^3_{sl})_0)\, \overline{\chi_c(s, (\bar j^3_{sl})_0)}  \, \delta_{\mathcal J_0 - \bar{\mathcal J}_0, 0} \, \delta(\mathcal J_0 + \bar{\mathcal J}_0)\ , 
\end{align}
where we introduced the character counting primaries of continuous $\mathfrak{sl}(2,\mathds R)_{n_5}$ representations
\begin{equation}
\chi_{\text{c}}(s,m) = q^{\frac{s^2}{\kappa_{sl}}} \, \xi^{\frac{2m}{\kappa_{sl}}} \ . 
\label{charact-cont}
\end{equation} 
The situation for current algebra descendants is less clear, in particular it has not been shown that their contributions can be written manifestly in terms of a sum over states that satisfy the zero mode of the vector null constraint.


\section{\texorpdfstring{$\boldsymbol{\big(\frac{\sltwo}{\uone}\!\times\!\frac{\sutwo}{\uone}\big)/\bZ_{\nfive}}$}{} coset orbifold} 
\label{sec:cosetorb}

We now come to a more intricate example of null gauging, namely the supersymmetric coset
\begin{equation}
\frac{SL(2,\mathds R)_{n_5} \times SU(2)_{n_5}}{U(1)_L \times U(1)_R}  \ .
\label{CB-coset}
\end{equation}
Now $H$ embeds in the numerator group, but is not a factor in it to be trivially removed by the additional gauge symmetry \eqref{second-gauge-sym}. We are going to compute the associated partition function 
\begin{equation}
\text{Tr}_R\Big[(-1)^F q^{L_0-\frac{c}{24}}\,  \bar q^{\bar L_0-\frac{c}{24}} \, e^{2 \pi i z J^{\cR}_0} \,  e^{-2 \pi i \bar z \bar J^{\cR}_0}\Big] \ , 
\end{equation}
where 
\begin{equation}
    J^{\cR} = J^{\cR}_{sl} + J^{\cR}_{su} \ , \qquad     \bar J^{\cR} = \bar J^{\cR}_{sl} + \bar J^{\cR}_{su} \ , 
\end{equation}
denote the holomorphic and anti-holomorphic $\mathcal R$-currents. Also in this case we consider the $R$ sector with insertions of $(-1)^F = (-1)^{F_L+F_R}$. As discussed already in Section~\ref{sec:standard=null}, see eq.~\eqref{cosetorb-equiv}, the coset CFT \eqref{CB-coset} is believed to be equivalent to the coset orbifold 
\begin{equation}
    \left(\frac{\sltwo_{n_5}}{\uone}\!\times\!\frac{\sutwo_{n_5}}{\uone}\right)\Bigr/\bZ_{\nfive} \ , 
    \label{coset-orbifold-5}
\end{equation}
playing a central role in string models of $n_5$ parallel NS5-branes and little string theory \rcite{Giveon:1999px,Giveon:1999tq}. The partition function of the coset orbifold~\eqref{coset-orbifold-5} was computed in~\rcite{Israel:2004ir,Eguchi:2004ik,Eguchi:2010cb,Giveon:2015raa}.
In Section \ref{sec:the coset orbifold} we show that the partition function of the null gauge coset CFT \eqref{CB-coset} exactly agrees with the known partition function \rcite{Israel:2004ir} of the coset orbifold \eqref{coset-orbifold-5}.  

\subsection{Null gauge coset}
\label{sec:null gauge coset}

Let us now derive the partition function for the null gauge coset~\eqref{CB-coset}, with embeddings 
\begin{equation}
\varepsilon_L^{sl} = \varepsilon_R^{sl} = -\frac{\sigma_3}{2} \ ,   \qquad \varepsilon_L^{su} = - \varepsilon_R^{su} = -\frac{\sigma_3}{2} \ .
\label{embeddings-CB}
\end{equation}
Following the general discussion of Section~\ref{sec:decoupling}, the partition function can be written as 
\begin{equation}
    Z^{\frac{\mathfrak{sl}(2,\mathds R) \times \mathfrak{su}(2)}{\mathfrak u(1)_L \times \mathfrak u(1)_R}}(\tau, z) = \int_0^1 \text d s_1 \int_0^1 \text ds_2 \,  Z_{gh} \, Z^{B}_{sl} \, Z^{B}_{su} \, Z^{F}_{sl}\, Z^{F}_{su} \ .  
\end{equation}
Since the embeddings \eqref{embeddings-CB} respectively equal the embeddings chosen in Sections \ref{sec:su2/u1} and \ref{sec:sl2/u1}, compare with eqs.~\eqref{SU(2)-U(1)-axial-embeddings} and \eqref{SL2-U1-embeddings}, it is easy to evaluate the various contributions, see eqs.~\eqref{Z-B-su}, \eqref{SU(2)-U(1)-fermions-pf}, \eqref{bosonic-sl2-pf}, \eqref{S-XY-sl}, \eqref{SL-U(1)-fermions-pf} with $\kappa_{sl} = \kappa_{su} = n_5$, 
\begin{align}
Z_{sl}^B &= Z^{GK}_{sl}(\tau,u + \tfrac{2}{n_5}z) \, \exp \left( - \mfrac{n_5+2}{\pi} \medint\int \text d^2 v \, |\partial_v(Y + \Phi[u + \tfrac{2}{n_5}z])|^2 \right) \ , \label{CB-boson-sector-sl}\\[.2cm]
Z^B_{su} & = e^{-\frac{\pi \, (n_5-2)}{\tau_2} (u_2-\frac{2}{n_5}z_2)^2}\exp \left( \mfrac{n_5-2}{\pi} \medint\int \text d^2 v \, |\partial_v(Y + \Phi[u-\tfrac{2}{n_5}z])|^2 \right) \mathcal Z_{B}^{\mathfrak{su}(2)}(\tau, u-\tfrac{2}{n_5}z) \ , \\[.2cm]
Z^F_{sl} &= e^{-\frac{2\pi}{\tau_2} (u_2 + \frac{n_5+2}{n_5}z_2)^2} \frac{| \theta_1(\tau, u + \tfrac{n_5+2}{n_5}z)|^2}{|\eta(\tau)|^2} \, \exp \left(\mfrac{2}{\pi} \medint\int \text d^2v \, \left| \partial_v (Y + \Phi[u + \tfrac{n_5+2}{n_5}z]) \right|^2 \right) \ , \\[.2cm]
Z^F_{su} &= e^{-\frac{2\pi}{\tau_2} (u_2 + \frac{n_5-2}{n_5}z_2)^2} \, \frac{\big| \theta_1(\tau, u + \tfrac{n_5-2}{n_5}z)\big|^2}{|\eta(\tau)|^2} \,  \exp \left(\mfrac{2}{\pi} \medint\int \text d^2v \,  \big| \partial_v (Y + \Phi[u + \tfrac{n_5-2}{n_5}z])\big|^2   \right) \ . \label{CB-fermion-sector-su}
\end{align}
Gathering all the terms in eqs.~\eqref{CB-boson-sector-sl}-\eqref{CB-fermion-sector-su}, together with the ghost contribution, after some algebra we obtain the null gauge coset partition function\footnote{Again, we will not be concerned with the overall normalization of the partition function.} 
\begin{align}
& Z^{\frac{\mathfrak{sl}(2,\mathds R) \times \mathfrak{su}(2)}{\mathfrak{u}(1)_L \times \mathfrak{u}(1)_R}}(\tau, z) = \tau_2 \, e^{\frac{4 \pi}{\tau_2} z_1^2} \int_0^1 \text d s_1 \int_0^1 \text ds_2 \, e^{-\frac{\pi (n_5+2)}{\tau_2}(u_2+\frac{2}{n_5}z_2)^2} \nonumber \\
& \hspace{70pt} \times \,  Z_{sl}^{GK}(\tau, u +\tfrac{2}{n_5}z ) \, \mathcal Z_B^{\mathfrak{su}(2)}(\tau, u-\tfrac{2}{n_5}z)  \, |\theta_1(\tau, u+\tfrac{n_5+2}{n_5}z)\, \theta_1(\tau, u+\tfrac{n_5-2}{n_5}z)|^2 \ ,
\label{pf-CB} 
\end{align}
where the partition functions $Z^{GK}_{sl}$ and $\mathcal Z_B^{\mathfrak{su}(2)}$ have been introduced in eqs.~\eqref{bosonic-sl2-pf} and \eqref{SU2-pf} respectively. 

\subsection{The coset orbifold}
\label{sec:the coset orbifold}

In this section we are going to show that the partition function \eqref{pf-CB} for the coset CFT \eqref{CB-coset} equals the partition function of the coset orbifold CFT \eqref{coset-orbifold-5}. The latter reads \cite{Israel:2004ir}
\begin{equation}
Z^{(\frac{\mathfrak{sl}(2, \mathds R)}{\mathfrak{u}(1)} \times \frac{\mathfrak{su}(2)}{\mathfrak{u}(1)})/\mathds Z_{n_5}}(\tau, z) = \mfrac{1}{2 \, n_5} \sum_{\alpha, \beta \in \mathds Z_{n_5}} Z_{\alpha, \beta}^{\frac{\mathfrak{sl}(2,\mathds R)}{u(1)}}(\tau, z) \, Z_{\alpha, \beta}^{\frac{\mathfrak{su}(2)}{u(1)}}( \tau, z) \ , 
\label{coset-orbifold-partition-function}
\end{equation}
where
\begin{multline}
 Z_{\alpha, \beta}^{\frac{\mathfrak{sl}(2,\mathds R)}{u(1)}}( \tau, z) = e^{\frac{2 \pi}{\tau_2}(\frac{n_5+2}{n_5}|z|^2 - \frac{n_5+4}{n_5}z_2^2)} \int_0^1 \text d s_1 \int_0^1 \text ds_2 \,  \left|\frac{\theta_1(\tau, u+\tfrac{n_5+2}{n_5}z)}{\theta_1(\tau, u+\tfrac{2}{n_5}z)} \right|^2 e^{-4 \pi \frac{u_2 z_2}{\tau_2}} \\
\times \, \sum_{m_1, m_2 \in \mathds Z} \exp \left( -\frac{\pi n_5}{\tau_2}\left|u + m_1 \tau + m_2 + \mfrac{\alpha \tau + \beta}{n_5}\right|^2 \right) 
\label{su(2)/u(1)-twisted-pf-text}
\end{multline}
and \rcite{Gepner:1986hr}
\begin{equation}
Z^{\frac{\mathfrak{su}(2)}{\mathfrak{u}(1)}}_{\alpha, \beta}(\tau,z) = \frac{e^{\frac{2 \pi}{\tau_2}\frac{n_5-2}{n_5}z_1^2}}{2} \sum_{j_{su}=0}^{n_5-2} \sum_{\ell = -n_5+1}^{n_5} e^{\frac{2 \pi i}{n_5}(\ell-\alpha)\beta} \, I^{j_{su}}_{-\ell+2\alpha}(\tau,z) \overline{I^{j_{su}}_{\ell}(\tau, z)}  \ , 
\end{equation}
see Appendix~\ref{app:su(2)} for the definition of the characters $I^{j_{su}}_\ell$. The exponential factor $e^{\frac{2 \pi}{\tau_2}\frac{n_5-2}{n_5}z_1^2}$ in eq.~\eqref{su(2)/u(1)-twisted-pf-text} is due to the relation between the trace partition function and the path integral partition function, see eq.~\eqref{pfs-relation}. It is a simple exercise to show that 
\begin{align}
   \sum_{m_1, m_2 \in \mathds Z} e^{-\frac{\pi n_5}{\tau_2} |u + m_1 \tau + m_2 |^2 } & = \sqrt{\frac{\tau_2}{n_5}} \, e^{-\frac{\pi n_5}{\tau_2}u_2^2}  \sum_{w,p \in \mathds Z} q^{n_5( \frac{p}{2n_5} - \frac{w}{2} )^2} \bar q^{n_5 ( \frac{p}{2n_5} + \frac{w}{2} )^2} \, y^{\frac{1}{2}(n_5w-p)} \, \bar y^{\frac{1}{2}(n_5w+p)} \\
   & = \sqrt{\frac{\tau_2}{n_5}} \, e^{-\frac{\pi n_5}{\tau_2}u_2^2} \, \sum_{\ell \in \mathds Z_{n_5}} \Theta^{(n_5)}_\ell(\tau, -u) \overline{\Theta^{(n_5)}_\ell(\tau, u)} \ ,  \label{Z_Y-to-Theta}
\end{align}
where in the first equality we used the Poisson resummation identity \rcite{DiFrancesco:1997nk}, while in the second equality we changed the summation variables according to 
\begin{equation}
p = n_5(m_L+m_R)+ \ell \ , \qquad w = m_R-m_L  \ , \qquad \text{with} \qquad \ell \in \mathds Z_{n_5} \ , \quad m_L, m_R \in \mathds Z  
\label{change-summation-labels-pw-to-ml}
\end{equation}
and used the definition \eqref{Theta-def}. In \eqref{change-summation-labels-pw-to-ml}, by $\ell \in \mathds Z_{n_5}$ we mean that $\ell$ can run over any set of $n_5$ consecutive integers. Using eq.~\eqref{Z_Y-to-Theta} we can then rewrite eq.~\eqref{su(2)/u(1)-twisted-pf-text} as 
\begin{multline}
 Z^{\frac{\mathfrak{sl}(2, \mathds R)}{\mathfrak{u}(1)}}_{\alpha, \beta}(\tau,z) = \sqrt{\frac{\tau_2}{n_5}} \, e^{\frac{2 \pi}{\tau_2}(\frac{n_5+2}{n_5}|z|^2 - \frac{n_5+4}{n_5}z_2^2)} \int_0^1 \text d s_1 \int_0^1 \text ds_2 \, \left|\frac{\theta_1(\tau, u+\tfrac{n_5+2}{n_5}z)}{\theta_1(\tau, u+\tfrac{2}{n_5}z)} \right|^2  \\
\times \, e^{-4 \pi \frac{u_2 z_2}{\tau_2}} e^{-\frac{\pi n_5}{\tau_2}(u_2 + \frac{\alpha \tau_2}{n_5})^2}\sum_{\ell \in \mathds Z_{n_5}} \Theta_\ell^{(n_5)}\left(\tau, -u -\tfrac{\alpha \tau + \beta}{n_5}\right) \overline{\Theta_\ell^{(n_5)}\left(\tau, u +\tfrac{\alpha \tau + \beta}{n_5}\right)} \ . 
\end{multline}
Recognising the $SL(2,\mathds R)$ partition function \eqref{bosonic-sl2-pf}, we find\footnote{We introduced the normalization constant $\mathcal N$ to parametrize our ignorance in the normalization of the SL$(2, \mathds R)$ partition function \eqref{bosonic-sl2-pf}. Since we have no control on the overall normalization, in the following we will reabsorb various factor of $n_5$ into $\mathcal N$. }
\begin{align}
 Z^{\frac{\mathfrak{sl}(2,\mathds R)}{\mathfrak{u}(1)}}_{\alpha, \beta}(\tau,z)  & = \mathcal N \,  \tau_2 \, e^{\frac{2 \pi}{\tau_2}\frac{n_5+2}{n_5}|z|^2} \int_0^1 \text d s_1 \int_0^1 \text ds_2 \, Z^{GK}_{sl}(\tau, u + \tfrac{2}{n_5}z) \, |\theta_1(\tau, u+\tfrac{n_5+2}{n_5}z)|^2 \nonumber \\
& \qquad \times \, e^{-\frac{\pi}{\tau_2}\left( 2\frac{(n_5+2)^2}{n_5^2}z_2^2 + (n_5+2)u_2^2+4\frac{n_5+2}{n_5} u_2 z_2 + 2 \alpha u_2 \tau_2 + \frac{\alpha^2 \tau_2^2}{n_5} \right)}\, \times \nonumber \\
& \qquad \times \, \sum_{\ell \in \mathds Z_{n_5}} \Theta_\ell^{(n_5)}\left(\tau, -u -\tfrac{\alpha \tau + \beta}{n_5}\right) \overline{\Theta_\ell^{(n_5)}\left(\tau, u +\tfrac{\alpha \tau + \beta}{n_5}\right)} \ , 
\end{align}
The coset orbifold partition function \eqref{coset-orbifold-partition-function} can thus be written as 
\begin{align}
Z^{(\frac{\mathfrak{sl}(2, \mathds R)}{\mathfrak{u}(1)} \times \frac{\mathfrak{su}(2)}{\mathfrak{u}(1)})/\mathds Z_{n_5}}(\tau, z) & =  \mathcal N \,  \tau_2 \, e^{\frac{4 \pi}{\tau_2} z_1^2} \sum_{\alpha \in \mathds Z_{n_5}} \sum_{j_{su}=0}^{n_5-2} \sum_{\ell' = -n_5+1}^{n_5} I^{j_{su}}_{-\ell'+2\alpha}(\tau,z) \overline{I^{j_{su}}_{\ell'}(\tau, z)}  \nonumber \\
& \quad  \times \, \int_0^1 \text d s_1 \int_0^1 \text ds_2 \, Z^{GK}_{sl}(\tau, u + \tfrac{2}{n_5}z) \, |\theta_1(\tau, u+\tfrac{n_5+2}{n_5}z)|^2\nonumber \\
& \quad \times \, e^{-\frac{\pi}{\tau_2} (n_5+2)(u_2+\frac{2}{n_5}z_2)^2-2\pi \alpha u_2 - \frac{\pi \alpha^2 \tau_2}{n_5} }  \nonumber \\
& \quad \times \, \sum_{\ell \in \mathds Z_{n_5}} \sum_{\beta \in \mathds Z_{n_5}} e^{\frac{2 \pi i}{n_5}(\ell'-\alpha)\beta} \Theta_\ell^{(n_5)}(\tau, -u -\tfrac{\alpha \tau + \beta}{n_5}) \overline{\Theta_\ell^{(n_5)} (\tau, u +\tfrac{\alpha \tau + \beta}{n_5})} \ . 
\label{Zorb-intermediate-1}
\end{align}
Consider the sum in the last line of \eqref{Zorb-intermediate-1}. Using the definition \eqref{Theta-def}, the representation of the Kronecker delta 
\begin{equation}
   \sum_{\beta \in \mathds Z_{n_5}} e^{\frac{2 \pi i \beta}{n_5}(\ell'-\alpha-\ell)} = n_5 \, \delta_{\alpha, \ell'-\ell} \ , 
\end{equation}
and the identities \eqref{Theta-id3}, \eqref{Theta-id2}, the last line of~\eqref{Zorb-intermediate-1} can be rewritten as
\begin{multline}
\sum_{\beta \in \mathds Z_{n_5}}  e^{\frac{2 \pi i}{n_5}(\ell'-\alpha)\beta} \Theta_\ell^{(n_5)} (\tau, -u -\tfrac{\alpha \tau + \beta}{n_5} ) \overline{\Theta_\ell^{(n_5)} (\tau, u +\tfrac{\alpha \tau + \beta}{n_5} )} \\
=   n_5 \, \delta_{\alpha, \ell'-\ell} \, e^{\frac{\pi \alpha^2 \tau_2}{n_5}} \, e^{2 \pi \alpha u_2} \, \Theta^{(n_5)}_{-\ell+\alpha}(\tau, u) \, \overline{\Theta_{\ell+\alpha}^{(n_5)}(\tau,u)} \ . 
\end{multline}
Substituting in eq.~\eqref{Zorb-intermediate-1} we find 
\begin{multline}
Z^{(\frac{\mathfrak{sl}(2, \mathds R)}{\mathfrak{u}(1)} \times \frac{\mathfrak{su}(2)}{\mathfrak{u}(1)})/\mathds Z_{n_5}}(\tau, z) = \mathcal N \,  \tau_2 \, e^{\frac{4 \pi}{\tau_2} z_1^2}  \int_0^1 \text d s_1 \int_0^1 \text ds_2 \, Z^{GK}_{sl}(\tau, u + \tfrac{2}{n_5}z) \, |\theta_1(\tau, u+\tfrac{n_5+2}{n_5}z)|^2 \\
 \times \, e^{-\frac{\pi (n_5+2)}{\tau_2} (u_2+\frac{2}{n_5}z_2)^2 } \, \sum_{j_{su}=0}^{n_5-2} \sum_{\ell' = -n_5+1}^{n_5} \overline{\Theta_{\ell'}^{(n_5)}(\tau,u) \, I^{j_{su}}_{\ell'}(\tau, z)} \sum_{\ell \in \mathds Z_{n_5}} \Theta^{(n_5)}_{-2\ell+\ell'}(\tau, u) \,   I^{j_{su}}_{-2\ell+\ell'}(\tau,z)\ . 
\end{multline}
We claim that the sum over $\ell$ can be rewritten as 
\begin{equation}
    \sum_{\ell \in \mathds Z_{n_5}} \Theta^{(n_5)}_{-2\ell+\ell'}(\tau, u) \,   I^{j_{su}}_{-2\ell+\ell'}(\tau,z) = \sum_{\ell = - n_5 +1}^{n_5} \Theta^{(n_5)}_{\ell}(\tau, u) \, I^{j_{su}}_{\ell}(\tau,z) \ . 
    \label{sum-l-identity}
\end{equation}
The proof slightly depends on the parity of $n_5$ and $j_{su}$. Let us first consider $n_5 \in 2 \mathds Z$ and $j_{su} \in 2 \mathds Z +1$. From eq.~\eqref{Ijl=0} it follows that we can restrict to $\ell' \in 2 \mathds Z$. Let us choose $\ell$ to run over 
\begin{equation}
    \frac{\ell' - n_5}{2} \leq \ell \leq \frac{n_5-2+\ell'}{2} \ . 
    \label{ell-interval-1}
\end{equation}
Notice that $\frac{\ell' - n_5}{2}, \frac{n_5-2+\ell'}{2} \in \mathds Z$ and that there are exactly $n_5$ integers in the interval \eqref{ell-interval-1}. Introducing a new sum label $\ell'' = -2 \ell + \ell'$ and using again eq.~\eqref{Ijl=0}, eq.~\eqref{sum-l-identity} follows. The argument for other choices of the parity of $n_5$ and $j$ is completely analogous. Hence, making use of eqs.~\eqref{SU2-pf} and \eqref{chi-theta1=ITheta-identity}, we finally obtain
\begin{multline}
Z^{(\frac{\mathfrak{sl}(2, \mathds R)}{\mathfrak{u}(1)} \times \frac{\mathfrak{su}(2)}{\mathfrak{u}(1)})/\mathds Z_{n_5}}(\tau, z) =   \mathcal N \, \tau_2 \, e^{\frac{4 \pi}{\tau_2} z_1^2} \int_0^1 \text d s_1 \int_0^1 \text ds_2  \, e^{-\frac{\pi (n_5+2)}{\tau_2} (u_2+\frac{2}{n_5}z_2)^2 } \, \times \\
\times \, Z^{GK}_{sl}(\tau, u + \tfrac{2}{n_5}z) \, \mathcal Z_B^{\mathfrak{su}(2)}(\tau, u - \tfrac{2}{n_5}z) \, |\theta_1(\tau, u+\tfrac{n_5+2}{n_5}z) \, \theta_1(\tau, u+\tfrac{n_5-2}{n_5}z)|^2 \ , 
\label{coset-orbifold-pf-final}
\end{multline}
which, up to the overall normalization, exactly reproduces the null gauged coset partition function~\eqref{pf-CB}.


\section{Supertube partition functions} 
\label{sec:supertube}

Finally, we come to a discussion of the generalized ``round supertube'' backgrounds of~\rcite{Martinec:2017ztd}.
These are null gauged models
\begin{equation}
\frac{G}{H} = 
\frac{SL(2, \mathds R)_{n_5} \times SU(2)_{n_5} \times \mathds R_t \times U(1)_y}{U(1)_L \times U(1)_R} \ , 
\label{st-coset}
\end{equation}
with the embeddings
\begin{equation}
\begin{aligned}
\varepsilon_L^{sl} &=  -\frac{\sigma_3}{2} \ , & \varepsilon_L^{su} &=  -\ell_2 \frac{\sigma_3}{2} \ , &  \varepsilon_L^{t} &= -\frac{\ell_3}{2} \ ,  &  \varepsilon_L^{y} &= \frac{\ell_4}{2} \ , \\
\varepsilon_R^{sl} &=  \frac{\sigma_3}{2} \ , & \varepsilon_R^{su} &=  -r_2 \frac{\sigma_3}{2} \ , &  \varepsilon_R^{t} &= -\frac{\ell_3}{2} \ ,  &  \varepsilon_{R}^y &= -\frac{r_4}{2} \ ,
\end{aligned}
\label{supertube-embeddings}
\end{equation}
satisfying the null conditions
\begin{equation}
n_5(-1+\ell_2^2) - \ell_3^2+\ell_4^2 = 0 \ , \qquad n_5(-1+r_2^2) - \ell_3^2+r_4^2 = 0 \ . 
\label{ST-null-constraints}
\end{equation}
Eq.~\eqref{ST-null-constraints} guarantees that the gauging is null, 
\begin{equation}
\begin{aligned}
    \sum_{i} \kappa_i \text{Tr}[(\varepsilon_L^i)^2] &=  -n_5 \, \text{Tr}\left[\left(\frac{\sigma_3}{2}\right)^2\right] + n_5 \ell_2^2 \, \text{Tr}\left[\left(\frac{\sigma_3}{2}\right)^2\right] - 2 \frac{\ell_3^2}{4} + 2 \frac{\ell_4^2}{4}  = 0 \ , \\
    \sum_{i} \kappa_i \text{Tr}[(\varepsilon_R^i)^2] &= -n_5 \, \text{Tr}\left[\left(\frac{\sigma_3}{2}\right)^2\right] + n_5 r_2^2 \, \text{Tr}\left[\left(\frac{\sigma_3}{2}\right)^2\right] - 2 \frac{\ell_3^2}{4} + 2 \frac{r_4^2}{4}  = 0 \ .
\end{aligned}
\end{equation}

Following the logic of Section~\ref{sec:decoupling}, the partition function  of the asymmetric coset \eqref{st-coset} can be written as
\begin{multline}
Z^{ST}(\tau, z) = Z_{tr}^F(\tau, z) \int_0^1 \text ds_1 \int_0^1 \text ds_2 \,  Z_{gh}(\tau)  \, Z^{F}_{sl}(\tau, u+\tfrac{n_5+2}{n_5}z) \, Z^{F}_{su}(\tau, u+\tfrac{n_5-2}{n_5}z) \\
Z^{B}_{sl}(\tau, u+\tfrac{2}{n_5}z) \, Z^{B}_{su}(\tau, u -\tfrac{2}{n_5}z) \, Z^{B}_t(\tau, u) \, Z^{B}_y(\tau, u) \ , 
\label{ST-pf-general}
\end{multline}
where, in order to include the dependence on the $\mathcal R$-charge chemical potential $z$, the chemical potential $u$ is shifted as in eqs.~\eqref{su2-chemical-potentials-relation} and \eqref{sl2-chemical-potentials-relation}. In eq.~\eqref{ST-pf-general}, $Z_{tr}^F(\tau, z)$ is the contribution of the two uncharged fermions in the ``trasverse space'', see the discussion in Section~\ref{sec:decoupling} and Appendix \ref{app:GWZW in superspace}. 
After briefly reviewing in Section \ref{sec:supertube-cosets} some properties of supertube cosets, in Section \ref{sec:ST-coset-pf} we compute the partition function of the supertube coset \eqref{st-coset} in its integral form while in Section \ref{sec:Evaluating the gauge field zero mode integrals} we decompose it into characters and explain how the gauge constraints emerge from the path integral perspective. 

\subsection{Supertube cosets}
\label{sec:supertube-cosets}

A convenient parametrization of the embedding parameters $\ell_i$ and $r_i$ is \cite{Bufalini:2021ndn}
\begin{align}
\label{li ri vals}
\begin{split}
\ell_2 &= \sfm+\sfn \in 2\mathds Z+1
~,~~~
r_2=-\sfm+\sfn \in 2\mathds Z+1
~,~~~
\\[.2cm]
\ell_3 &= r_3 =  -\sqrt{\sfk^2 R_y^2+\frac{\sfp^2}{R_y^2}+n_5(\sfm^2+\sfn^2-1)} ~,
\\[.2cm]
\ell_4 &= -{\sfk} R_y + \frac{\sfp}{R_y}
~,~~~
r_4 = {\sfk} R_y + \frac{\sfp}{R_y} ~,
~~~~~~~~
\sfp = n_5\frac{\sfm \sfn }{\sfk }~,
\end{split}
\end{align}
where $\sfm, \sfn, \sfk, \sfp \in \mathds Z$. For $\sfm=1,\sfn=0$, 
the model describes 1/2-BPS states of NS5-branes and fundamental strings in which the NS5-brane source lies at the origin in $\sltwo$ and along a great circle in $\sutwo$~\rcite{Lunin:2001fv}.  For $\sfm-\sfn=1,\sfn\ne0$, the background is 1/4-BPS and describes an NS5-F1 bound state carrying a macroscopic angular momentum along $\sutwo_L$~\rcite{Giusto:2004id,Giusto:2012yz}.  Finally, general values of $\sfm,\sfn$ describe non-supersymmetric NS5-F1 states having both left and right $\sutwo$ angular momentum~\rcite{Jejjala:2005yu}.

One sees all these features by choosing a gauge that fixes two of the coordinates of $G$, and then integrating out the gauge field to arrive at an effective sigma model action from which one reads off the metric and $B$-field.  The schematic form of the gauge field dependence in the action~\eqref{asym-act-A} is
\be
\int \big( \cA \bar\cJ + \bar\cA \cJ + \Sigma \cA\bar\cA
\big)~,
\ee
leading to a term in the effective action $\frac{\cJ\bar\cJ}{\Sigma}$.
The momentum, angular momentum and string winding charges carried by the background arise from various cross-terms between the $\uone_y$, $\sutwo$ and $\mathds R_t$ contributions to the null currents $\cJ,\bar\cJ$; see \eg~\rcite{Martinec:2017ztd} for further details.

\subsection{The coset partition function}
\label{sec:ST-coset-pf}

In this section we compute the partition function \eqref{ST-pf-general} of the coset CFT \eqref{st-coset}. 

\subsubsection{Fermionic sector}

Let us begin by considering the contributions in the first line of \eqref{ST-pf-general}. 

\paragraph{$\boldsymbol{SL(2,\mathds R)}$:} The contribution of $\mathfrak{sl}(2, \mathds R)$ fermions has already been computed in Section~\ref{sec:sl2/u1} and directly follows from eq.~\eqref{SL-U(1)-fermions-pf},
\begin{equation}
Z^{F}_{sl}(\tau, u+\tfrac{n_5+2}{n_5}z) = e^{-\frac{2\pi}{\tau_2} (u_2 + \frac{n_5+2}{n_5}z_2)^2} \frac{\big| \theta_1(\tau, u +\frac{n_5+2}{n_5} z)\big|^2}{|\eta(\tau)|^2} \, \exp \Bigl(\tfrac{2}{\pi} \medint\int \text d^2v \left| \partial_v \big(Y + \Phi[u+\tfrac{n_5+2}{n_5}z]\big) \right|^2 \Bigr) \ . 
\label{St-pf-SL2F}
\end{equation}

\paragraph{$\boldsymbol{SU(2)}$:}  The $SU(2)$ fermionic action 
\begin{equation}
S^F_{{su}} = \medint\int \mfrac{\text d^2v}{2 \pi} \left[  \psi^+_{su}(\partial_{\bar v} + \ell_2 A_{\bar v} )\psi^-_{su} + \psi^-_{su}(\partial_{\bar v} - \ell_2 A_{\bar v} )\psi^+_{su} + \tilde \psi^+_{su}(\partial_v + r_2 A_v ) \tilde \psi^-_{su} + \tilde \psi^-_{su}(\partial_v - r_2 A_v ) \tilde \psi^+_{su} \right] 
\label{SU(2)-fermion-action-lr-f}
\end{equation}
is gauge invariant under 
\begin{equation}
\begin{aligned}
\psi^\pm_{su}  \mapsto e^{\pm \, \ell_2 \,  \gamma} \, \psi^\pm_{su} \quad &, \qquad 
A_v  \mapsto A_v + \partial_v\gamma  \ ,  
\\
\tilde \psi^\pm_{su}  \mapsto e^{\pm \, r_2 \, \gamma} \, \tilde \psi^\pm_{su} \quad &, \qquad 
A_{\bar v}  \mapsto A_{\bar v} + \partial_{\bar v} \gamma  \ . 
\end{aligned}
\label{psi-gauge-trsf-lr-f}
\end{equation}
As one would expect, for $\ell_2=r_2=1$ eq.~\eqref{SU(2)-fermion-action-lr-f} reduces to eq.~\eqref{SU(2)-fermionic-action}. In order to preserve gauge invariance at the quantum level, we assume the fermion path integral measure to be invariant under ``rotations'' of the form \eqref{psi-gauge-trsf-lr-f}.
Performing a gauge transformation with parameter $\gamma= - X$, we can then eliminate $X$ from the action \eqref{psi-gauge-trsf-lr-f}  without picking up any anomaly factor. Following the same punchline of Section~\ref{sec:su(2)/u(1)-fermions}, we introduce the fermions
\begin{equation}
\begin{aligned}
\eta^+ & = e^{\ell_2(-iY-i \Phi)} \psi^+_{su} \ , & \eta^- &= e^{\ell_2(+iY+i \Phi)} \psi^-_{su} \ , \\
\tilde \eta^+ & = e^{r_2(iY+i \Phi)} \tilde \psi^+_{su} \ , & \tilde \eta^- &= e^{r_2(-iY-i \Phi)} \tilde \psi^-_{su} \ . 
\end{aligned}
\label{twisted-fermions-lr-f}%
\end{equation}
As it was the case in Section~\ref{sec:su(2)/u(1)-fermions}, this transformation produces an anomaly, 
\begin{equation}
\mathcal{D} \psi^+_{su} \, \mathcal{D} \psi^-_{su} \, \mathcal{D} \tilde \psi^+_{su} \, \mathcal{D} \tilde \psi^-_{su}  = \mathcal{D} \eta^+ \, \mathcal{D} \eta^- \, \mathcal{D} \tilde \eta^+ \, \mathcal{D} \tilde \eta^- \, e^{\check{\mathfrak A}} \ , 
\label{axial-anomaly-lr}
\end{equation}
which can formally be computed by promoting 
\begin{equation}
\begin{pmatrix} \ 1 \  & 0 \\ 0 & -1 \  \end{pmatrix} \mapsto \begin{pmatrix} \ \ell_2 \  & 0 \\ 0 & -r_2 \  \end{pmatrix}  
\end{equation}
in \eqref{su2-fermion-anomalous-rotation} and in the various formulae for the axial anomaly. We find 
\begin{equation}
\check{\mathfrak A} = \frac{i}{\pi} \int \text d^2v \, (\ell_2 \, \partial_v \alpha_\ell \, A_{\bar v} - r_2 \,  \partial_{\bar v} \alpha_r \,  A_v)  = \frac{\ell^2_2 + r^2_2}{\pi} \int \text d^2v \, |\partial_v(Y + \Phi[u])|^2  \ , 
\end{equation} 
where 
\begin{equation}
\alpha_\ell = \ell_2 (-Y - \Phi) \ , \qquad \alpha_r = -r_2 (Y + \Phi) \ . 
\end{equation}
Let us recall that, as also discussed in Section \ref{sec:su2/u1}, one should shift the chemical potentials according to
\begin{equation}
\ell_2 u \, \mapsto \, \ell_2 u + \tfrac{n_5-2}{n_5} z \ ,  \qquad r_2 \bar u \, \mapsto \, r_2 \bar  u + \tfrac{n_5-2}{n_5} \bar z \ . 
\end{equation}
As a result, the anomaly reads
\begin{align}
\check{\mathfrak A} &  = \frac{\ell_2^2+r_2^2}{\pi} \int \text d^2v \, |\partial_v (Y+ \Phi[u])|^2 + \frac{2}{\pi} \int \text d^2v \, |\partial_v \Phi[\tfrac{n_5-2}{n_5}z]|^2 \nonumber \\
& \qquad + \frac{\ell_2+r_2}{\pi} \int \text d^2v \, \left[ \partial_v \Phi[\tfrac{n_5-2}{n_5}z] \,  \partial_{\bar v} (Y + \Phi[u]) + \partial_{\bar v} \Phi[\tfrac{n_5-2}{n_5}z] \,  \partial_v (Y + \Phi[u])\right]
\end{align} 
The path integral for the action \eqref{SU(2)-fermion-action-lr-f} thus amounts to   
\begin{equation}
Z^F_{su} = e^{\check{\mathfrak A}} \, Z_\eta \ , 
\label{decoupled-fermion-path-integral-f}
\end{equation}
where we introduced the notation
\begin{equation}
Z_\eta = \int \mathcal{D} \eta^+  \mathcal{D} \eta^- \mathcal{D} \tilde \eta^+  \mathcal{D} \tilde \eta^- \, \exp \left(-\int \frac{\text d^2v}{2 \pi} \left[ \eta^+ \partial_{\bar v} \eta^- + \eta^- \partial_{\bar v} \eta^+ + \tilde \eta^+ \partial_{\bar v} \tilde \eta^- + \tilde \eta^- \partial_{\bar v} \tilde \eta^+ \right] \right) \ , 
\label{Z-eta-lr}
\end{equation}
and the fermions $\eta^{\pm}, \tilde \eta^{\pm}$ obey the boundary conditions
\begin{equation}
\begin{aligned}
\eta^\pm(v+2 \pi, \bar v + 2 \pi) &= e^{\pm 2 \pi i \ell_2 s_1} \, \eta^{\pm}(v,\bar v) \ , \\
\eta^\pm(v+2 \pi \tau, \bar v + 2 \pi \bar \tau) &= e^{\mp 2 \pi i \ell_2 s_2} \, \eta^{\pm}(v,\bar v) \ , \\
\tilde \eta^\pm(v+2 \pi, \bar v + 2 \pi) &= e^{\mp 2 \pi i r_2 s_1} \, \tilde \eta^{\pm}(v,\bar v) \ , \\
\tilde \eta^\pm(v+2 \pi \tau, \bar v + 2 \pi \bar \tau) &= e^{\pm 2 \pi i r_2 s_2} \, \tilde \eta^{\pm}(v,\bar v) \ . 
\end{aligned}
\end{equation}
Following \rcite{Alvarez-Gaume:1986rcs}, the partition function \eqref{Z-eta-lr} reads 
\begin{equation}
Z_\eta = e^{\pi i (\ell_2^2 - r_2^2) s_1 s_2} \, e^{ \frac{\pi i u_2^2}{\tau_2^2}(\ell^2_2 \tau - r^2_2 \bar \tau)} \, \frac{\theta_1(\tau, \ell_2 u) \, \overline{\theta_1(\tau, r_2 u)}}{|\eta(\tau)|^2} \ . 
\label{Z-eta2}
\end{equation}
Let us comment on the phase $e^{\pi i (\ell_2^2 - r_2^2) s_1 s_2}$. This cannot be fixed by computing the partition function as a trace over the Hilbert space, and does not follow from the analysis of \rcite{Alvarez-Gaume:1986rcs}. We fix it by comparing with the free boson at the free fermion radius, see eq.~\eqref{pf-U(1)-free-fermion-radius}.%
\footnote{Different Theta functions appear in eqs. \eqref{Z-eta} and \eqref{pf-U(1)-free-fermion-radius}, due to a different choice of spin structure.} Using that 
\begin{equation}
e^{\pi i (\ell_2^2 - r_2^2) s_1 s_2}\Biggl|_{\ell_2 u \, \mapsto \,  \ell_2 u + \frac{n_5-2}{n_5}z, \, r_2 u \, \mapsto \, r_2 u + \frac{n_5-2}{n_5}z} = e^{\pi i (\ell_2^2 - r_2^2) s_1 s_2} \, e^{\pi i \frac{n_5-2}{n_5}(\ell_2-r_2)(s_1(z_1-\frac{z_2 \tau_1}{\tau_2}) + s_2 \frac{z_2}{\tau_2})} \ , 
\end{equation}
we finally obtain
\begin{align}
 Z^F_{su} &  = \frac{e^{\pi i (\ell_2^2 - r_2^2) s_1 s_2}}{|\eta(\tau)|^2} \, e^{\pi i\frac{n_5-2}{n_5}(\ell_2-r_2)(s_1(z_1-\frac{z_2 \tau_1}{\tau_2}) + s_2 \frac{z_2}{\tau_2})} \, e^{ \frac{\pi i (\ell_2 u_2+\frac{n_5-2}{n_5}z_2)^2}{\tau_2^2} \tau} \, e^{ -\frac{\pi i (r_2 u_2+\frac{n_5-2}{n_5}z_2)^2}{\tau_2^2} \bar \tau} \nonumber \\
& \times \, \theta_1(\tau, \ell_2 \, u + \tfrac{n_5-2}{n_5}z) \overline{\theta_1(\tau, r_2 \, u + \tfrac{n_5-2}{n_5} z)}  \nonumber \\
& \times \, \exp \left(\frac{\ell_2^2 + r_2^2}{\pi} \int \text d^2v \,  |\partial_v \left(Y + \Phi[u] \right)|^2  + \frac{2}{\pi} \int \text d^2v |\partial_v \Phi[\tfrac{n_5-2}{n_5}z]|^2 \right)  \nonumber \\
& \times \, \exp \left( \frac{\ell_2+r_2}{\pi} \int \text d^2v \left[ \partial_v \Phi[\tfrac{n_5-2}{n_5}z] \, \partial_{\bar v} (Y + \Phi[u]) + \partial_{\bar v} \Phi[\tfrac{n_5-2}{n_5}z] \, \partial_v (Y + \Phi[u])\right] \right) \ . 
\label{St-pf-SU2F}
\end{align}

\paragraph{Uncharged transverse fermions:} The two uncharged fermions in the ``transverse'' space that are not eliminated by the gauging have partition function
\begin{equation}
    Z_{tr}^F = \frac{|\theta_1(\tau, z)|^2}{|\eta(\tau)|^2} \ . 
\label{trans fermions}
\end{equation}

\subsubsection{Bosonic sector}

Let us then consider the various contributions in the second line of \eqref{ST-pf-general}. 

\paragraph{$\boldsymbol{SL(2,\mathds R)}$:} The contribution from $SL(2,\mathds R)_{n_5}$ has already been computed in Section \ref{sec:sl2/u1} and follows from eqs.~\eqref{SL2-partition-function} and \eqref{S-XY-sl}. It reads
\begin{equation}
Z^{B}_{sl}(\tau, u+\tfrac{2}{n_5}z)  = \frac{e^{ \frac{2 \pi (u_2+\frac{2}{n_5}z_2)^2}{\tau_2}} \,}{\sqrt{\tau_2} \, |\theta_1(\tau,u +\frac{2}{n_5}z)|^2} \, \exp \Bigl( - \mfrac{n_5+2}{\pi} \int \text d^2 v \, \bigl| \partial_v(Y + \Phi[u +\tfrac{2}{n_5}z]) \bigr|^2 \Bigr)   \ . 
\label{St-pf-SL2B}
\end{equation}

\paragraph{$\boldsymbol{U(1)}_{\boldsymbol y}$ and $\mathds R_{\boldsymbol t}$:} Let us consider the contributions \eqref{decoupling} for the $U(1)_y$ boson. A short computation gives
\begin{equation}
S_{g_y}  =  \int \frac{\text d^2v}{\pi} \, \partial_v \left(y + r_4 \Phi[u] \right)  \partial_{\bar v} \left( y -\ell_4 \Phi[u] \right) + \frac{(\ell_4+r_4)^2}{4 \pi}  \int \text d^2v \, |\partial_v \Phi[u]|^2  
\end{equation}
and 
\begin{equation}
S^{X,Y}_y = - \frac{\ell_4^2+r_4^2}{2 \pi} \int \text d^2v \, | \partial_v(\Phi[u] + Y)|^2 \ . 
\end{equation}
Using eqs.~\eqref{asymm-U(1)y-operator} and \eqref{alpha=1/2} we obtain 
\begin{multline}
Z^B_y(\tau, u) = \frac{e^{\frac{\pi i}{2} (\ell_4^2 - r_4^2) s_1 s_2} \, e^{\frac{\pi i u_2^2}{2 \tau_2^2}(\ell_4^2 \tau - r_4^2 \bar \tau)} }{ |\eta(\tau)|^2}  \, \exp \left(\frac{\ell^2_4 + r^2_4}{2 \pi} \int \text d^2v \, | \partial_v Y + \partial_v \Phi[u]|^2 \right)  \\
\times \, \sum_{w,n \in \mathds Z} q^{\frac{1}{4} \left( \frac{n}{R_y} + wR_y \right)^2} \bar q^{\frac{1}{4} \left(\frac{n}{R_y} - wR_y \right)^2} \, e^{\pi i \ell_4 (\frac{n}{R_y}+wR_y)u} \, e^{-\pi i r_4 (\frac{n}{R_y}-wR_y)\bar u } \ . 
\label{St-pf-yB}
\end{multline}
Similarly, we find 
\begin{equation}
Z^B_t(\tau,u) = \frac{e^{\frac{\pi \, \ell^2_3 \,  u_2^2}{\tau_2}} }{|\eta(\tau)|^2} \, \exp \left( -\frac{\ell_3^2}{\pi} \int \text d^2v \, |\partial_v(Y + \Phi[u])|^2 \right) \, \int dE  \, q^{-\frac{E^2}{4}} \, \bar q^{-\frac{E^2}{4}} \, e^{-2 \pi u_2 \ell_3 E} \ . 
\label{St-pf-tB}
\end{equation}

\paragraph{$\boldsymbol{SU(2)}$:} Here the computation becomes more involved. For $S_{g_{su}}$, using eq.~\eqref{SU2-LF-Shgh} we obtain
\begin{align}
 \int \mathcal D g_{su} \, e^{-S_{g_{su}}} & = \int \mathcal D g_{su} \, \exp \left(-S^\WZW(a^{-\varepsilon_L^{su}} \, g_{su} \, a^{\dagger \varepsilon_R^{su}}) \right)\Bigr|_{\ell_2 u \, \mapsto \, \ell_2 u - \frac{2}{n_5}z, \, r_2 u \, \mapsto \, r_2 u - \frac{2}{n_5}z } \nonumber \\
& =  \, e^{-\pi i\frac{n_5-2}{n_5}(\ell_2-r_2)(s_1(z_1-\frac{z_2 \tau_1}{\tau_2}) + s_2 \frac{z_2}{\tau_2})} \, e^{ \frac{(n_5-2)\pi \, i \, (\ell_2 \, u_2-\frac{2}{n_5}z_2)^2}{2\tau_2^2} \tau} \,  e^{ -\frac{(n_5-2)\pi \, i \, (r_2 \, u_2-\frac{2}{n_5}z_2)^2}{2\tau_2^2}  \bar \tau} \,\nonumber  \\
& \quad \times \, e^{\pi i \frac{n_5-2}{2}(\ell_2^2 - r_2^2) s_1 s_2} \, \mathcal Z_B^{\mathfrak{su}(2)}(\tau, \bar \tau, \ell_2 u-\tfrac{2}{n_5}z, r_2 \bar u-\tfrac{2}{n_5}\bar z) \ , 
\label{SU(2)-supertube-pf}
\end{align}
where we shifted the chemical potentials as described in Section~\ref{sec:su2/u1}, see eq.~\eqref{su2-chemical-potentials-relation}. Let us now proceed with the computation of the two terms entering eq.~\eqref{decoupling-b}. The first one gives
\begin{align}
- & S^\WZW(a^{-\varepsilon_L^{su}} \, \Omega^{-\varepsilon_L^{su}} \, \Omega^{\dagger -\varepsilon_R^{su}} \, a^{\dagger -\varepsilon_R^{su}})\Bigr|_{\ell_2 u \, \mapsto \, \ell_2 u - \frac{2}{n_5}z, \, r_2 u \, \mapsto \, r_2 u - \frac{2}{n_5}z } \nonumber \\
& = -\frac{(n_5-2)(\ell_2-r_2)^2}{4 \pi} \int \text d^2v \left( |\partial_v X|^2 + |\partial_v (Y+ \Phi[u])|^2 \right) \nonumber \\
& \quad + \frac{(n_5-2)(\ell_2-r_2)}{4\pi} \int \text d^2v \left(  \partial_v \Phi[-\tfrac{2}{n_5}z] \partial_{\bar v}(Y+ \Phi[u]) - \partial_{\bar v} \Phi[-\tfrac{2}{n_5}z] \partial_v(Y+ \Phi[u]) \right)  \ , 
\label{SU(2)-supertube-second-term} 
\end{align}
while for the second we find
\begin{align}
& \frac{n_5-2}{2\pi} \int \text d^2v \, \text{Tr} \Bigl[ \varepsilon_L^{su} \, \partial_v a \, a^{-1}\, \varepsilon_L^{su} \partial_{\bar v} a \, a^{-1} + \varepsilon_R^{su} \, \partial_v a^\dagger a^{\dagger-1} \, \varepsilon_R^{su} \, \partial_{\bar v} a^\dagger a^{\dagger -1} \nonumber \\
& \hspace{120pt} -2 \, \varepsilon_L^{su} \, \partial_{\bar v} a \, a^{-1} \varepsilon_R^{su} \, \partial_v a^\dagger a^{\dagger -1} \Bigr]\Bigl|_{\ell_2 u \,  \mapsto \, \ell_2 u -\frac{2}{n_5}z, \, r_2 u \,  \mapsto \, r_2 u -\frac{2}{n_5}z } \nonumber \\
& = \frac{(n_5-2)(\ell_2-r_2)^2}{4 \pi} \int \text d^2v \, |\partial_v X |^2 - \frac{(n_5-2)(\ell_2+r_2)^2}{4\pi} \int \text d^2v \, |\partial_v(Y + \Phi[u])|^2 \nonumber \\
& \quad -\frac{(n_5-2)}{\pi} \int \text d^2v \, |\partial_v \Phi[-\tfrac{2}{k}z]|^2 -\frac{(n_5-2)(r_2 + 3 \ell_2)}{4 \pi} \int \text d^2v \, \partial_v \Phi[-\tfrac{2}{n_5}z] \partial_{\bar v} (Y + \Phi[u]) \nonumber \\
& \quad -\frac{(n_5-2)(3r_2 + \ell_2)}{4 \pi} \int \text d^2v \, \partial_{\bar v} \Phi[-\tfrac{2}{n_5}z] \partial_v (Y + \Phi[u]) \ . 
\label{SU(2)-supertube-third-term} 
\end{align}
Assembling eqs.~\eqref{SU(2)-supertube-pf}, \eqref{SU(2)-supertube-second-term} and \eqref{SU(2)-supertube-third-term} we finally obtain
\begin{align}
Z^{B}_{su}&(\tau, u -\tfrac{2}{n_5}z) = e^{\pi i \frac{n_5-2}{2} (\ell_2^2 - r_2^2) s_1 s_2} \, e^{-\pi i\frac{n_5-2}{n_5}(\ell_2-r_2)(s_1(z_1-\frac{z_2 \tau_1}{\tau_2}) + s_2 \frac{z_2}{\tau_2})} \nonumber \\
& \times \, e^{ \frac{(n_5-2)\pi \, i \, (\ell_2 \, u_2-\frac{2}{n_5}z_2)^2}{2\tau_2^2} \tau} \,  e^{ -\frac{(n_5-2)\pi \, i \, (r_2 \, u_2-\frac{2}{n_5}z_2)^2}{2\tau_2^2}  \bar \tau} \, \mathcal Z_B^{\mathfrak{su}(2)}\left(\tau, \bar \tau, \ell_2 \, u-\tfrac{2}{n_5}z, r_2 \, \bar u-\tfrac{2}{n_5} \bar z \right) \nonumber \\
& \times \, \exp \left( \mfrac{(n_5-2)}{2\pi} \int \text d^2v \, \left[(\ell_2^2+r_2^2) |\partial_v Y + \partial_v \Phi[u]|^2 + 2|\partial_v \Phi[-\tfrac{2}{n_5}z]|^2 \right] \right)\nonumber \\
& \times \, \exp \left(\mfrac{(n_5-2)(\ell_2+r_2)}{2 \pi} \int \text d^2v \left[ \partial_v \Phi[-\tfrac{2}{n_5}z] \partial_{\bar v} (Y + \Phi[u]) + \partial_{\bar  v} \Phi[-\tfrac{2}{n_5}z] \partial_v (Y + \Phi[u])\right] \right) \ . 
\label{St-pf-SU2B}
\end{align}

\subsubsection{Assembling the various contributions}

Let us assemble all the contributions listed in eq.~\eqref{ST-pf-general} and computed in the previous sections. From eqs.~\eqref{St-pf-SL2F}, \eqref{St-pf-SU2F}-\eqref{St-pf-SL2B}, \eqref{St-pf-yB}, \eqref{St-pf-tB} and \eqref{St-pf-SU2B} we obtain the partition function of the supertube~\eqref{ST-pf-general}:\footnote{Notice that the last line in \eqref{St-pf-SU2F} cancels against the last line in \eqref{St-pf-SU2B}. }
\begin{multline}
Z^{ST}(\tau, z) = \frac{|\theta_1(\tau, z)|^2}{\sqrt{\tau_2} \, |\eta(\tau)|^{10}} \int_0^1 \text ds_1 \int_0^1 \text ds_2 \, \mathcal I_{u} \, \mathcal I_Y \,  Z_{gh}(\tau, \bar \tau) \, \mathcal Z_{B}^{\mathfrak{su}(2)}\left(\tau, \bar \tau, \ell_2 \,  u-\tfrac{2}{n_5}z , r_2 \, \bar u-\tfrac{2}{n_5} \bar z \right) \\
\times \, \int_{-\infty}^\infty \text dE \, q^{-\frac{E^2}{4}} \, \bar q^{-\frac{E^2}{4}} \, e^{-2 \pi u_2 \ell_3 E} \, \sum_{w,n \in \mathds Z} q^{\frac{1}{4} \left( \frac{n}{R_y} + wR_y \right)^2} \bar q^{\frac{1}{4} \left(\frac{n}{R_y} - wR_y \right)^2} \, e^{\pi i \ell_4 (\frac{n}{R_y}+wR_y)u} \, e^{-\pi i r_4 (\frac{n}{R_y}-wR_y) \bar u} \\
\times \, \frac{| \theta_1(\tau, u +\tfrac{n_5+2}{n_5} z)|^2 \, \theta_1(\tau, \ell_2 u + \tfrac{n_5-2}{n_5}z) \, \overline{\theta_1(\tau, r_2 u + \tfrac{n_5-2}{n_5} z)}}{|\theta_1(\tau,u +\frac{2}{n_5}z)|^2} \ ,  
\end{multline}
where 
\begin{align}
\mathcal I_u & =  e^{2 \pi \frac{(u_2+\frac{2}{n_5 }z_2)^2}{\tau_2}} \,  e^{-\frac{2\pi}{\tau_2} (u_2 + \frac{n_5 +2}{n_5 }z_2)^2} \, e^{\frac{\pi \, \ell^2_3 \,  u_2^2}{\tau_2}} \, e^{\frac{\pi i u_2^2}{2 \tau_2^2}(\ell_4^2 \tau - r_4^2 \bar \tau)} \nonumber  \\
& \qquad  e^{ \frac{(n_5 -2)\pi \, i \, (\ell_2 \, u_2-\frac{2}{n_5 }z_2)^2}{2\tau_2^2} \tau} \,  e^{ -\frac{(n_5 -2)\pi \, i \, (r_2 \, u_2-\frac{2}{n_5 }z_2)^2}{2\tau_2^2}  \bar \tau} \, e^{ \frac{\pi i (\ell_2 u_2+\frac{n_5 -2}{n_5 }z_2)^2}{\tau_2^2} \tau} \, e^{ -\frac{\pi i (r_2 u_2+\frac{n_5 -2}{n_5 }z_2)^2}{\tau_2^2} \bar \tau} \nonumber \\
& \qquad e^{\pi i \frac{n_5 -2}{2}(\ell_2^2 - r_2^2) s_1 s_2} \, e^{\pi i (\ell_2^2 - r_2^2) s_1 s_2} \, e^{\frac{\pi i}{2} (\ell_4^2 - r_4^2) s_1 s_2} \label{ST-Iu} 
\end{align}
and 
\begin{align}
\mathcal I_Y = \exp \Biggl( \int \text d^2v \Bigl[ & - \mfrac{n_5+2}{\pi}  |\partial_v(Y + \Phi[u +\tfrac{2}{n_5}z])|^2  + \mfrac{n_5-2}{2 \pi} (\ell_2^2+r_2^2)\, \big|\partial_v (Y + \Phi[u])\big|^2  \nonumber \\
& -\mfrac{\ell_3^2}{\pi} |\partial_v(Y + \Phi[u])|^2 + \mfrac{\ell_4^2 + r_4^2}{2 \pi} \, |\partial_v(Y + \Phi[u])|^2 \nonumber \\
&  + \mfrac{2}{\pi} \left| \partial_v (Y + \Phi[u+\tfrac{n_5+2}{n_5}z]) \right|^2  + \mfrac{\ell_2^2 + r_2^2}{\pi} |\partial_v \left(Y + \Phi[u] \right)|^2  \nonumber \\
& + \mfrac{n_5-2}{\pi} |\partial_v \Phi[-\tfrac{2}{n_5}z]|^2 + \mfrac{2}{\pi} |\partial_v \Phi[\tfrac{n_5-2}{n_5}z]|^2 \Bigr] \Biggr) \ . \label{ST-m} 
\end{align}
Using the null constraints \eqref{ST-null-constraints}, eq.~\eqref{ST-Iu} simplifies to 
\begin{equation}
\mathcal I_u = e^{-\frac{4 \pi}{\tau_2}z_2^2} \, e^{-\frac{4 \pi z_2^2}{n_5 \tau_2}} \, e^{-\pi n_5 \tau_2 s_1^2} \, e^{-4 \pi z_2 s_1} \ ,
\end{equation} 
while eq.~\eqref{ST-m} reduces to
\begin{align}
\mathcal I_Y = \exp \left( \frac{4 \pi |z|^2}{\tau_2} \right)  \ . 
\end{align} 
With these simplifications, the supertube partition function reads
\begin{subequations}
\begin{tcolorbox}[ams equation]
    Z^{ST}(\tau, z) = \frac{\sqrt{\tau_2} \, |\theta_1(\tau, z)|^2}{|\eta(\tau)|^6} \, e^{ \frac{4 \pi z_1^2}{\tau_2} } \, e^{ -\frac{4 \pi z_2^2}{n_5 \tau_2} } \, \int_0^1 \text ds_1 \int_0^1 \text ds_2 \, \mathcal I(\tau, u, z) \ , 
\end{tcolorbox}
\noindent where we introduced the notation
\begin{tcolorbox}[ams align]
& \nonumber \\
\mathcal I(\tau, u, z) =  & \, e^{-\pi n_5 \tau_2 s_1^2} \, e^{-4 \pi z_2 s_1} \, \mathcal Z_{B}^{\mathfrak{su}(2)}\left(\tau, \bar \tau, \ell_2 u-\tfrac{2}{n_5}z , r_2 \bar u-\tfrac{2}{n_5} \bar z \right) \, \int_{-\infty}^\infty \text dE \, q^{-\frac{E^2}{4}} \, \bar q^{-\frac{E^2}{4}} \, e^{-2 \pi u_2 \ell_3 E} \nonumber \\
&  \times \, \frac{| \theta_1(\tau, u +\tfrac{n_5+2}{n_5} z)|^2 \, \theta_1(\tau, \ell_2 u + \tfrac{n_5-2}{n_5}z) \, \overline{\theta_1(\tau, r_2 u + \tfrac{n_5-2}{n_5} z)}}{|\theta_1(\tau,u +\frac{2}{n_5}z)|^2} \nonumber \\
&  \times \, \sum_{w,n \in \mathds Z} q^{\frac{1}{4} \left( \frac{n}{R_y} + w R_y \right)^2} \bar q^{\frac{1}{4} \left(\frac{n}{R_y} - w R_y \right)^2} \, e^{\pi i \ell_4 (\frac{n}{R_y}+w R_y)u} \, e^{-\pi i r_4 (\frac{n}{R_y}-w R_y )\bar u} \ .
\label{ST-pf-final-integrand}
\end{tcolorbox}
    \label{ST-pf-final} 
\end{subequations}

\subsection{Evaluating the gauge field zero mode integrals}
\label{sec:Evaluating the gauge field zero mode integrals}

Let us now perform the integral over $s_1$ and $s_2$ in \eqref{ST-pf-final} following \rcite{Maldacena:2000kv, Hanany:2002ev}.
Formally, one would expect that the role of the Wilson lines for the gauge field is to enforce the zero mode parts of the gauge constraints. While we will see that this is eventually the correct picture, as reviewed in Section~\ref{sec:zeromodeints-sl2} there are a few subtleties arising from the fact that the action is quadratic rather than linear in the gauge fields.

\paragraph{Integral over $\boldsymbol{s_2}$:} Let us start by carrying out the integral over $s_2$ in eq.~\eqref{ST-pf-final}. We remind the reader that $u$ is related to $s_1$ and $s_2$ by eq.~\eqref{u-def}. We are going to see momentarily that the integral gives rise to a Kronecker delta, enforcing the difference of the left and right null conditions. 

In order to make the dependence on $s_2$ manifest, let us expand the various Theta functions entering \eqref{ST-pf-final}. Using the identity \eqref{theta1-denom} we can express the Theta functions in the denominator of \eqref{ST-pf-final-integrand} as 
\begin{multline}
\frac{1}{|\theta_1(\tau,u +\frac{2}{n_5 }z)|^2} =  \frac{1}{|\eta(\tau)|^6} \sum_{r, \bar r \in \mathds Z} S_r(\tau) \overline{S_{\bar r}(\tau)} \, e^{2 \pi i s_1 \tau_1 (r-\bar r)} \, e^{-2 \pi s_1 \tau_2 (r+\bar r +1)}  \\
\times \,  e^{\frac{4 \pi i}{n_5 }z_1 (r-\bar r) } \, e^{-\frac{4 \pi}{n_5 } z_2 (r + \bar r +1)}  \, e^{2 \pi i (r-\bar r )s_2} \ . 
\end{multline}
Using \eqref{Theta-def}, \eqref{SU2-pf} and \eqref{parafermion-character-decomposition}, the $\mathfrak{su}(2)$ partition function entering \eqref{ST-pf-final-integrand} can be expanded as 
\begin{align}
& \mathcal Z_{B}^{\mathfrak{su}(2)}\left(\tau, \bar \tau, \ell_2  u-\tfrac{2}{n_5 }z , r_2 \bar u-\tfrac{2}{n_5 } \bar z \right) \nonumber \\
& \qquad = \frac{1}{|\eta(\tau)|^2} \, \sum_{j_{su}=0}^{n_5 -2} \sum_{l, l' = 3-n_5 }^{n_5 -2} \chi^{\frac{\mathfrak{su}(2)}{\mathfrak u(1)}}_{B,j_{su},l}(\tau) \, \overline{\chi^{\frac{\mathfrak{su}(2)}{\mathfrak u(1)}}_{B,j_{su},l'}(\tau)} \sum_{n_b^{su}, \bar n_b^{su} \in \mathds Z} q^{(n_5 -2)(n_b^{su}+\frac{l}{2(n_5 -2)})^2} \, \bar q^{(n_5 -2)(\bar n_b^{su}+\frac{l'}{2(n_5 -2)})^2} \nonumber \\
& \qquad \quad \times \, e^{2 \pi i (s_1 \tau_1 + s_2)[\ell_2( (n_5 -2)n_b^{su}+\frac{l}{2}) - r_2((n_5 -2) \bar n_b^{su} + \frac{l'}{2}) ]} \, e^{-4 \pi i \frac{n_5 -2}{n_5 }[z (n_b^{su} + \frac{l}{2(n_5 -2)})- \bar z (\bar n_b^{su} + \frac{l'}{2(n_5 -2)})]} \nonumber \\
& \qquad \quad \times \, e^{-2 \pi (n_5 -2) s_1 \tau_2 (\ell_2 n_b^{su} + r_2 \bar n_b^{su} + \frac{\ell_2 l+r_2 l'}{2(n_5 -2)})} \ . 
\end{align}
The other $\theta_1$ functions entering eq.~\eqref{ST-pf-final-integrand} can be expressed in terms of their series definition \eqref{theta-1-def}. Assembling all the contributions and making use of the quantization conditions \eqref{li ri vals}, the supertube partition function \eqref{ST-pf-final} can be rewritten as 
\begin{align}
Z^{ST}(\tau, z) &= \frac{\sqrt{\tau_2} \, |\theta_1(\tau, z)|^2}{|\eta(\tau)|^{14}} \, e^{ \frac{4 \pi z_1^2}{\tau_2} } \, e^{ -\frac{4 \pi z_2^2}{n_5  \, \tau_2} } \int_{-\infty}^\infty \text d E \sum_{w,n \in \mathds Z} \ \sum_{r, \bar r \in \mathds Z} \ \sum_{n_f^{sl}, \bar n_f^{sl} \in \mathds Z + \frac{1}{2} } \  \sum_{n_f^{su}, \bar n_f^{su} \in \mathds Z + \frac{1}{2} } \nonumber \\
& \qquad \times \, \sum_{j_{su}=0}^{n_5 -2} \ \sum_{l, l' = 3-n_5 }^{n_5 -2} \ \sum_{n_b^{su}, \bar n_b^{su} \in \mathds Z} \mathcal I_{sl} \, \mathcal I_{t,y} \, \mathcal I_\theta \, \mathcal I_{su} \, \int_0^1 \text ds_1 \, \mathcal I_{s_1} \int_0^1 \text ds_2 \, \mathcal I_{s_2}   \ , 
\label{ZST-open}
\end{align}
where 
\begin{subequations}
\begin{align}
\mathcal I_{sl} & = S_r(\tau) \overline{S_{\bar r}(\tau)} \, e^{\frac{4 \pi i}{n_5 }z_1 (r-\bar r) } \, e^{-\frac{4 \pi}{n_5 } z_2 (r + \bar r +1)} \ , \\[.2cm]
\mathcal I_{t,y} & =  q^{-\frac{E^2}{4}} \, \bar q^{-\frac{E^2}{4}} \, q^{\frac{1}{4}(\frac{n}{R_y}+w R_y)^2} \bar q^{\frac{1}{4}(\frac{n}{R_y}-w R_y)^2} \ , \\[.2cm]
\mathcal I_{\theta} & = (-1)^{n_f^{sl} + \bar n_f^{sl} + n_f^{su} + \bar n_f^{su}} q^{\frac{1}{2}(n_f^{sl})^2+\frac{1}{2}(n_f^{su})^2} \bar q^{\frac{1}{2}(\bar n_f^{sl})^2+\frac{1}{2}(\bar n_f^{sl})^2} \, e^{\frac{2 \pi i (n_5 +2)}{n_5 }(n_f^{sl} z - \bar n_f^{sl} \bar z )+\frac{2 \pi i (n_5 -2)}{n_5 }(n_f^{su} z - \bar n_f^{sl} \bar z )} \ ,  \\
\mathcal I_{su} & = \chi^{\frac{\mathfrak{su}(2)}{\mathfrak u(1)}}_{B,j_{su},l}(\tau) \, \overline{\chi^{\frac{\mathfrak{su}(2)}{\mathfrak u(1)}}_{B,j_{su},l'}(\tau)} \, \, q^{(n_5 -2)(n_b^{su}+\frac{l}{2(n_5 -2)})^2} \, \bar q^{(n_5 -2)(\bar n_b^{su}+\frac{l'}{2(n_5 -2)})^2} \nonumber \\[.2cm]
& \hskip 7cm \times \, e^{-4 \pi i \frac{n_5 -2}{n_5 }[z (n_b^{su} + \frac{l}{2(n_5 -2)})-\bar z (\bar n_b^{su} + \frac{l'}{2(n_5 -2)})]}  \ , 
\end{align}
 \label{Isl-Isu}%
\end{subequations}
and 
\begin{align}
\mathcal I_{s_1} & = \, e^{-\pi n_5 \tau_2 s_1^2 -4 \pi z_2 s_1} \, e^{2 \pi i s_1 \tau_1 (r-\bar r + \frac{\ell_4 p_L - r_4 p_R}{2} + n_f^{sl}-\bar n_f^{sl}+\ell_2 \, n_f^{su}-r_2 \, \bar n_f^{su}+ \ell_2((n_5-2) n_b^{su}+\frac{l}{2}) - r_2 ((n_5-2)  \bar n_b^{su}+\frac{l'}{2}) )} \nonumber \\[.2cm]
& \hskip 0.7cm \times \, e^{-2 \pi s_1 \tau_2 (r+\bar r +1 +\ell_3 E + \frac{\ell_4 p_L + r_4 p_R}{2} + n_f^{sl}+\bar n_f^{sl}+\ell_2 \, n_f^{su}+r_2 \, \bar n_f^{su} + (n_5-2)\ell_2 n_b^{su} + (n_5-2)r_2 \bar n_b^{su} + \frac{\ell_2 l+r_2 l'}{2})} \ , \\[.2cm]
\mathcal I_{s_2} & = e^{2 \pi i s_2 (r-\bar r + \frac{\ell_4 p_L - r_4 p_R}{2} + n_f^{sl}-\bar n_f^{sl}+\ell_2 \, n_f^{su}-r_2 \, \bar n_f^{su}+ \ell_2((n_5-2) n_b^{su}+\frac{l}{2}) - r_2 ((n_5-2)  \bar n_b^{su}+\frac{l'}{2}) ) } \label{Is2} \ . 
\end{align}
Notice that the supertube partition function \eqref{ZST-open} depends on $s_2$ only through $\mathcal I_{s_2}$, see eq.~\eqref{Is2}. The quantization of $l_i,r_i$, see eq.~\eqref{li ri vals}, and eq.~\eqref{parf-ch=0} imply
\begin{equation}
\frac{1}{2}(\ell_2 l - r_2 l') = \frac{1}{4}(\ell_2-r_2)(l+l') + \frac{1}{4}(\ell_2+r_2)(l-l') \in \mathds Z 
\end{equation}
and 
\begin{equation}
\ell_2 \, n_f^{su} - r_2 \, \bar n_f^{su} \in \mathds Z \ , \qquad \ell_4 p_L - r_4 p_R  \in 2 \mathds Z \ . 
\end{equation}
We can then use the Kronecker delta representation \eqref{Kr-delta} and evaluate the integral over $s_2$, 
\begin{equation}
\label{axialnullzm}
    \int_0^1 \text ds_2 \, \mathcal I_{s_2} = \delta_{\mathcal J_0 - \bar{\mathcal J}_0, 0} \ , 
\end{equation}
where making the identifications
\begin{subequations}
\begin{equation}
(j^3_{sl})_0 - (\bar j^3_{sl})_0 = r-\bar r  \ , \qquad  (j^3_{f,sl})_0 = n_f^{sl}  \ , \qquad (\bar j^3_{f,sl})_0 = \bar n_f^{sl}  \ ,  
\end{equation}
\begin{equation}
(j^3_{su})_0 = (n_5-2) n_b^{su} + \frac{l}{2} \ , \quad  ( \bar j^3_{su})_0 = (n_5-2) \bar n_b^{su} + \frac{l'}{2} \ , \quad  (j^3_{f,su})_0 = n_f^{su} \ , \quad  ( \bar j^3_{f,su})_0 = \bar n_f^{su} \ ,
\end{equation} 
\label{s2-int-identif}%
\end{subequations}
$\mathcal J_0 - \bar{\mathcal J}_0$ is the difference of the null current zero modes 
\begin{equation}
\begin{aligned}
\mathcal J_0 &= (J^3_{sl})^{~}_0 + \ell_2 (J^3_{su})^{~}_0 + \ell_3 \frac{E}{2} + \frac{\ell_4}{2} \, p^{~}_L \ , \\
\bar \cJ_0 &= (\bar J^3_{sl})^{~}_0 + r_2 (\bar J^3_{su})^{~}_0 + \ell_3 \frac{E}{2} + \frac{r_4}{2} \, p^{~}_R \ . 
\end{aligned}
\label{ST-null-curr}
\end{equation}
Thus, we see that the effect of the $s_1$ integral is indeed to impose the zero mode of the axial null gauge constraints.

\paragraph{Integral over $\boldsymbol{s_1}$:}

Following~\rcite{Maldacena:2000kv,Hanany:2002ev}, a manipulation of the remaining $s_1$ integral enables us to rewrite the partition function as a sum over discrete and continuous series representations of $\sltwo$.

After carrying out the integral over $s_2$ we are left with 
\begin{align}
Z^{ST}(\tau, z) &= \frac{\sqrt{\tau_2} \, |\theta_1(\tau, z)|^2}{|\eta(\tau)|^{14}} \, e^{ \frac{4 \pi z_1^2}{\tau_2} } \, e^{ -\frac{4 \pi z_2^2}{n_5  \, \tau_2} } \int_{-\infty}^\infty \text d E \sum_{w,n \in \mathds Z} \ \sum_{r, \bar r \in \mathds Z} \ \sum_{n_f^{sl}, \bar n_f^{sl} \in \mathds Z + \frac{1}{2} } \ \sum_{n_f^{su}, \bar n_f^{su} \in \mathds Z + \frac{1}{2} } \nonumber \\
& \qquad \times \, \sum_{j_{su}=0}^{n_5 -2} \ \sum_{l, l' = 3-n_5 }^{n_5 -2} \ \sum_{n_b^{su}, \bar n_b^{su} \in \mathds Z} \mathcal I_{sl} \, \mathcal I_{t,y} \, \mathcal I_\theta \, \mathcal I_{su} \, \delta_{\mathcal J_0 - \bar{\mathcal J}_0,0}  \,  \int_0^1 \text ds_1 \, \mathcal I_{s_1}   \ , 
\end{align}
where $\mathcal I_{s_1}$ collapsed to 
\begin{equation}
\mathcal I_{s_1} = e^{-\pi n_5 \tau_2 s_1^2} \, e^{-4 \pi z_2 s_1}  e^{-2 \pi s_1 \tau_2 \mathcal K} \ ,
\label{I-s1-1}
\end{equation}
and we introduced the shorthand notation
\begin{subequations}
\begin{align}
\mathcal K &= r+\bar r +1 + \mathcal F + \bar{\mathcal F} \ , \\
\mathcal F &= \mathcal J_0- (j^3_{sl})_0 =  \tfrac{1}{2}\ell_3 E +\frac{\ell_4 \, p_L}{2} + n_f^{sl}+\ell_2 \, n_f^{su} + (n_5-2)\ell_2 n_b^{su} + \tfrac{1}{2}\ell_2 l \ , \\
\bar{\mathcal F} &= \bar{\mathcal J}_0 - (\bar j^3_{sl})_0 = \tfrac{1}{2}\ell_3 E + \frac{r_4 \, p_R}{2} + \bar n_f^{sl}+r_2 \, \bar n_f^{su} + (n_5-2)r_2 \bar n_b^{su} + \tfrac{1}{2}r_2 l' \ . 
\end{align}
\label{K-F-barF}%
\end{subequations}
Let us now proceed as in Section \ref{sec:zeromodeints-sl2} and complete the square in eq.~\eqref{I-s1-1}, 
\begin{align}
\int_0^1 \text d s_1 \, \mathcal I_{s_1} & =  e^{\frac{4 \pi z_2^2}{n_5 \tau_2}} \int_0^1 \text ds_1 \, \exp \left(-\pi n_5 \tau_2(s_1+\tfrac{2 z_2}{n_5 \tau_2})^2 -2 \pi s_1 \tau_2 \mathcal K \right) \nonumber \\
& = e^{\frac{4 \pi z_2^2}{n_5 \tau_2}+\frac{4 \pi z_2}{n_5} \mathcal K} \int_{0}^{1} \text ds_1 \, \exp \left(-\pi n_5 \tau_2 s_1^{2} -2 \pi s_1 \tau_2 \mathcal K \right) \ , 
\end{align} 
where in the second equality we exploited the periodicity properties discussed in Appendix~\ref{app:period-ST}, see eq.~\eqref{ST-integral-periodicity}. Using the identity \eqref{c-linearize} it is easy to evaluate the integral over $s_1$. We are left with 
\begin{equation}
\int_0^1 \text d s_1 \, \mathcal I_{s_1} = -\frac{e^{\frac{4 \pi z_2^2}{n_5 \tau_2}+\frac{4 \pi z_2}{n_5} \mathcal K}}{2 \pi \sqrt{n_5 \tau_2}} \Biggl( \int_{-\infty}^\infty \text dc \, \, \frac{e^{-\frac{\pi \, c^2}{n_5 \tau_2}} \, e^{-2\pi(i c + \tau_2 \mathcal K)}}{i c + \tau_2 \mathcal K} - \int_{-\infty}^\infty \text dc \, \, \frac{e^{-\frac{\pi \, c^2}{n_5 \tau_2}}}{i c + \tau_2 \mathcal K} \Biggr) \ . 
\label{int-to-split}
\end{equation}
Consider the first integral in the right-hand-side of eq.~\eqref{int-to-split}  and similarly to what we did in Section \ref{sec:zeromodeints-sl2} let us shift the integration contour vertically from $\text{Im} \, c = 0$ to  $\text{Im} \, c =- i n_5 \tau_2$, see Figure \ref{fig:int-contours}. We obtain 
\begin{equation}
\int_0^1 \text d s_1 \, \mathcal I_{s_1} = \mathcal I_{dis} + \mathcal I_{con} \ , 
\label{Ish+Ilo}
\end{equation}
with 
\begin{align}
\mathcal I_{dis} & = -\frac{e^{\frac{4 \pi z_2^2}{n_5 \tau_2}} \, e^{\frac{4 \pi z_2}{n_5}  \mathcal K} }{2 \pi \sqrt{n_5 \tau_2}} \, \oint_{\Gamma_{dis}} \text dc \, \, \frac{e^{-\frac{\pi \, c^2}{n_5 \tau_2}} \, e^{-2\pi(i c + \tau_2 \mathcal K)}}{i c + \tau_2 \mathcal K} \ , 
\label{short-string-int} \\
\mathcal I_{con} & = -\frac{e^{\frac{4 \pi z_2^2}{n_5 \tau_2}} \, e^{\frac{4 \pi z_2}{n_5} \mathcal K}}{2 \pi \sqrt{n_5 \tau_2}} \, \Biggl( \int_{-\infty-i \tau_2 n_5}^{\infty-i \tau_2 n_5} \text dc \, \, \frac{e^{-\frac{\pi \, c^2}{n_5 \tau_2}} \, e^{-2\pi(i c + \tau_2 \mathcal K)}}{i c + \tau_2 \mathcal K} - \int_{-\infty}^\infty \text dc \, \, \frac{e^{-\frac{\pi \, c^2}{n_5 \tau_2}}}{i c + \tau_2 \mathcal K} \Biggr) \ ,
\label{long-string-int}
\end{align}
where the integral in \eqref{short-string-int} is evaluated along the clockwise contour $\Gamma_{dis}$ encircling the region 
\begin{equation}
   -\tau_2 n_5 < \text{Im } c < 0 \ ,  
\label{poles-region-c}
\end{equation}
see Figure~\ref{fig:int-contours} with $\kappa_{sl} = n_5$. Similarly to Section~\ref{sec:zeromodeints-sl2}, the split of integration contours induces for the partition function the split 
\begin{equation}
    Z^{ST}(\tau, z) =  Z^{ST}_{dis}(\tau, z) + Z^{ST}_{con}(\tau, z) \ . 
    \label{ST-pf-long+short}
\end{equation}
We are going to see that, as the notation should suggest, $Z^{ST}_{dis}$ will reproduce the short string spectrum while $ Z^{ST}_{con}(\tau, z)$ captures the long string spectrum. Let us analyze these two contributions one after the other. 

\paragraph{Short string spectrum:} We start from the short string contribution to the partition function and evaluate the contour integral \eqref{short-string-int} by means of Cauchy's theorem. Rewriting $c$ as in eq.~\eqref{c-to-jsl}, eq.~\eqref{short-string-int} reads 
\begin{equation}
\mathcal I_{dis}  = i\frac{e^{\frac{4 \pi z_2^2}{n_5 \tau_2}+\frac{4 \pi z_2}{n_5}  \mathcal K} }{\pi \sqrt{n_5 \tau_2}} \, \oint_{\Gamma_{dis}} \text dj_{sl} \, \, \frac{e^{\frac{\pi \, \tau_2 \, (2j_{sl}-1)^2}{n_5}} \, e^{-2\pi \tau_2(2j_{sl}-1 +\mathcal K)}}{2j_{sl} -1 + \mathcal K} \ . 
\label{I-short-1}
\end{equation}
Exactly as in Section \ref{sec:zeromodeints-sl2}, poles only appear for real $j_{sl}$ and in the region corresponding to the Maldacena-Ooguri bound. Also in this case, the denominator of the integrand in \eqref{I-short-1} equals the sum of the null current zero modes~\eqref{ST-null-curr}, 
\begin{equation}
2j_{sl} -1 + \mathcal K = \mathcal J_0 + \bar{\mathcal J}_0 \ , 
\end{equation}
where we used eq.~\eqref{K-F-barF} and made the identifications 
\begin{equation}
(j^3_{sl})_0  = r+j_{sl}  \ ,  \qquad (\bar j^3_{sl})_0 = \bar r+j_{sl}  \ . 
\label{J3-id-dis}
\end{equation}
These identifications are consistent with eq.~\eqref{s2-int-identif} and provide the correct quantization of quantum numbers for the discrete representation of $\mathfrak{sl}(2,\mathds R)$. We can then rewrite the integral $\mathcal I_{dis}$ as
\begin{align}
\mathcal I_{dis} = -\frac{2 \, e^{\frac{4 \pi z_2^2}{n_5 \tau_2}}}{\sqrt{n_5 \tau_2}} \, (q \bar q)^{-\frac{1}{4n_5}} \, \int_{\frac{1}{2}}^{\frac{n_5 +1}{2}} \text dj_{sl} \, \delta(\mathcal J_0 + \mathcal J_0) \, q^{-\frac{j_{sl}(j_{sl}-1)}{n_5 }} \, \bar q^{-\frac{j_{sl}(j_{sl}-1)}{n_5 }} \, e^{-\frac{4\pi(2j_{sl}-1) \, z_2}{n_5 } }  \ . 
\end{align}
Thus, the short string contribution to the partition function reads
\begin{align}
& Z^{ST}_{dis}(\tau, z) = \frac{e^{ \frac{4 \pi z_1^2}{\tau_2} } \, |\theta_1(\tau, z)|^2}{|\eta(\tau)|^{14}}  \, \int_{-\infty}^{\infty} \text d E \sum_{w,n \in \mathds Z} \  \sum_{n_f^{sl}, \bar n_f^{sl} \in \mathds Z + \frac{1}{2} } \ \sum_{n_f^{su}, \bar n_f^{su} \in \mathds Z + \frac{1}{2} } \  \sum_{j_{su}=0}^{n_5 -2} \ \sum_{l, l' = 3-n_5 }^{n_5-2} \sum_{n_b^{su}, \bar n_b^{su} \in \mathds Z} \nonumber \\
&  \quad \times \,  (q\bar q)^{-\frac{1}{4n_5}} \, \int_{\frac{1}{2}}^{\frac{n_5 +1}{2}} \text dj_{sl}  \sum_{r, \bar r \in \mathds Z} (q\bar q)^{-\frac{j_{sl}(j_{sl}-1)}{n_5}} \, S_r(\tau) \, \overline{S_{\bar r}(\tau)} \, e^{2 \pi i z \, \frac{2}{n_5} (j^3_{sl})_0 } e^{-2 \pi i \bar z \, \frac{2}{n_5} (\bar j^3_{sl})_0 } \nonumber \\
& \quad \times \, \mathcal I_{\theta} \, \mathcal I_{t,y}  \, \mathcal I_{su} \, \delta_{\mathcal J_0 - \bar{\mathcal J}_0,0} \, \delta(\mathcal J_0 + \mathcal J_0) \ , 
\label{ZST-short-string-characters}
\end{align}
where we reabsorbed some $n_5$ factors into the overall normalization and observed that in terms of the identifications~\eqref{J3-id-dis} we have 
\begin{align}
\mathcal I_{t,y} & =  q^{-\frac{E^2}{4}} \, \bar q^{-\frac{E^2}{4}} \, q^{\frac{1}{4}(\frac{n}{R_y}+w R_y)^2} \bar q^{\frac{1}{4}(\frac{n}{R_y}-w R_y)^2} \ , \\[.2cm]
\mathcal I_{\theta} & = (-1)^{F} q^{\frac{1}{2}((j^3_{f,sl})_0)^2+\frac{1}{2}((j^3_{f,su})_0)^2} \bar q^{\frac{1}{2}((\bar j^3_{f,sl})_0)^2+\frac{1}{2}((\bar j^3_{f,sl})_0)^2} \nonumber \\
& \hskip 2cm \times \, e^{2 \pi i z  \frac{n_5 +2}{n_5}(j^3_{f,sl})_0} \, e^{- 2 \pi i \bar z \frac{n_5 +2}{n_5} (\bar j^3_{f,sl})_0} e^{2 \pi i z \frac{n_5 -2}{n_5} (j^3_{f,su})_0} e^{- 2 \pi i  \bar z \frac{n_5 -2}{n_5} (\bar j^3_{f,sl})_0 } \ ,  \\[0.2cm]
\mathcal I_{su} & = \chi^{\frac{\mathfrak{su}(2)}{\mathfrak u(1)}}_{B,j_{su},l}(\tau) \, \overline{\chi^{\frac{\mathfrak{su}(2)}{\mathfrak u(1)}}_{B,j_{su},l'}(\tau)} \, \, q^{\frac{(j^3_{su})_0^2}{(n_5 -2)}} \, \bar q^{\frac{(\bar j^3_{su})_0^2}{(n_5 -2)}} \, e^{-2 \pi i z \frac{2}{n_5} (j^3_{su})_0} \, e^{2 \pi i \bar z \frac{2}{n_5 } (j^3_{su})_0}  \ . 
\end{align}
Using eqs.~\eqref{JBFR-rel-su} and \eqref{JBFR-rel-sl}, it is now manifest that the discrete part of the partition function can be written as a trace over a constrained Hilbert space
\begin{equation}
    Z^{ST}_{dis}(\tau, z) = e^{ \frac{4 \pi z_1^2}{\tau_2} } \, \text{Tr}_{\mathcal H \otimes \bar{\mathcal H}} \, q^{L_0 -\frac{c}{24} } \, \bar q^{\bar L_0 -\frac{c}{24} } \, e^{2 \pi i z J^{\mathcal R}_0} \, e^{-2 \pi i \bar z \bar J^{\mathcal R}_0} \Bigl|_{\mathcal J_0 = \bar{\mathcal J}_0=0} \ .
    \label{Z-ST-trace}
\end{equation}
The holomorphic Hilbert space $\mathcal H$ (and similarly for $\bar{\mathcal H}$) is defined by tensoring $\mathcal D^+_{j_{sl}}$ representations of $\mathfrak{sl}(2,\mathds R)_{n_5}$ with representations of the other factors entering the numerator $G$ of the coset theory~\eqref{st-coset}. The exponential prefactor in eq.~\eqref{Z-ST-trace} is the expected factor relating path integral and trace partition functions, see \cite{Kraus:2006nb} and Appendix~\ref{app:different-pfs}. Eq.~\eqref{Z-ST-trace} confirms that the partition function \eqref{ST-pf-final} reproduces the expected short string spectrum, confirming the analysis carried out in \rcite{Martinec:2018nco, Bufalini:2021ndn} in the operator formalism.

\paragraph{Long string spectrum:} Let us now consider the second term in \eqref{ST-pf-long+short} and show how the long string spectrum emerges after manipulating the integral $\mathcal I_{con}$ defined in eq.~\eqref{long-string-int}.  Performing a simple change of variable in the first integral, we can write eq.~\eqref{long-string-int} as the sum of the two contributions, 
\begin{equation}
\mathcal I_{con} = \mathcal I_{con}^- + \mathcal I_{con}^+ \ ,     
\end{equation}
defined as
\begin{align}
\mathcal I_{con}^- & = -\frac{e^{\frac{4 \pi z_2^2}{n_5  \tau_2}} \, e^{\frac{4 \pi z_2}{n_5} \mathcal K}}{2 \pi \sqrt{n_5  \tau_2}} \, \int_{-\infty}^{\infty} \text dc \, \, \frac{e^{-\frac{\pi \, c^2}{n_5  \tau_2}} \, e^{- \pi \tau_2 n_5  -2\pi \tau_2 \mathcal K}}{i c +n_5  \tau_2 + \tau_2 \mathcal K}\ , \label{Ip} \\
\mathcal I_{con}^+ & =  \frac{e^{\frac{4 \pi z_2^2}{n_5 \tau_2}} \, e^{\frac{4 \pi z_2}{n_5 } \mathcal K}}{2 \pi \sqrt{n_5  \tau_2}} \, \int_{-\infty}^\infty \text dc \, \, \frac{e^{-\frac{\pi \, c^2}{n_5  \tau_2}}}{i c + \tau_2 \mathcal K} \ . \label{Im} 
\end{align}
Following the strategy reviewed in Section~\ref{sec:zeromodeints-sl2}, we want to manipulate $\mathcal I_{con}^-$ so that the numerator of the integrand becomes similar to the one in $\mathcal I_{con}^+$. In order to do so, let us split the long string contribution to the partition function \eqref{ST-pf-long+short} as 
\begin{equation}
Z^{ST}_{con}(\tau, z) = Z^{ST, +}_{con}(\tau, z) + Z^{ST,-}_{con}(\tau, z) \ , 
\label{Zst+-}
\end{equation}
where we defined
\begin{align}
Z^{ST,\pm}_{con}(\tau, z) &= \frac{\sqrt{\tau_2} \, |\theta_1(\tau, z)|^2}{|\eta(\tau)|^{14}} \, e^{ \frac{4 \pi z_1^2}{\tau_2} } \, e^{- \frac{4 \pi z_2^2}{n_5 \tau_2} } \int_{-\infty}^\infty \text dE \sum_{w,n \in \mathds Z} \ \sum_{r, \bar r \in \mathds Z} \ \sum_{n_f^{sl}, \bar n_f^{sl} \in \mathds Z + \frac{1}{2} } \ \sum_{n_f^{su}, \bar n_f^{su} \in \mathds Z + \frac{1}{2} } \nonumber \\
& \qquad \times \, \sum_{j_{su}=0}^{n_5-2} \ \sum_{l, l' = 3-n_5}^{n_5-2} \mathcal I_{sl} \ \sum_{n_b^{su}, \bar n_b^{su} \in \mathds Z} \, \mathcal I_{t,y} \, \mathcal I_{\theta} \, \mathcal I_{su} \, \delta_{\mathcal J_0 - \bar{\mathcal J}_0,0} \, \, \mathcal I_{con}^\pm \,,
\label{Zm-long}
\end{align}
and $\mathcal I_{sl}$, $\mathcal I_{t,y}$, $\mathcal I_{\theta}$ and $\mathcal I_{su}$ are still given by eqs.~\eqref{Isl-Isu}. In order to cancel the factor $e^{- \pi \tau_2 n_5 -2\pi \tau_2 \mathcal K}$ in the numerator of $\mathcal I^-_{con}$, let us shift the various summation labels entering eq.~\eqref{Zm-long} according to  
\begin{subequations}
\begin{equation}
w \to w+\text k \ , \quad  n \to n - \text p \ , \quad  n_f^{sl} \to  n_f^{sl} - 1 \ , \quad \bar n_f^{sl} \to  \bar n_f^{sl} - 1 \ , \quad E\to E + \ell_3 \ ,  
\end{equation}
\begin{equation}
n_f^{su} \to  n_f^{su} - \ell_2 \ , \quad  \bar n_f^{su}  \to  \bar n_f^{su} - r_2 \ ,  \quad  n_b^{su}  \to  n_b^{su} - \frac{\ell_2}{2} \ , \quad \bar n_b^{su}  \to  \bar n_b^{su} - \frac{r_2}{2} \ . 
\end{equation}
\label{sum-shifts}%
\end{subequations}
Notice that due to the null conditions \eqref{ST-null-constraints} the Kronecker delta $\delta_{\mathcal J_0-\bar{\mathcal J}_0,0}$ is  unaffected by the shifts above. Using once more the null conditions \eqref{ST-null-constraints} and exploiting the presence of the Kronecker delta, it is tedious but straightforward to check that the net effect of these shifts is producing the extra term
\begin{equation}
q^r \, \bar q^{\bar r}\, e^{\pi \tau_2 n_5 + 2 \pi \tau_2 \mathcal K} \ , 
\end{equation}
and adding $-(n_5+2)\tau_2$ to the denominator of the integrand. Eq.~\eqref{Zm-long} for $Z^{ST,-}_{con}(\tau, z)$ thus becomes
\begin{align}
Z^{ST,-}_{con}(\tau, z) &= -\frac{e^{ \frac{4 \pi z_1^2}{\tau_2} } \, |\theta_1(\tau, z)|^2}{2 \pi \sqrt{n_5} \, |\eta(\tau)|^{14}} \, \int_{-\infty}^\infty \text d E \sum_{w,n \in \mathds Z} \ \sum_{n_f^{sl}, \bar n_f^{sl} \in \mathds Z + \frac{1}{2} } \ \sum_{n_f^{su}, \bar n_f^{su} \in \mathds Z + \frac{1}{2} }   \nonumber\\
& \quad \times \, \sum_{j_{su}=0}^{n_5-2} \ \sum_{l, l' = 3-n_5}^{n_5-2}  \ \sum_{n_b^{su}, \bar n_b^{su} \in \mathds Z} \ \sum_{r, \bar r \in \mathds Z} S_r(\tau)  \, \overline{S_{\bar r}(\tau)} \, e^{-\frac{4 \pi i z_1}{n_5} (r-\bar r)} \, \mathcal{I}_{t,y} \, \mathcal I_\theta \, \mathcal I_{su} \nonumber \\
& \quad \times \, \delta_{\mathcal J_0 - \bar{\mathcal J}_0,0} \, \, e^{\frac{4 \pi z_2}{n_5}  (\mathcal F + \bar{\mathcal F} )} \int_{-\infty}^{\infty} \text dc \, \frac{e^{-\frac{\pi \, c^2}{n_5 \tau_2}} }{i c - \tau_2(1+r+\bar r) + \tau_2 ( \mathcal F + \bar{\mathcal F})} \,,
\end{align}
where we used eqs.~\eqref{K-F-barF} to rewrite $\mathcal K$ in terms of $\mathcal F$ and $\bar{\mathcal F}$ and the identity \eqref{rmr-id} to change the sign of the summation labels $r$ and $\bar r$. Notice that using eqs.~\eqref{s2-int-identif} and \eqref{K-F-barF} the difference of null currents can be written as 
\begin{equation}
    \mathcal J_0 - \bar{\mathcal J}_0 =  \mathcal F - \bar{\mathcal F} -(r - \bar r)
\end{equation}
and hence 
\begin{equation}
    \delta_{\mathcal J_0 - \bar{\mathcal J}_0,0} \, e^{-\frac{4 \pi i z_1}{n_5} (r-\bar r)} = \delta_{\mathcal J_0 - \bar{\mathcal J}_0,0} \, e^{-\frac{4 \pi i z_1}{n_5} (\mathcal F-\bar{\mathcal F})} \ . 
\end{equation}
The long string contribution to the partition function \eqref{Zst+-} then reads
\begin{align}
Z^{ST}_{con}(\tau, z) &= \frac{e^{ \frac{4 \pi z_1^2}{\tau_2} } \, |\theta_1(\tau, z)|^2 }{2 \pi \sqrt{n_5} \,  |\eta(\tau)|^{14}} \, \int_{-\infty}^\infty \text dE \sum_{w,n \in \mathds Z} \ \sum_{n_f^{sl}, \bar n_f^{sl} \in \mathds Z + \frac{1}{2} } \ \sum_{n_f^{su}, \bar n_f^{su} \in \mathds Z + \frac{1}{2} }  \nonumber\\
& \times \,  \sum_{j_{su}=0}^{n_5-2} \ \sum_{l, l' = 3-n_5}^{n_5-2} \ \sum_{n_b^{su}, \bar n_b^{su} \in \mathds Z} \ \sum_{r, \bar r \in \mathds Z} S_r(\tau)  \, \overline{S_{\bar r}(\tau)} \, \mathcal I_{t,y} \, \mathcal I_\theta \, \mathcal I_{su} \, \xi^{-\frac{2}{n_5}\mathcal F} \, \bar \xi^{-\frac{2}{n_5} \bar{\mathcal F}} \, \delta_{\mathcal J_0 - \bar{\mathcal J}_0,0} \nonumber \\
&  \times \, \int_{-\infty}^{\infty} \text dc \, \, e^{-\frac{\pi \, c^2}{n_5 \tau_2}} \, \biggl( \mfrac{1}{i c +\tau_2(r + \bar r +1)+ \tau_2 \mathcal F + \tau_2 \bar{\mathcal F}} -\mfrac{1}{i c - \tau_2(r + \bar r +1) + \tau_2 \mathcal F + \tau_2 \bar{\mathcal F}} \biggr) \ , 
\label{Z-ST-con-c}
\end{align}
where we introduced the chemical potentials
\begin{equation}
 \xi = e^{2 \pi i z} \ , \qquad  \bar \xi = e^{-2 \pi i \bar z} \ . 
\end{equation}
Let us now perform the change of variable $c = 2 \tau_2 s$ in \eqref{Z-ST-con-c}. We find 
\begin{align}
Z^{ST}_{con}&(\tau, z)= \frac{e^{ \frac{4 \pi z_1^2}{\tau_2} } \, |\theta_1(\tau, z)|^2}{\pi \sqrt{n_5} \,  |\eta(\tau)|^{14}} \, \int_{-\infty}^\infty \text dE \sum_{w,n \in \mathds Z} \ \sum_{n_f^{sl}, \bar n_f^{sl} \in \mathds Z + \frac{1}{2} } \ \sum_{n_f^{su}, \bar n_f^{su} \in \mathds Z + \frac{1}{2} }  \nonumber\\
& \times \, \sum_{j_{su}=0}^{n_5-2} \  \sum_{l, l' = 3-n_5}^{n_5-2} \ \sum_{n_b^{su}, \bar n_b^{su} \in \mathds Z}  \, \mathcal I_{t,y} \, \mathcal I_\theta \, \mathcal I_{su} \, \xi^{-\frac{2}{n_5}\mathcal F} \, \bar \xi^{-\frac{2}{n_5} \bar{\mathcal F}} \, \int_{-\infty}^{\infty} \text ds \, \, (q \bar q)^\frac{s^2}{n_5}  \nonumber \\
& \times \, \sum_{r, \bar r =0}^\infty \Biggl\{ \left( \mfrac{1}{2is + r + \bar r +1 + \mathcal F + \bar{\mathcal F}} -\mfrac{1}{2is - r - \bar r -1 + \mathcal F + \bar{\mathcal F}} \right) \left(S_r  \, \overline{S_{\bar r}} -(1-S_r)(1-\overline{S_{\bar r}}) \right) \nonumber \\
& \qquad \ \ + \left( \mfrac{ 1}{2is + r - \bar r  + \mathcal F + \bar{\mathcal F}} -\mfrac{1}{2is - r + \bar r + \mathcal F + \bar{\mathcal F}} \right)\left(S_r  (1-\overline{S_{\bar r}}) -(1-S_r)\overline{S_{\bar r}}) \right)\Biggr\} \delta_{\mathcal J_0 - \bar{\mathcal J}_0,0} \ , 
\end{align}
where we split the sums over $r, \bar r$ over positive and negative integers, and made use of the identity \eqref{S-r-1=1-Sr}. Here we proceed as in Section~\ref{sec:zeromodeints-sl2} and consider the contribution of $\mathfrak{sl}(2, \mathds R)$ primaries by setting $S_r(\tau) \to 1$. Introducing the density of states 
\begin{equation}
\rho(s) = \sum_{r_+ =0}^\infty  \left( \frac{1}{2is + r_+ +1 + \mathcal F + \bar{\mathcal F} } +\frac{1}{2is +r_+ +1 - \mathcal F - \bar{\mathcal F} } \right) \ , 
\label{density-states}
\end{equation} 
this sector of the partition function reads 
\begin{align}
Z^{ST}_{con}(\tau, z)&\Big|_{\mathfrak{sl}(2, \mathds R) \text{ prim.}} = \frac{e^{ \frac{4 \pi z_1^2}{\tau_2} } \, |\theta_1(\tau, z)|^2}{\pi \sqrt{n_5} \,  |\eta(\tau)|^{14}} \, \int_{-\infty}^\infty \text dE \sum_{w,n \in \mathds Z} \ \sum_{n_f^{sl}, \bar n_f^{sl} \in \mathds Z + \frac{1}{2} } \ \sum_{n_f^{su}, \bar n_f^{su} \in \mathds Z + \frac{1}{2} }  \nonumber\\
& \times \, \sum_{j_{su}=0}^{n_5-2} \ \sum_{l, l' = 3-n_5}^{n_5-2} \ \sum_{n_b^{su}, \bar n_b^{su} \in \mathds Z}  \, \mathcal I_{t,y} \, \mathcal I_\theta \, \mathcal I_{su} \, \xi^{-\frac{2}{n_5}\mathcal F} \, \bar \xi^{-\frac{2}{n_5} \bar{\mathcal F}} \, \int_{-\infty}^{\infty} \text ds \, \, (q \bar q)^\frac{s^2}{n_5}  \, \rho(s) \ . 
\end{align}
Making use of eq.~\eqref{charact-cont}, we can rewrite the partition function as 
\begin{align}
Z^{ST}_{con}(\tau, z)&\Big|_{\mathfrak{sl}(2, \mathds R) \text{ prim.}} = \frac{e^{ \frac{4 \pi z_1^2}{\tau_2} } \, |\theta_1(\tau, z)|^2 }{\pi \sqrt{n_5} \,  |\eta(\tau)|^{14}} \, \int_{-\infty}^\infty \text dE \sum_{w,n \in \mathds Z} \ \sum_{n_f^{sl}, \bar n_f^{sl} \in \mathds Z + \frac{1}{2} } \ \sum_{n_f^{su}, \bar n_f^{su} \in \mathds Z + \frac{1}{2} }  \nonumber\\
& \times \, \sum_{j_{su}=0}^{n_5-2} \ \sum_{l, l' = 3-n_5}^{n_5-2} \ \sum_{n_b^{su}, \bar n_b^{su} \in \mathds Z}  \, \mathcal I_{t,y} \, \mathcal I_\theta \, \mathcal I_{su} \int_{-\infty}^{\infty} \text ds \, \rho(s) \, \chi_{\text c}(s, -\mathcal F) \overline{\chi_{\text c}(s, - \bar{\mathcal F})} \ . 
\label{pri-long-pf}
\end{align}
Notice that $-\mathcal F$ is the solution of 
\begin{equation}
\mathcal J_0 = \mathcal F + (j^3_{sl})_0 = 0 \ , 
\end{equation}
viewed as a linear equation in $(j^3_{sl})_0$. Similarly for $-\bar{\mathcal F}$. We can then write the characters in \eqref{pri-long-pf} as 
\begin{equation}
\chi_{\text{c}}(s,-\mathcal F) = \chi_{\text{c}}\left(s,(j^3_{sl})_0 \right)\Biggl|_{\mathcal J \equiv \mathcal F + (j^3_{sl})_0 = 0} \ . 
\end{equation}
Reabsorbing some numerical factors into the overall normalization, we can rewrite eq.~\eqref{pri-long-pf} as
\begin{align}
Z^{ST}_{con}(\tau, z)&\Big|_{\mathfrak{sl}(2, \mathds R) \text{ prim.}} = \frac{e^{ \frac{4 \pi z_1^2}{\tau_2} } \, |\theta_1(\tau, z)|^2 }{|\eta(\tau)|^{14}} \, \int_{-\infty}^\infty \text dE \sum_{w,n \in \mathds Z} \ \sum_{n_f^{sl}, \bar n_f^{sl} \in \mathds Z + \frac{1}{2} } \ \sum_{n_f^{su}, \bar n_f^{su} \in \mathds Z + \frac{1}{2} }  \nonumber\\
& \times \, \sum_{j_{su}=0}^{n_5-2} \ \sum_{l, l' = 3-n_5}^{n_5-2} \ \sum_{n_b^{su}, \bar n_b^{su} \in \mathds Z}  \,  \int_{-\infty}^{\infty} \text d \bigl((j^3_{sl})_0+ (\bar j^3_{sl})_0 \bigr) \sum_{(j^3_{sl})_0 - (\bar j^3_{sl})_0 \in \mathds Z} \nonumber \\
& \times \, \mathcal I_{t,y} \, \mathcal I_\theta \, \mathcal I_{su} \int_{-\infty}^{\infty} \text ds \, \rho(s) \, \chi_{\text{c}}(s,(j^3_{sl})_0) \, \,  \overline{\chi_{\text{c}}(s,(\bar j^3_{sl})_0}) \, \delta_{\mathcal J_0 - \bar{\mathcal J}_0,0} \, \delta(\mathcal J_0 + \bar{\mathcal J}_0) \ , 
\label{Z-ST-decomp-long}
\end{align}
making manifest how the gauge constraints emerge from the path integral.


\section{Discussion} 
\label{sec:discussion}

In this work, we have computed the torus partition function in a wide variety of gauged Wess-Zumino-Witten models.  We have shown how known results for simple cosets $G/H$ can be reproduced by gauging null isometries in $G\times \tilde H$, where $\tilde H$ is a copy of $H$ whose action has the opposite sign.  We illustrated this equivalence by computing the partition functions in both representations and demonstrating their equality for the simple cases
\be
\frac\sutwo\uone \simeq \frac{\sutwo\times \bR_t}{\uone_L\times\uone_R}
~~,~~~~
\frac\sltwo\uone \simeq \frac{\sltwo\times\uone}{\uone_L\times\uone_R} \ . 
\ee

The null-gauged theory also matches a standard operator approach to $G/H$ coset models~\rcite{Karabali:1989dk}, in which one works in the Hilbert space of $G\times \tilde H$, and imposes a BRST constraint that equates the excitations of $H\subset G$ to those of $\tilde H$, {\it separately} among left- and right-movers.  This BRST formalism is precisely that of the null-gauged model.  That one arrives at the same partition function as the $G/H$ coset model via this approach provides confirmation of the equivalence $\frac{G}{H} = \frac{G\times\tilde H}{H_L\times H_R}$.

We furthermore demonstrated the equivalence of the partition functions for the cosets\footnote{The null-gauged formulation was discussed in~\rcite{Israel:2004ir}, but the partition function was evaluated only in the coset orbifold description.}
\be
\bigg(\frac{\sltwo_{n_5}}{\uone}\times\frac{\sutwo_{n_5}}{\uone}\bigg)\Big/\bZ_{n_5} \simeq \frac{\sltwo_{n_5} \times \sutwo_{n_5}}{\uone_L\times\uone_R} \ , 
\ee
describing $n_5$ NS5-branes distributed in a $\bZ_{n_5}$-symmetric fashion in a transverse plane.  

Having established the equivalence of the null-gauged models to conventional coset CFT's, we evaluated the partition function of null-gauged models for the supertube geometries \eqref{MMgeoms}, for which there is no alternative construction.  The fact that we were able to establish the equivalence to standard cosets in the warm-up exercises and the decomposition into characters we carried out gives us confidence that eq.~\eqref{ST-pf-final} is the right result. Note that in arriving at this result, we have made no specialization to any particular supertube background; it applies to any of them, even the non-supersymmetric JMaRT solution.  It was argued in~\rcite{Martinec:2018nco} that the fivebrane decoupling limit of JMaRT is stable at the level of perturbative string theory; here we see confirmation of this result, in that there are no pathologies in the one-loop partition function.

While we have focused on the computation of the partition function, our methods are closely related to the computation of the elliptic genus, which has for instance been computed for coset orbifold constructions of fivebrane backgrounds in~\rcite{Eguchi:2003yy,Eguchi:2004ik,Eguchi:2004yi, Giveon:2015raa}.  The elliptic genus provides a topological invariant of a given string background, in which the right-movers are in a (worldsheet) supersymmetric ground state while the left-movers are arbitrarily excited.  These states embed in string theory as BPS states once we compactify an additional direction on a circle~\cite{Harvey:2013mda}, see also~\rcite{Giveon:2015raa}.  String states with momentum $p$ and winding $w$ on $\bS^1$ must satisfy the Virasoro level-matching constraint
\be
N_L-N_R = pw ~;
\ee
BPS perturbative string states having $N_R=0$ then have some nonzero $N_L$ that is counted by the elliptic genus of the worldsheet CFT.

This embedding is not quite available for the supertube.  A candidate for the circle would be $\bS^1_y$, except it is not entirely physical since the gauge orbits are partly along $y$.  Nevertheless we can identify the analogous states here as those which have nonzero $n_y w_y$ while also having no right-moving oscillator excitation beyond that needed to compensate the Casimir energy ($-\half$ in the NS sector, $0$ in the R sector).  Then $n_yw_y$ forces left-moving oscillator excitation, and the partition function in this sector plays the role of the elliptic genus of the supertube.  As in the coset orbifold that appears in double-scaled little string theory~\rcite{Giveon:1999px,Giveon:1999tq}, there are contributions to this quantity from both the discrete series and from the bottom of the long string continuum, if one follows the same manipulations performed there~\rcite{Eguchi:2010cb,Giveon:2015raa}.  One also finds an analogous appearance of mock-modularity in the partition sums involved.

Our supertube partition functions mirror a similar analysis of perturbative strings in the Euclidean BTZ black hole background~\rcite{Maldacena:2000kv}.  More recently, this partition function has been revisited in~\rcite{Ashok:2022vdz} (see also~\rcite{Ashok:2020dnc}) with an eye toward extracting a stringy quasinormal mode spectrum.  It would be interesting to see whether a similar analysis can be performed for the supertube backgrounds, and whether such modes might be related to the absorption of winding strings into the background.

Also of interest would be to investigate torus correlation functions.  The two-point function of light string vertex operators on the torus computes the OPE coefficients for the emission/absorption of such excitations onto a highly excited string (if we look at the appropriate limit of the amplitude).  It would be interesting to compare such an amplitude to the absorption/emission amplitude onto the long string inside the fivebrane that describes BPS black hole microstates~\rcite{Callan:1996dv,Das:1996wn,Maldacena:1996ix}, in particular for supertubes with a deep throat with large redshift to the cap of the geometry.

Finally, the supertube cosets~\eqref{MMgeoms} have a modulus $R_y$ that descends from a marginal deformation of the parent WZW model.%
\footnote{Naively, the circular array of fivebranes~\eqref{DSLST} also has a modulus, namely the radius of the circle, which controls the value of the dilaton at the tip of the $\frac\sltwo\uone$ ``cigar''.  However, in the worldsheet theory this parameter is a scaling parameter rather than a modulus per se.  One way to see this feature is through the quantum equivalence of this coset CFT to the $\cN=2$ Liouville CFT~\rcite{Giveon:1999px,Giveon:1999tq}, in which the corresponding parameter is the coefficient of the exponential Liouville superpotential.  This parameter can be shifted away by a field redefinition, whose effect is a (KPZ) scaling relation among correlation functions~\rcite{Knizhnik:1988ak,David:1988hj,Distler:1988jt} of a single theory, rather than a moduli space of CFT's.  While one might call this parameter a modulus, it is a rather trivial one, in that the spectrum of the theory does not depend on it, and the OPE coefficients only depend on it via an overall coefficient.}
The moduli of WZW models consist of current-current deformations $\cC_{ab}J^a\bar J^b$; as long as such a deformation is invariant under the gauge group $H$, it leads to a moduli space of the coset theory.
Also, any continuous parameter(s) specifying inequivalent embeddings of the gauge group results in a moduli space of CFT's, see for example~\rcite{Giveon:1993ph}.  Since, as we have seen, null gauging provides a general framework for describing such CFT's, the methods developed above can be extended to a computation of their partition functions.

For instance, the deformation of $AdS_3$ to an asymptotically linear dilaton background%
\footnote{Often mischaracterized as a ``single-trace $T\bar T$ deformation'' of the dual CFT~-- this property only holds at the symmetric orbifold locus in the CFT moduli space.} 
has been recast as an example of null gauging~\rcite{Chakraborty:2019mdf}.
The computation of the worldsheet partition function of these backgrounds, exactly as a function of the deformation parameter, should be a minor modification of the computations described in Sections~\ref{sec:su2/u1} and \ref{sec:sl2/u1}.%
\footnote{See \rcite{Israel:2003ry} and \rcite{Chakraborty:2024mls} for alternative approaches to this computation.}



\vspace{1mm}

\section*{Acknowledgements}

We thank 
Davide Bufalini, Soumangsu Chakraborty, Lorenz Eberhardt, Ji Hoon Lee, Nicolas Kovensky, Stefano Massai 
and
David Turton
for discussions.
The work of EJM is supported in part by DOE grant DE-SC0009924. AD acknowledges support from the Mafalda \& Reinhard Oehme Fellowship.

\vskip 3cm

\appendix 

\section{Theta functions}
\label{app:theta}

In this appendix we spell out our conventions for Theta functions and collect various properties they obey. We define  
\begin{equation}
\Theta^{(\kappa)}_j(\tau,z) = \sum_{n \in \mathds Z} q^{\kappa(n+\frac{j}{2\kappa})^2}\xi^{k(n+\frac{j}{2\kappa})} \ , \qquad q = e^{2 \pi i \tau} \ , \quad  \xi = e^{2 \pi i z} \ , 
\label{Theta-def}
\end{equation}
and 
\begin{subequations}
\begin{align}
\theta_1(\tau, z) & = i \sum_{n = - \infty}^\infty (-1)^n q^{\frac{1}{2}(n-\frac{1}{2})^2}\xi^{n-\frac{1}{2}}  = 2 \sin(\pi z) \, q^{\frac{1}{12}} \, \eta(\tau) \, \prod_{m=1}^\infty (1-\xi q^m) (1-\xi^{-1}q^m) \ , \label{theta-1-def} \\
\theta_2 (\tau, z) & = \sum_{n = - \infty}^\infty q^{\frac{1}{2}(n-\frac{1}{2})^2}\xi^{n-\frac{1}{2}}  = 2 \cos(\pi z) \, q^{\frac{1}{12}} \, \eta(\tau) \, \prod_{m=1}^\infty (1+\xi q^m) (1+\xi^{-1}q^m) \ ,  \label{theta-2-def} \\
\theta_3 (\tau, z) & = \sum_{n = - \infty}^\infty q^{\frac{n^2}{2}}\xi^n  = q^{-\frac{1}{24}} \, \eta(\tau) \, \prod_{m=1}^\infty (1+\xi q^{m-\frac{1}{2}}) (1+\xi^{-1}q^{m-\frac{1}{2}}) \ , \label{theta-3-def} \\
\theta_4(\tau, z) & = \sum_{n = - \infty}^\infty (-1)^n q^{\frac{n^2}{2}}\xi^n  = q^{-\frac{1}{24}} \, \eta(\tau) \, \prod_{m=1}^\infty (1-\xi q^{m-\frac{1}{2}}) (1-\xi^{-1}q^{m-\frac{1}{2}}) \ ,  \label{theta-4-def} 
\end{align}
\end{subequations}
where 
\begin{equation}
\eta(\tau) = q^{1/24} \prod_{n=1}^\infty (1-q^n) \ . 
\label{Dedekind-eta}
\end{equation}
Let us now list a number of relation among Theta functions, which will be used in the main text, see \eg~\rcite{Eguchi:2003yy,Eguchi:2010cb}:
\begin{align}
\theta_1(\tau, z) & = \Theta^{(2)}_{1}(\tau, z) - \Theta^{(2)}_{-1}(\tau, z) \ ,   \label{Theta1m1totheta} \\[.2cm]
\Theta^{(\kappa)}_\ell(\tau, -z) & = \Theta_{-\ell}^{(\kappa)}(\tau,z) \ , \label{Theta-id3} \\[.2cm]
\Theta^{(\kappa)}_{\ell+2\kappa}(\tau, z) & = \Theta_{\ell}^{(\kappa)}(\tau,z) \ , \label{Theta-id4} \\[.2cm] 
\Theta_\ell^{(\kappa)}\left(\tau, z+\mfrac{m_1}{\kappa}\tau\right) & = q^{-\frac{\alpha^2}{4 \kappa}}\, e^{-\pi i m_1 z} \, \Theta^{(\kappa)}_{\ell+m_1}(\tau, z) \ , \qquad \forall \, m_1 \in \mathds Z \ , \label{Theta-id2} \\
\theta_1(\tau, z+ m_1 \tau + m_2) & = (-1)^{m_1 + m_2} q^{-\frac{m_1^2}{2}} e^{-2 \pi i m_1 z} \theta_1(\tau,z) \ , \quad \forall \, m_1, m_2 \in \mathds Z \ , \label{theta1-period} \\[.2cm]
\Theta^{(\kappa)}_\ell(\tau, z + m_1 \tau + m_2) & = q^{-\frac{k m_1^2}{4}} \, e^{-\pi i \kappa m_1 z} \, e^{\pi i \ell m_2 } \, \Theta_{\ell + m_1 \kappa}^{(\kappa)}(\tau, z) \ , \qquad \forall \, m_1,  m_2 \in \mathds Z \ , \label{Theta-period} 
\end{align}
In Section~\ref{sec:su2/u1} we will also need the following identity,\footnote{See \rcite{Elitzur:1989nr, Chang2019} for similar derivations.}
\begin{equation}
\int_0^1 \text d s_1 \int_0^1 \text d s_2 \, e^{-\frac{\pi \kappa}{\tau_2}u_2^2} \, \Theta_\ell^{(k)}(\tau, u) \overline{\Theta_{\ell'}^{(k)}(\tau, u)} =  \, \delta_{\ell, \ell'} \sum_{n \in \mathds Z} \int_0^{1} \text d s_1 \, e^{-\pi \, \kappa \, \tau_2 \left( s_1 + 2 (n+\frac{\ell}{2 \kappa}) \right)^2 } \ , 
\label{Theta2-integral}
\end{equation}
where $\ell, \ell' = -\kappa+1, \dots, \kappa$ and $u = s_1 \tau + s_2$. Let us derive it. From the definitions \eqref{Theta-def} and \eqref{u-def} we have 
\begin{align}
 e^{-\frac{\pi \kappa}{\tau_2}u_2^2} & \, \Theta_\ell^{(\kappa)}(\tau, u) \overline{\Theta_{\ell'}^{(\kappa)}(\tau, u)} \nonumber \\
 & = \sum_{n,n' \in \mathds Z} \exp\Bigl[ -\pi \kappa \Bigl(\tau_2 \, s_1^2 + 2 \tau_2 (n + \tfrac{\ell}{2 \kappa})^2 + 2 \tau_2 (n' + \tfrac{\ell'}{2 \kappa})^2 \Bigr) + 2\pi  i \kappa \tau_1 \Bigl( (n + \tfrac{\ell}{2\kappa})^2 - (n' + \tfrac{\ell'}{2\kappa})^2 \Bigr)\Bigr]  \nonumber \\
& \hspace{40pt} \times \, \exp \Bigl[-2 \pi \kappa s_1 \tau_2  \left( n + \tfrac{\ell}{2\kappa} + n' + \tfrac{\ell'}{2\kappa} \right) + 2 \pi i \kappa s_1 \tau_1 \left( n + \tfrac{\ell}{2\kappa} - n' - \tfrac{\ell'}{2\kappa} \right)\Bigr] \nonumber \\
& \hspace{40pt} \times \, \exp \Bigl[2 \pi i \kappa s_2 \Bigl( (n + \tfrac{\ell}{2\kappa}) - (n' + \tfrac{\ell'}{2\kappa}) \Bigr)\Bigr] \ . 
\label{theta-theta-opened}
\end{align}
Using the Kronecker delta representation
\begin{equation}
\int_0^1 e^{2 \pi i x s} \text d s = \delta_{x,0} \ , 
\label{Kr-delta}
\end{equation}
we can evaluate the integral over $s_2$ and the last line in \eqref{theta-theta-opened} gives
\begin{equation}
\int_0^1 \text d s_2 \, \exp \Bigl[2 \pi i \kappa s_2 \Bigl( (n + \tfrac{\ell}{2\kappa}) - (n' + \tfrac{\ell'}{2\kappa}) \Bigr)\Bigr]= \delta_{\kappa(n-n')+\frac{\ell-\ell'}{2},0} = \delta_{n,n'} \, \delta_{\ell,\ell'} \ , 
\end{equation}
where the second equality follows from $-\kappa+1 \leq \ell, \ell' \leq \kappa$. Enforcing the delta functions and completing the square one obtains eq.~\eqref{Theta2-integral} as claimed. 

\section{Character formulae of the \texorpdfstring{$\boldsymbol{SU(2)}$}{} 
WZW model and related coset theories}
\label{app:su(2)}

Let us review some character formulae for the $SU(2)$ WZW model, the parafermion coset and the $\mathcal N =2$ minimal model $\frac{SU(2)}{U(1)}$.   

\paragraph{Bosonic $\mathfrak{su}\boldsymbol{(2)_{\kappa-2}}$ WZW model} The characters of the bosonic $SU(2)$ WZW model at level $\kappa -2$ read \rcite{DiFrancesco:1997nk}
\begin{equation}
\chi^{\mathfrak{su}(2)}_{B,j}(\tau, z_B) = \frac{\Theta_{j +1}^{(\kappa)}(\tau, z_B) - \Theta_{-j -1}^{(\kappa)}(\tau, z_B)}{\Theta_1^{(2)}(\tau, z_B) - \Theta_{-1}^{(2)}(\tau, z_B)} \ , \qquad j = 0 \,, \dots \,, \kappa -2 \ ,  
\label{su(2)B-characters}
\end{equation}
where Theta functions have been defined in eq.~\eqref{Theta-def}, the suffix $B$ stands for bosonic, as opposed to the supersymmetric $SU(2)$ WZW theory and $z_B$ is the chemical potential associated to the $SU(2)$ Cartan $j_0^3$ (similarly for the anti-holomoprhic sector). The diagonal modular invariant partition function is
\begin{equation}
\mathcal Z_{B}^{\mathfrak{su}(2)}(\tau, z_B) = \text{Tr}\left[ q^{L_0-\frac{1}{24}} \bar q^{\bar L_0-\frac{1}{24}} e^{2 \pi i z_B \, j_0^3} e^{-2 \pi i \bar z_B \, \bar j_0^3} \right] = \sum_{j = 0}^{\kappa -2} \left|\chi^{\mathfrak{su}(2)}_{B,j}(\tau, z_B)\right|^2 \ . 
\label{SU2-pf}
\end{equation}
Let us study some periodicity properties of the characters \eqref{su(2)B-characters}, which are used in the main text. Consider the denominator in eq.~\eqref{su(2)B-characters}. It follows from \eqref{Theta1m1totheta} and \eqref{theta1-period} that for any $m_1, m_2 \in \mathds Z$, 
\begin{multline}
\Theta_1^{(2)}(\tau, z + m_1 \tau + m_2) - \Theta_{-1}^{(2)}(\tau, z+ m_1 \tau + m_2) \\
= (-1)^{m_1+m_2} \, q^{-\frac{m_1^2}{2}} \,  e^{-2 \pi i m_1 z } \, \left(\Theta_1^{(2)}(\tau, z) - \Theta_{-1}^{(2)}(\tau, z)\right) \ . 
\end{multline}
Let us now consider the numerator of \eqref{su(2)B-characters}. We first consider $m_1 \in 2 \mathds Z$. From \eqref{Theta-id4} and \eqref{Theta-period} it follows 
\begin{align}
\Theta^{(\kappa)}_{j+1}(\tau, z + m_1 \tau + m_2) & = (-1)^{(j+1)m_2} q^{-\frac{\kappa m_1^2}{4}} \, e^{-\pi i \kappa m_1 z} \, \Theta^{(\kappa)}_{j+1}(\tau, z) \ , \\
\Theta^{(\kappa)}_{-j-1}(\tau, z + m_1 \tau + m_2) & = (-1)^{(j+1)m_2} q^{-\frac{\kappa m_1^2}{4}} \, e^{-\pi i \kappa m_1 z} \, \Theta^{(\kappa)}_{-j-1}(\tau, z)
\end{align}
and hence 
\begin{equation}
\chi^{\mathfrak{su}(2)}_{B,j}(\tau, z + m_1 \tau + m_2)=  (-1)^{j m_2} \, q^{-\frac{(\kappa -2) m_1^2}{4}} \, e^{-\pi i (\kappa -2)m_1 z } \, \chi^{\mathfrak{su}(2)}_{B,j}(\tau, z) \ . 
\label{su2-ch-period-even}
\end{equation}
Let us now consider $m_1 \in 2 \mathds Z + 1$. Using again \eqref{Theta-id4} and \eqref{Theta-period} we find 
\begin{equation}
\Theta_{j+1}^{(\kappa)}(\tau, z + m_1 \tau + m_2) = (-1)^{(j+1)m_2}  q^{-\frac{\kappa m_1^2}{4}} \, e^{-\pi i \kappa m_1 z} \, \Theta_{-1-(\kappa-2-j)}^{(\kappa)}(\tau, z) 
\end{equation}
and similarly 
\begin{equation}
\Theta_{-j-1}^{(\kappa)}(\tau, z + m_1 \tau + m_2) = (-1)^{(j+1)m_2} q^{-\frac{\kappa m_1^2}{4}} \, e^{-\pi i \kappa m_1 z} \, \Theta_{1+(\kappa-2-j)}^{(\kappa)}(\tau, z) \ . 
\end{equation}
It follows that
\begin{equation}
\chi^{\mathfrak{su}(2)}_{B,j}(\tau, z_B + m_1 \tau + m_2)=  (-1)^{(j+1) m_2} \, q^{-\frac{(\kappa-2) m_1^2}{4}} \, e^{-\pi i (\kappa-2)m_1 z_B } \, \chi^{\mathfrak{su}(2)}_{B,\kappa-2-j}(\tau, z_B) \ . 
\label{su2-ch-period-odd}
\end{equation}

\paragraph{The parafermion theory} The $\mathfrak{su}(2)_{\kappa-2}$ characters \eqref{su(2)B-characters} can be decomposed as 
\begin{equation}
\chi^{\mathfrak{su}(2)}_{B,j}(\tau, z_B) = \sum_{\ell = 3-\kappa}^{\kappa -2} \chi_{B,j, \ell}^{\frac{\mathfrak{su}(2)}{\mathfrak{u}(1)}}(\tau) \,  \frac{\Theta_\ell^{(\kappa-2)}(\tau,z_B)}{\eta(\tau)} \ , 
\label{parafermion-character-decomposition}
\end{equation}
where $\eta(\tau)$ is the Dedekind eta function \eqref{Dedekind-eta} and we introduced the characters $\chi_{B,j, \ell}^{\frac{\mathfrak{su}(2)}{\mathfrak{u}(1)}}(\tau)$ of the parafermion coset $\frac{SU(2)_{\kappa-2}}{ U(1)}$.
The parafermion characters vanish for $j+\ell \in 2 \mathds Z + 1$ \cite{DiFrancesco:1997nk}, 
\begin{align}
 \chi_{B,j, \ell}^{\frac{\mathfrak{su}(2)}{\mathfrak{u}(1)}}(\tau) & = 0 \ ,  \qquad  j + \ell \text{ odd} \ , \label{parf-ch=0} 
\end{align}
and obey the identifications \rcite{DiFrancesco:1997nk}
\begin{equation}
\chi_{B, \, j, \ell}^{\frac{\mathfrak{su}(2)}{\mathfrak{u}(1)}}(\tau) = \chi_{B, \, j, -\ell}^{\frac{\mathfrak{su}(2)}{\mathfrak{u}(1)}}(\tau) = \chi_{B, \, j, \ell+2k_B}^{\frac{\mathfrak{su}(2)}{\mathfrak{u}(1)}}(\tau) = \chi_{B, \, k_B-j, \ell+k_B}^{\frac{\mathfrak{su}(2)}{\mathfrak{u}(1)}}(\tau) \ .
\label{parafermion-ch-id} 
\end{equation}
The diagonal invariant partition function for the parafermion coset is \rcite{Gepner:1986hr}
\begin{equation}
\mathcal Z_{B}^{\frac{\mathfrak{su}(2)}{\mathfrak{u}(1)}}(\tau) = \frac{1}{2} \sum_{j = 0}^{\kappa-2} \sum_{\ell = 3-\kappa}^{\kappa-2} \Bigl| \chi_{B, \, j, \ell}^{\frac{\mathfrak{su}(2)}{\mathfrak{u}(1)}}(\tau)\Bigr|^2 \ , 
\label{parafermion-pf}
\end{equation}
where the factor of $\frac{1}{2}$ is due to the identification $\chi_{B, \, j, \ell}^{\frac{\mathfrak{su}(2)}{\mathfrak{u}(1)}}(\tau) = \chi_{B,\, k-j, \ell+\kappa}^{\frac{\mathfrak{su}(2)}{\mathfrak{u}(1)}}(\tau)$. 

\smallskip

\paragraph{The $\boldsymbol{\mathcal N =2}$ minimal model} The characters $\chi^{\frac{\mathfrak{su}(2)}{\mathfrak{u}(1)}}_{S, \, j,\ell,s}(\tau,z)$ of the $\mathcal N =2$ minimal model $\frac{SU(2)}{U(1)}$ are defined by \rcite{Gepner:1987qi, Eguchi:2004ik},
\begin{equation}
\chi^{\mathfrak{su}(2)}_{B,j}(\tau, z_B) \, \Theta_s^{(2)}(\tau,z_F) = \sum_{\ell = -\kappa+1}^{\kappa} \chi^{\frac{\mathfrak{su}(2)}{\mathfrak{u}(1)}}_{S, \, j,\ell,s}(\tau,z) \, \Theta^{(\kappa)}_{\ell}(\tau, u) \ , \qquad s= -1,0,1,2 \ ,
\label{su(2)-pf-decomposition}
\end{equation} 
where the superscript S stands for $\mathcal N=2$ supersymmetric, as opposed to the bosonic parafermion theory. Let us remind that in \eqref{su(2)-pf-decomposition} the chemical potentials are related by eq.~\eqref{su2-chemical-potentials-relation}. The characters vanish for \rcite{Gepner:1987qi}
\begin{equation}
\chi^{\frac{\mathfrak{su}(2)}{\mathfrak{u}(1)}}_{S,j,\ell,s}(\tau,z) = 0 \ , \qquad j + \ell + s \in 2\mathds{Z}+1 \ ,
\label{chiS=0}
\end{equation}
and obey the identification \rcite{Gepner:1987qi, Eguchi:2004ik}
\begin{equation}
\chi^{\frac{\mathfrak{su}(2)}{\mathfrak{u}(1)}}_{S, \, \kappa-j-2, \, \ell+\kappa, \, s+2}(\tau, z) = \chi^{\frac{\mathfrak{su}(2)}{\mathfrak{u}(1)}}_{S, \, j, \ell, s}(\tau,z)
\label{super-pf-ch-identifications}
\end{equation}
From eq.~\eqref{su(2)-pf-decomposition} for $s=-1$ and $s=1$ and eq.~\eqref{Theta1m1totheta}, it follows in particular
\begin{equation}
\chi^{\mathfrak{su}(2)}_{B,j}(\tau, z_B) \, \theta_1(\tau, z_F) = \sum_{\ell = -\kappa+1}^\kappa I_\ell^j(\tau, z) \Theta^{(\kappa)}_\ell(\tau, u) \ , 
\label{chi-theta1=ITheta-identity}
\end{equation}
where 
\begin{equation}
I^j_\ell(\tau, z) = \chi^{\frac{\mathfrak{su}(2)}{\mathfrak{u}(1)}}_{S, \, j,\ell,1}(\tau,z) - \chi^{\frac{\mathfrak{su}(2)}{\mathfrak{u}(1)}}_{S, \, j,\ell,-1}(\tau,z)
\label{susy-su(2)-tilde-R-ch}
\end{equation}
are the $\mathcal N=2$ coset characters in the $\tilde{\text R}$ sector. It follows from eq.~\eqref{super-pf-ch-identifications} that they obey 
\begin{equation}
I^j_\ell(\tau,z) = - I^{\kappa-j-2}_{\ell + \kappa}(\tau,z) = I^{j}_{\ell -2 \kappa}(\tau,z)  \ . 
\label{Is-identification}
\end{equation} 
From eqs.~\eqref{chiS=0} and \eqref{susy-su(2)-tilde-R-ch} we also find that 
\begin{equation}
    I^j_\ell(\tau, z) = 0 \ , \qquad \text{for} \quad j+\ell \in 2 \mathds Z \ . 
    \label{Ijl=0}
\end{equation}
The diagonal modular invariant partition function of the $\mathcal N =2$ minimal model in the $\tilde{\text{R}}$ sector reads~\rcite{Gepner:1986hr, Cappelli:1986hf}
\begin{equation}
\mathcal Z_{S}^{\frac{\mathfrak{su}(2)}{\mathfrak{u}(1)}}(\tau,z) = \frac{1}{2} \sum_{j = 0}^{\kappa-2} \sum_{\ell = -\kappa+1}^{\kappa} I^j_\ell(\tau,z) \overline{I^j_\ell(\tau, z)} \ . 
\label{super-pf-pf}
\end{equation}

\section{The \texorpdfstring{$\boldsymbol{U(1)}$}{} partition function}
\label{app:u(1)y-pf}

In this appendix we discuss partition functions of free bosons for symmetric and asymmetric gauging. 

\subsection{Symmetric gauging}
\label{app:u(1)y-pf-symmetric}

Consider a compact boson of radius $R$ on the torus, with boundary conditions 
\begin{subequations}
\begin{align}
X(\sigma_1+2 \pi, \sigma_2) & = X(\sigma_1, \sigma_2) + 2 \pi R \, m_1 + 2 \pi s_1 \label{U(1)y-bc-1} \ , \\
X(\sigma_1, \sigma_2 +2 \pi) & = X(\sigma_1, \sigma_2) - 2 \pi R \, m_2 - 2 \pi s_2 \label{U(1)y-bc-2} \ , 
\end{align}
\label{U(1)y-bc}%
\end{subequations}
where $\sigma_1, \sigma_2 \in [0, 2 \pi]$ parametrize the torus and $m_1, m_2 \in \mathds Z$. Let us compute the partition function from both the operator formalism and the path integral and see that the two results agree. 

\paragraph{Path integral formalism} 
Let us recall that for $s_1 = s_2 =0$, the partition function 
\begin{equation}
Z(\tau, u=0) =  \int \mathcal D X \, e^{- S} \ ,  \qquad \text{with} \qquad S = \frac{1}{\pi} \int \text d^2 v \, \partial_v X \partial_{\bar v} X  \ , 
\label{e-S}
\end{equation}
can be computed in the path integral formalism and reads \rcite{DiFrancesco:1997nk} 
\begin{equation}
Z(\tau, u =0) = \frac{1}{ |\eta(\tau)|^2} \sum_{w,p \in \mathds Z} q^{\frac{1}{4}\left( \frac{p}{R} + R w \right)^2} \bar q^{\frac{1}{4}\left( \frac{p}{R} - R w \right)^2} \ . 
\end{equation}
Let us now consider the derivation for generic $s_1$ and $s_2$. In order to enforce \emph{both} the boundary conditions \eqref{U(1)y-bc}, we introduce the free boson
\begin{equation}
X'(\sigma_1, \sigma_2) = X(\sigma_1, \sigma_2)  - \Phi(\sigma_1, \sigma_2) \ , \qquad \text{with} \quad \Phi(\sigma_1, \sigma_2) = -\sigma_1 s_1 + \sigma_2 s_2 \ . 
\end{equation} 
We can then repeat the derivation of \rcite{DiFrancesco:1997nk} with the free boson $X'(\sigma_1, \sigma_2)$ in place of $X(\sigma_1, \sigma_2)$. The action is now 
\begin{equation}
S = \frac{1}{\pi} \int \text d^2 v \, \partial_v X' \partial_{\bar v} X' =  \frac{1}{\pi} \int \text d^2 v \, \partial_v (X - \Phi) \partial_{\bar v} (X-\Phi)   \ , 
\label{S-U(1)-symm-path-int}
\end{equation}
and one obtains 
\begin{equation}
Z(\tau, \bar \tau, u, \bar u)  =  \frac{R}{\sqrt{\tau_2 } \, |\eta(\tau)|^2} \sum_{m_1, m_2 \in \mathds Z} \exp \left( -\frac{\pi R^2}{\tau_2} \left|m_1 \tau + m_2 + \frac{u}{R}\right|^2 \right) \ . 
\label{path-integral-U(1)y-pf}
\end{equation}
We remind the reader that $u$ is related to $s_1$ and $s_2$ by eq.~\eqref{u-def}. Poisson resumming eq.~\eqref{path-integral-U(1)y-pf} we find
\begin{equation}
Z(\tau, u) = \frac{e^{-\frac{\pi}{\tau_2}u_2^2} }{ |\eta(\tau)|^2} \sum_{w,p \in \mathds Z} q^{\frac{1}{4} \left( \frac{p}{R} + wR \right)^2} \bar q^{\frac{1}{4} \left( \frac{p}{R} - wR \right)^2} \, e^{\pi i u(\frac{p}{R}+wR)} \, e^{\pi i \bar u(\frac{p}{R}-wR)} \ . 
\label{U1y-pf-path-int}
\end{equation}
Since
\begin{equation}
u_2 = s_1 \tau_2 \ , 
\end{equation}
the partition function can equivalently be written as 
\begin{equation}
Z(\tau, u)  = \frac{q^\frac{s_1^2}{4} \bar q^\frac{s_1^2}{4}}{ |\eta(\tau)|^2} \sum_{w, \, p \, \in \, \mathds Z} q^{\frac{1}{4} \left( \frac{p}{R} + wR \right)^2} \bar q^{\frac{1}{4} \left( \frac{p}{R} - wR \right)^2} \, e^{\pi i u(\frac{p}{R}+Rw)} \, e^{\pi i \bar u(\frac{p}{R}-wR)} \ . 
\label{Z-U(1)-symmetric-path-int}
\end{equation}

\paragraph{Operator formalism} In the operator formalism we compute the partition function as a trace over the Hilbert space. Let us start by decomposing the boson $X$ into holomorphic and anti-holomorphic contributions, 
\begin{equation}
X = X_L(v) + X_R(\bar v) \ , \qquad \partial_{\bar v} X_L(v) = 0 \ , \qquad \partial_v X_R(\bar v) = 0 \ . 
\end{equation}
Since $\partial_v X_L(v)$ and $\partial_{\bar v} X_R(\bar v)$ 
are periodic, we obtain the Fourier series, 
\begin{equation}
\begin{split}
X_L(v) & = a_L + b_L v -i \sum_{n \neq 0} \frac{\alpha_n}{n} e^{i n v} \ , \\
X_R(\bar v) & = a_R +  b_R \bar v - i \sum_{n \neq 0} \frac{\bar \alpha_n}{n} e^{i n \bar v} \ . 
\end{split}
\end{equation}
Enforcing the boundary condition \eqref{U(1)y-bc-1} we find, 
\begin{equation}
b_L + b_R =  m_1 R + s_1 \ , 
\end{equation}
that we solve as 
\begin{equation}
b_L = \frac{p_L + s_1}{2} \ ,  \qquad b_R = \frac{-p_R + s_1}{2} \ , \qquad p_L - p_R = 2 m_1 R \ . 
\label{bl-br}
\end{equation}
In the operator formalism eq.~\eqref{U(1)y-bc-2} is implemented as a twist in the trace partition function. The partition sum we wish to compute is then 
\begin{equation}
\text{Tr}_{\mathcal H}[g q^{H} \bar g \bar q^{\bar H}] = \text{Tr}_{\mathcal H_L}[g q^{H}] \, \text{Tr}_{\mathcal H_R}[\bar g \bar q^{\bar H}]  \ , 
\end{equation}
where the twists read
\begin{equation}
g = e^{\pi i (p_L + \alpha s_1) s_2} \ , \qquad \bar g = e^{\pi i (p_R-\alpha s_1) s_2} \ ,
\label{g-twists}
\end{equation}
for some $\alpha \in \mathds R$. The precise value of $\alpha$ in \eqref{g-twists} does not matter here, since it cancels when assembling holomorphic and anti-holomorphic sectors.
The Hamiltonian for the holomorphic sector is
\begin{equation}
H = \frac{1}{2 \pi} \int_0^{2 \pi} \text d \sigma_1 \, \partial_v X(\sigma_1, \sigma_2 = 0) \, \partial_v X(\sigma_1, \sigma_2 = 0) -\frac{1}{24}= b_L^2 -\frac{1}{24} + \sum_{n>0} \alpha_n \alpha_{-n} \ , 
\end{equation}
and similarly for the anti-holomorphic sector. The holomorphic contribution to the partition function thus reads 
\begin{equation}
Tr_{\mathcal H_L}[g q^{H}] = \frac{1}{\eta(\tau)} \, \sum_{p_L} q^{\frac{(p_L+s_1)^2}{4}} e^{\pi i p_L s_2} = q^{\frac{s_1^2}{4}} \sum_{p_L} q^{\frac{p_L^2}{4}} e^{\pi i p_L u} \ . 
\end{equation}
Putting together holomorphic and anti-holomorphic sectors, 
\begin{equation}
Z(\tau, \bar \tau, u , \bar u) = Tr_{\mathcal H}[g q^{H} \bar g \bar q^{\bar H}] = \frac{q^{\frac{s_1^2}{4}} \bar q^{\frac{s_1^2}{4}}}{|\eta(\tau)|^2} \sum_{p_L, p_R} q^{\frac{p_L^2}{4}} \, \bar q^{\frac{p_R^2}{4}} \, e^{\pi i p_L u} \, e^{\pi i p_R \bar u} \ . 
\label{pf-operator-U(1)y}
\end{equation}
Let us specify the lattice of the momenta. From eq.~\eqref{bl-br} it follows
\begin{equation}
p_L - p_R \in 2 R \, \mathds Z \ . 
\label{pL+pR}
\end{equation}
Furthermore, we have
\begin{equation}
Z(\tau, \bar \tau, u , \bar u) = Z(\tau+1, \bar \tau+1, u , \bar u) = e^{\frac{\pi i (p_L^2 - p_R^2)}{2}} \, Z(\tau, \bar \tau, u , \bar u)  \ , 
\end{equation}
and hence modular invariance implies
\begin{equation}
(p_L-p_R)(p_L+p_R) \in 4 \mathds Z \ . 
\end{equation}
Using \eqref{pL+pR} we find 
\begin{equation}
p_L + p_R \in \frac{2}{R} \mathds Z \ .
\end{equation}
We thus find that the momenta obey the usual quantization, 
\begin{equation}
p_L = \frac{n}{R} + w R \ , \qquad  p_R = \frac{n}{R} - w R \ , 
\end{equation}
and the partition function reads
\begin{equation}
Z(\tau, u)  = \frac{q^\frac{s_1^2}{4} \bar q^\frac{s_1^2}{4}}{ |\eta(\tau)|^2} \sum_{w, \, p \, \in \, \mathds Z} q^{\frac{1}{4} \left( \frac{p}{R} + wR \right)^2} \bar q^{\frac{1}{4} \left( \frac{p}{R} - wR \right)^2} \, e^{\pi i u(\frac{p}{R}+Rw)} \, e^{\pi i \bar u(\frac{p}{R}-Rw)} \ , 
\label{Z-U(1)-symmetric-oper}
\end{equation}
which agrees with eq.~\eqref{Z-U(1)-symmetric-path-int}. As we review in Appendix~\ref{app:different-pfs}, in the operator formalism the partition function is often defined omitting the prefactor $q^\frac{s_1^2}{4} \bar q^\frac{s_1^2}{4} = e^{-\frac{\pi}{\tau_2}u_2^2}$ in \eqref{Z-U(1)-symmetric-oper}. In fact, the free compact boson partition function is often defined as
\begin{equation}
\mathcal Z(\tau, u)  = \frac{1}{ |\eta(\tau)|^2} \sum_{w, \, p \, \in \, \mathds Z} q^{\frac{1}{4} \left( \frac{p}{R} + wR \right)^2} \bar q^{\frac{1}{4} \left( \frac{p}{R} - wR \right)^2} \, e^{\pi i u(\frac{p}{R}+Rw)} \, e^{\pi i \bar u(\frac{p}{R}-Rw)} \ . 
\end{equation}
This results in different modular properties. 

\subsection{Asymmetric gauging}

Let us now consider a slight variation of eq.~\eqref{S-U(1)-symm-path-int} and compute
\begin{equation}
Z(\tau, u) = \int \mathcal D X e^{-S_1 - S_2} \ , \label{S-U(1)-asymm-path-int}
\end{equation}
with
\begin{equation}
\begin{aligned}
S_1 & = \frac{1}{\pi} \int \text d^2 v \, \partial_v (X - r \Phi) \, \partial_{\bar v} (X- \ell \Phi) \ ,  \\
S_2 & =  \frac{(\ell -r)^2}{4 \pi} \int \text d^2v \, \partial_v \Phi \, \partial_{\bar v} \Phi   \ .  
\end{aligned}
\end{equation}%
and $\ell, r \in \mathds R$. Notice that for $\ell=r$, eq.~\eqref{S-U(1)-asymm-path-int} reduces to \eqref{S-U(1)-symm-path-int}. We will also see that the same partition function can be computed in the operator formalism and that the two results agree. 

\paragraph{Path integral formalism} The contribution of $S_2$ is easy to compute, 
\begin{align}
S_2 = \frac{(\ell -r)^2}{4 \pi} \int \text d^2v \, \left(\frac{-i \bar u}{2 \tau_2} \right) \left(\frac{i u}{2 \tau_2} \right) = \frac{\pi \, (\ell -r)^2 \, |u|^2}{4 \, \tau_2} \ . 
\end{align}
Since 
\begin{equation}
\partial_v \Phi = -\frac{i \bar u}{2 \tau_2} \ , \qquad \partial_{\bar v} \Phi = \frac{i u}{2 \tau_2} \ , 
\end{equation}
the contribution of $S_1$ simply follows from the discussion of Appendix~\ref{app:u(1)y-pf-symmetric} with the replacements 
\begin{equation}
u \mapsto \ell u \ , \qquad \bar u \mapsto r \bar u \ . 
\end{equation}
We obtain
\begin{equation}
Z(\tau,  u)  = \frac{R \, e^{-\frac{\pi \, (\ell -r)^2 \, |u|^2}{4 \, \tau_2}}}{\sqrt{\tau_2 } \, |\eta(\tau)|^2} \sum_{m_1, m_2 \in \mathds Z} \exp \Bigl( -\mfrac{\pi R^2}{\tau_2} \left(m_1 \tau + m_2 + \mfrac{\ell \, u}{R}\right)\left(m_1 \bar \tau + m_2 + \mfrac{r \, \bar u}{R}\right) \Bigr) \ . 
\label{Z-U(1)-asymmetric-path-int}
\end{equation}

\paragraph{Operator formalism} Since in the operator formalism holomorphic and anti-holomorphic characters are first independently computed and then assembled to define the partition function, a natural notion of asymmetric gauging is mapping $s_1$ and $s_2$ to 
\begin{equation}
s_1 \mapsto \ell s_1 \ , \qquad s_2 \mapsto \ell s_2  \ , 
\end{equation}
in the holomorphic sector and 
\begin{equation}
s_1 \mapsto r s_1 \ , \qquad s_2 \mapsto r s_2  \ , 
\end{equation}
in the anti-holomorphic sector. This notion of gauging is also the one adopted in asymmetric orbifolds, see \eg~\rcite{Aoki:2004sm}. Eq.~\eqref{Z-U(1)-symmetric-oper} is replaced by 
\begin{equation}
Z(\tau, u) = \frac{e^{\pi i \alpha (\ell^2 - r^2) s_1 s_2}}{ |\eta(\tau)|^2}  \, e^{\frac{\pi i u_2^2}{2 \tau_2^2}(\ell^2 \tau - r^2 \bar \tau)} \sum_{w,p \in \mathds Z} q^{\frac{1}{4} \left( \frac{p}{R} + wR \right)^2} \bar q^{\frac{1}{4} \left(\frac{p}{R} - wR \right)^2} \, e^{\pi i \ell (\frac{p}{R}+Rw)u} \, e^{\pi i r (\frac{p}{R}-Rw)\bar u} \ , 
\label{asymm-U(1)y-operator}
\end{equation}
where the extra factor $e^{\pi i \alpha (\ell^2 - r^2) s_1 s_2}$ is due to the phases in \eqref{g-twists}, no longer cancelling against each other. Poisson resumming eq.~\eqref{asymm-U(1)y-operator} we obtain
\begin{multline}
Z(\tau, u) = \frac{R}{\sqrt{\tau_2} \, |\eta(\tau)|^2} \, e^{\pi i \alpha (\ell^2 - r^2) s_1 s_2} \, e^{\frac{\pi i u_2^2}{2 \tau_2^2}(\ell^2 \tau - r^2 \bar \tau)} \, e^{-\frac{\pi}{4 \tau_2}(\ell u - r \bar u)^2 } \\
\sum_{w, m \in \mathds Z} \exp \Bigl( -\mfrac{\pi R^2}{\tau_2}\left(m + w \tau + \mfrac{\ell u}{R} \right)\left(m + w \bar \tau + \mfrac{r \bar u}{R} \right) \Bigr) \ . 
\label{Z-U(1)-asymmetric-oper}
\end{multline}
Using eq.~\eqref{u-def} it is not difficult to check that for 
\begin{equation}
\alpha = \frac{1}{2} \ , 
\label{alpha=1/2}
\end{equation}
one has 
\begin{equation}
\pi i \alpha (\ell^2 - r^2) s_1 s_2 + \frac{\pi i u_2^2}{2 \tau_2^2}(\ell^2 \tau - r^2 \bar \tau) - \frac{\pi}{4 \tau_2}(\ell u - r \bar u)^2  = - \frac{\pi (\ell-r)^2 |u|^2}{4 \tau_2} \ , 
\end{equation}
and hence eqs.~\eqref{Z-U(1)-asymmetric-path-int} and \eqref{Z-U(1)-asymmetric-oper} agree.

\paragraph{Free-fermion radius} At the free-fermion radius $R=\sqrt 2$ (in our conventions $\alpha'=1$) we have~\rcite{Ginsparg:1988ui}
\begin{equation}
\sum_{w,p \in \mathds Z} q^{\frac{1}{4} \left( \frac{p}{R} + wR \right)^2} \bar q^{\frac{1}{4} \left( \frac{p}{R} - wR \right)^2} \, e^{\pi i \ell (\frac{p}{R}+wR)u} \, e^{\pi i r (\frac{p}{R}-wR)\bar u} = \frac{1}{2} \sum_{i=1}^4 \theta_i(\tau, \tfrac{\ell}{\sqrt 2} u) \overline{\theta_i(\tau, \tfrac{r}{\sqrt 2} u)} \ , 
\end{equation}
and the partition function can be rewritten as
\begin{equation}
Z(\tau, u) = \frac{e^{\frac{\pi i}{2}(\ell^2 - r^2) s_1 s_2}}{ 2 |\eta(\tau)|^2} \, e^{\frac{\pi i u_2^2}{2 \tau_2^2}(\ell^2 \tau - r^2 \bar \tau)} \sum_{i=1}^4 \theta_i(\tau, \tfrac{\ell}{\sqrt 2} u) \overline{\theta_i(\tau, \tfrac{r}{\sqrt 2} u)} \ . 
\label{pf-U(1)-free-fermion-radius}
\end{equation}

\section{Different frames for the partition function}
\label{app:different-pfs}

There are two different notions of ``partition function'' in 2D CFT \rcite{Kraus:2006nb}. One, which we will refer to as \emph{trace} partition function, is the character formula 
\begin{equation}
\mathcal Z(\tau, u) = \text{Tr}\left[ q^{L_0-\frac{1}{24}} \bar q^{\bar L_0-\frac{1}{24}} e^{2 \pi i u J} e^{-2 \pi i \bar u \bar J} \right] \ , 
\end{equation}  
where $u$ (respectively $\bar u$) is the chemical potential associated to the current $J$ (respectively $\bar J$). In the examples discussed in the main text, $J$ will be identified with either the $\mathcal R$-charge or the momentum of some free boson. The trace partition function tipically obeys a modular transformation of the schematic form
\begin{equation}
\mathcal Z\left(-\frac{1}{\tau}, \frac{u}{\tau} \right) = e^{\frac{\pi i \, \hat c \, u^2}{\tau}} e^{-\frac{\pi i \, \hat c \, \bar u^2}{\bar \tau}} \, \mathcal Z(\tau, u ) \ , 
\label{modular-trsf-trace}
\end{equation}
for some $\hat c \in \mathds R$. The second notion of ``partition function'', which we are going to call \emph{path integral} partition function, is given by
\begin{equation}
Z(\tau, u) = \int \mathcal Dg \, e^{-S}  \ , 
\end{equation}
where $S$ is the action of the model under consideration, coupled to background gauge fields for the current $J$. In the path integral, a modular transformation is implemented as a change of coordinates followed by a Weyl transformation and hence the path integral partition function is automatically modular invariant, 
\begin{equation}
Z\Big(-\frac{1}{\tau}, \frac{u}{\tau} \,\Big) = Z(\tau, u) \ . 
\end{equation}
To avoid confusion, in the following as well as in the main text, we denote path integral partition functions by $Z$ and reserve the curly character $\mathcal Z$ for trace partition functions.  
Given the different modular transformations, it is clear that these two notions of partition function do not coincide exactly. They are typically related as \rcite{Kraus:2006nb}
\begin{equation}
Z(\tau, u, \bar u)= e^{\frac{2 \pi \hat c}{\tau_2}u_1^2} \mathcal Z(\tau, u, \bar u)\ , 
\label{pfs-relation}
\end{equation}
where, for example, 
\begin{equation}
    \hat c = \frac{1}{2} 
    \label{c-hat-u(1)}
\end{equation}
for a compact boson and we saw in Section~\ref{sec:su2/u1} that for the $\mathcal N=2$ minimal model defined by the coset $\frac{SU(2)_\kappa}{U(1)}$ we have
\begin{equation}
\hat c = \frac{\kappa -2}{\kappa} \ .     
\end{equation}
In fact, one can check that the factor $e^{\frac{2 \pi \hat c}{\tau_2}u_1^2}$ in eq.~\eqref{pfs-relation} offsets the exponential factors in \eqref{modular-trsf-trace}.

\subsection*{Relating path integral and trace partition functions for the $\boldsymbol{SU(2)}$ WZW model}

In this section we derive some relations between trace and path integral partition functions for the $SU(2)$ WZW model. The discussion is not entirely trivial since, to the best of our knowledge, we are not aware of a path integral derivation of the partition function of the $SU(2)$ WZW model. 

\paragraph{Symmetric embeddings} We define the path integral partition function as 
\begin{equation}
    Z_{B}^{\mathfrak{su}(2)}(\tau, z_B) = \int \mathcal D g_{su} \, \exp \left(-(\kappa-2)S_{su}^B(g_{su},a[z_B],a[z_B]^\dagger) \right) \ , 
    \label{su2-path-integral-pf}
\end{equation}
see eq.~\eqref{asym-act-a}, with embeddings 
\begin{equation}
\varepsilon_L^{su} = \varepsilon_R^{su} \equiv \varepsilon = -\frac{\sigma_3}{2} \ , 
\label{SU2-embeddings-appendix}
\end{equation} 
The trace partition function of the bosonic $SU(2)_{\kappa-2}$ model,
\begin{equation}
\mathcal Z_{B}^{\mathfrak{su}(2)}(\tau, z_B) = \text{Tr}\left[ q^{L_0-\frac{1}{24}} \bar q^{\bar L_0-\frac{1}{24}} e^{2 \pi i z_B \, j_0^3} e^{-2 \pi i \bar z_B \, \bar j_0^3} \right] \ , 
\end{equation}
see eq.~\eqref{SU2-pf}, obeys the modular transformation~\eqref{modular-trsf-trace} with 
\begin{equation}
    \hat c = \frac{\kappa-2}{2} \ . 
    \label{c-hat-su(2)}
\end{equation}
According to the discussion of the previous section, we would then expect that\footnote{We denote by $(z_B)_1$ and $(z_B)_2$ respectively the real and imaginary part of $z_B$. }
\begin{equation}
 Z_{B}^{\mathfrak{su}(2)}(\tau, z_B) = e^{\frac{\pi (\kappa-2)}{\tau_2}(z_B)_1^2} \, \mathcal Z_{B}^{\mathfrak{su}(2)}(\tau, z_B) \ . 
\label{ZZ-relation-1}
\end{equation}
Unfortunately, we cannot verify eq.~\eqref{ZZ-relation-1} directly, by computing left- and right-hand-side independently. In fact, we are not aware of a path integral derivation of the $SU(2)$ partition function. However, a few consistency checks on eq.~\eqref{ZZ-relation-1} can be made. For $\kappa=4$, the bosonic $SU(2)$ WZW model is at level 2 and is equivalent to three free fermions transforming in the adjoint representation. In fact, making use of eq.~\eqref{SU(2)-U(1)-fermions-pf} with $z=0$ and $Y=0$, we reproduce eq.~\eqref{ZZ-relation-1} with $\kappa=4$. Similarly, for $\kappa = 3$ the $SU(2)$ WZW model is at level 1 and is equivalent to a compact boson. In fact, eq.~\eqref{c-hat-su(2)} reduces to eq.~\eqref{c-hat-u(1)}. 

From eq.~\eqref{S-decoupled} we find 
\begin{multline}
Z_{B}^{\mathfrak{su}(2)}(\tau, z_B)  = \int \mathcal D g_{su} \, e^{-(\kappa-2)S_{su}^B(g_{su}, \, a[z_B], \,a[z_B]^\dagger) } \\
= e^{\frac{\pi (\kappa-2)}{\tau_2}|z_B|^2} \int \mathcal D g_{su} \, \exp \left(-(\kappa-2)S^\WZW(a[z_B]^{-\varepsilon} \, g_{su} \,  a[z_B]^{\dagger \varepsilon}) \right) \ , 
\label{ZZ-relation-2}
\end{multline}
Equating eqs.~\eqref{ZZ-relation-1} and \eqref{ZZ-relation-2} we obtain 
\begin{equation}
\int \mathcal D g_{su} \, e^{-(\kappa-2)S^\WZW(a[z_B]^{-\varepsilon}\, g_{su} \, a[z_B]^{\dagger\varepsilon})} = e^{-\frac{\pi (k-2) (z_B)_2^2}{\tau_2}} \mathcal Z_{B}^{\mathfrak{su}(2)}(\tau, z_B) \ , 
\label{S(hgh)-su2}
\end{equation}
which will be used in the main text. 

\paragraph{Asymmetric gauging} We are interested in generalizing \eqref{S(hgh)-su2} to the asymmetric embeddings
\begin{equation}
\varepsilon_L^{su} = - \ell \frac{\sigma_3}{2} \ , \qquad  \varepsilon_R^{su} = - r \frac{\sigma_3}{2} \ , 
\label{su(2)-embeddings}
\end{equation}
with $\ell, r \in \mathds R$. This time, to relate the path integral and the trace partition function we cannot rely on modular invariance: this is supposed to hold only after assembling together all the factors involved in the gauging. However, we can proceed as follows. Let us assume that the correct generalization of \eqref{ZZ-relation-1} to asymmetric embeddings \eqref{su(2)-embeddings} reads 
\begin{equation}
 Z_{B, \ell, r}^{\mathfrak{su}(2)}(\tau, u) = e^{f(\kappa, \ell, r, \tau, u, \bar u)} \, \mathcal Z_{B}^{\mathfrak{su}(2)}(\tau, \bar \tau, \ell \,  u, r \, \bar u) \ . 
\label{ZZ-relation-3}
\end{equation}
where $f(\kappa, \ell, r, \tau, u, \bar u)$ is some unknown function of $\ell, r,\tau, u, \bar u$ and $\kappa$. In eq.~\eqref{ZZ-relation-3}, the left-hand-side is defined as
\begin{equation}
     Z_{B, \ell, r}^{\mathfrak{su}(2)}(\tau, u) = \int \mathcal D g_{su} \, \exp \left(-(\kappa-2)S_{su}^B(g_{su},a[u],a[u]^\dagger) \right) \ , 
\end{equation}
with the embeddings \eqref{su(2)-embeddings}, while by $\mathcal Z_{B}^{\mathfrak{su}(2)}(\tau, \bar \tau, \ell \,  u, r \, \bar u)$ we denote the partition function \eqref{SU2-pf} evaluated at $z_B= \ell \, u$ and $\bar z_B = r \, \bar u$. Some useful information about the function $f(\kappa, \ell, r, u, \bar u)$ comes from requiring that at $\kappa = 4$ and $\kappa = 3$ eq.~\eqref{ZZ-relation-3} respectively reproduces the free fermion and free boson asymmetric gauging results of Section~\ref{sec:supertube}. For the free fermions, setting $Y=0$ and $z=0$ in eq.~\eqref{St-pf-SU2F}, we find
\begin{equation}
f(4, \ell, r, \tau, u, \bar u)  = \pi i (\ell^2 - r^2) s_1 s_2 + \frac{\pi i u_2^2}{\tau_2^2}(\ell^2 \tau - r^2 \bar \tau) + \frac{\pi(\ell^2 + r^2)|u|^2}{\tau_2}  \ ,
\label{f(4)}
\end{equation}
Similarly, by setting $Y=0$ in eq.~\eqref{St-pf-yB} we obtain
\begin{equation}
f(3, \ell, r, \tau, u, \bar u)  = \frac{\pi i}{2} (\ell^2 - r^2) s_1 s_2 + \frac{\pi i u_2^2}{2 \tau_2^2}(\ell^2 \tau - r^2 \bar \tau) +\frac{\pi(\ell^2 + r^2)|u|^2}{2 \tau_2} = \frac{1}{2} f(4, \ell, r, \tau, u, \bar u) \ . 
\label{f(3)}
\end{equation}
In light of equations \eqref{f(4)} and \eqref{f(3)}, we find it reasonable to assume that $f(\kappa, \ell, r, u, \bar u)$ is linear in $\kappa$ and hence we deduce
\begin{equation}
Z_{B, \ell, r}^{\mathfrak{su}(2)}(\tau, u) = e^{\pi i \frac{\kappa-2}{2} (\ell^2 - r^2) s_1 s_2} \, e^{ \frac{(\kappa-2)\pi \, i \, u_2^2}{2\tau_2^2}(\ell^2 \tau - r^2 \bar \tau)} \, e^{\frac{\pi(\kappa-2)(\ell^2 + r^2)}{2 \tau_2} |u|^2} \, \mathcal Z_{B}^{\mathfrak{su}(2)}(\tau, \bar \tau, \ell \,  u, r \, \bar u) \ . 
\end{equation}
Using eq.~\eqref{S-decoupled} with
\begin{equation}
(\kappa-2) S^{X,Y}_{su} = -\frac{\pi(\kappa-2)}{2 \tau_2}(\ell^2+r^2) |u|^2 \ , 
\end{equation}
we finally obtain 
\begin{equation}
\int \mathcal D g_{su} \,e^{-S^\WZW(a[u]^{-\varepsilon_L^{su}} \, g_{su} \, a[u]^{\dagger \varepsilon_R^{su}}) }  = e^{\pi i \frac{\kappa-2}{2} (\ell^2 - r^2) s_1 s_2} \, e^{ \frac{(\kappa-2)\pi \, i \, u_2^2}{2\tau_2^2}(\ell^2 \tau - r^2 \bar \tau)} \,  \mathcal Z_{B}^{\mathfrak{su}(2)}(\tau, \bar \tau, \ell \,  u, r \, \bar u) \ . 
\label{SU2-LF-Shgh}
\end{equation}

\section{GWZW in superspace}
\label{app:GWZW in superspace}

Following \cite{Figueroa-OFarrill:1996xek}, in this appendix we expand in components the $\mathcal N =1$ GWZW action and extend their analysis to asymmetric and null gauging.

\subsection{Symmetric gauging}

The $\mathcal N=1$ vector GWZW action for a simple group $G$ is defined by promoting the fields in \eqref{vector-action} to superfields \cite{Abdalla:1984ef, DiVecchia:1984nyg, Figueroa-OFarrill:1996xek},
\begin{equation}
    I[\mathbb G, \mathbb A, \bar{\mathbb A}] = I[\mathbb G] + \frac{1}{2 \pi} \int \text d \theta \text d \bar \theta \int \text d^2 v \, \text{Tr} \left( - \mathbb J \bar{\mathbb A} - \mathbb A \bar{\mathbb J} -  \mathbb A \bar{\mathbb A}  + \mathbb A \mathbb G^{-1} \bar{\mathbb A} \mathbb G \right) \ , 
    \label{symmetric-ss-action}
\end{equation}
where $I[\mathbb G]$ is the $\mathcal N=1$ WZW action, also defined by promoting the fields in \eqref{WZW-action} to superfields. Once decomposed into components it reads \cite{Figueroa-OFarrill:1996xek}
\begin{equation}
    I[\mathbb G] = S^\WZW(g) + \int \frac{\text d^2v}{2\pi} \psi \bar \nabla \psi + \int\frac{\text d^2v}{2\pi} \bar \psi \nabla \bar \psi \ , 
\end{equation}
where covariant derivatives are defined by 
\begin{equation}
\nabla = \partial_{v}- [\partial_{v}g g^{-1}, \cdot \ ] \ , \qquad \bar \nabla = \partial_{\bar v}+ [g^{-1} \partial_{\bar v}g , \cdot \ ] \ , 
\label{nabladef}
\end{equation}
and here and in the following $[ \ \, , \ \, ]$ denotes the generalised commutator, \ie~$[\mathcal O_1,\mathcal O_2] = \mathcal O_1 \mathcal O_2 - \mathcal O_2 \mathcal O_1$ when $\mathcal O_1$ or $\mathcal O_2$ are Grassmann even and $[\mathcal O_1,\mathcal O_2] =  \mathcal O_1 \mathcal O_2 + \mathcal O_2 \mathcal O_1$ when both $ \mathcal O_1$ and $\mathcal O_2 $ are Grassmann odd. 
We parametrize the superfield $\mathbb G$ as 
\begin{align}
    \mathbb G &= g + \theta g \psi + \bar \theta \bar \psi g + \theta \bar \theta (g a - g \psi g^{-1} \bar \psi g) \ , \\
    \mathbb G^{-1} &= g^{-1} - \theta \psi g^{-1} - \bar \theta g^{-1} \bar \psi - \theta \bar \theta (a g^{-1} - g^{-1} \bar \psi g \psi g^{-1}) \ . \label{Gm1}
\end{align}
The supercurrents read 
\begin{align}    
    \mathbb J &= - D \mathbb G \mathbb G^{-1}  = - g \psi g^{-1} - \theta(\partial_{v}g g^{-1} + g \psi^2 g^{-1}) - \bar \theta g a g^{-1} - \theta \bar \theta (\nabla \bar \psi + g [a, \psi] g^{-1}) \ , \\
    \bar{\mathbb J} &= \mathbb G^{-1} \bar D \mathbb G= g^{-1} \bar \psi g - \theta a + \bar \theta (g^{-1} \partial_{\bar v}g - g^{-1} \bar \psi^2 g) - \theta \bar \theta(\bar \nabla \psi + [a, g^{-1}\bar \psi g]) \ , 
\end{align}
where we introduced the usual superderivatives
\begin{equation}
    D = \frac{\partial}{\partial \theta} + \theta \partial_{v}\ , \qquad  \bar D = \frac{\partial}{\partial \bar \theta} + \bar \theta \partial_{\bar v} \ . 
\end{equation}
Let us parametrize the super gauge field as 
\begin{subequations}
\begin{align}
 \mathbb A &= \rho + \theta A + \bar \theta \bar B + \theta \bar \theta \lambda \ , \\
   \bar{\mathbb A} &= \bar \rho + \theta B + \bar \theta \bar A + \theta \bar \theta \bar \lambda \ .     
\end{align}
\label{mathbb-A-bar-A}%
\end{subequations}
Expanding in components and integrating out auxiliary fields, the action \eqref{symmetric-ss-action} can be rewritten as~\cite{Figueroa-OFarrill:1996xek}
\begin{equation}
    I[\mathbb G, \mathbb A, \bar{\mathbb A}] = S^V(g, \mathcal A, \bar{\mathcal A}) + \int \frac{\text d^2v}{2\pi} \left(  \Psi|_{\mathfrak h^\perp} \, \partial_{\bar v} \Psi|_{\mathfrak h^\perp} - \Psi|_{\mathfrak h^\perp} \, [\mathcal{\bar A},  \Psi|_{\mathfrak h^\perp}]
    + \bar \Psi|_{\mathfrak h^\perp} \, \partial_{v}\bar \Psi|_{\mathfrak h^\perp} - \bar \Psi|_{\mathfrak h^\perp} \, [\mathcal A, \bar \Psi|_{\mathfrak h^\perp}] \right) \ , 
\label{compos-ss-wzw-action-9}%
\end{equation}
where we defined
\begin{equation}
    \Psi|_{\mathfrak h^\perp} = g(\psi + \rho)g^{-1}|_{\mathfrak h^\perp} \ , \qquad \bar \Psi|_{\mathfrak h^\perp} = g^{-1}(\bar \psi - \bar \rho)g|_{\mathfrak h^\perp} \ , 
    \label{coupled-fermions}
\end{equation}
and 
\begin{equation}
    \mathcal A = A - \rho^2 \ , \qquad    \bar{\mathcal A} = \bar A - \bar \rho^2 \ ,
    \label{mathcalAdef}
\end{equation}
while $I_B[g, \mathcal A, \bar{ \mathcal A}]$ denotes the bosonic vector gauged WZW action
\begin{equation}
    S^V(g, \mathcal A, \bar{ \mathcal A}) = S^\WZW(g) +\frac{1}{\pi}\int \text d^2v \, \text{Tr}( \partial_{v}g g^{-1}\bar{ \mathcal A}  - \mathcal A g^{-1} \partial_{\bar v}g  - \mathcal A \bar{ \mathcal A} + \mathcal A g^{-1}\bar{ \mathcal A} g) \ .  
    \label{IBgAA}
\end{equation}
In eqs.~\eqref{compos-ss-wzw-action-9} and \eqref{coupled-fermions}, $\mathfrak h^\perp$ denotes the orthogonal complement of $\mathfrak h$, 
\begin{equation}
    \mathfrak h = \mathfrak h \oplus \mathfrak h^\perp \ ,  
\end{equation}
and \eg~the notation $\Psi|_{\mathfrak h^\perp}$ denotes the projection of $\Psi$ onto $\mathfrak h^\perp$.

\subsection{Asymmetric gauging}

Let us now repeat the analysis of the previous section for asymmetric gauging. The asymmetric $\mathcal N=1$ GWZW action reads
\begin{equation}
    I[\mathbb G, \mathbb A, \bar{\mathbb A}] = I[\mathbb G] - \int \text d \theta \text d \bar \theta \int \frac{\text d^2v}{\pi} \, \text{Tr} \Bigl( \mathbb J \bar{\mathbb A}_L + \mathbb A_R \bar{\mathbb J} + \mfrac{1}{2}\mathbb A_L \bar{\mathbb A}_L + \mfrac{1}{2} \mathbb A_R \bar{\mathbb A}_R  - \mathbb A_R \mathbb G^{-1} \bar{\mathbb A}_L \mathbb G \Bigr) \ , 
    \label{symmetric-ss-action-LR}
\end{equation}
where in order to lighten the notation we omitted the sum over components entering \eg~eq.~\eqref{asym-act-A}. 
The short-hand notation $\bar{\mathbb A}_{L,R}$ and $\mathbb A_{L,R}$ stands for 
\begin{equation}
    \bar{\mathbb A}_{L,R} \equiv \epsilon_{L,R}( \bar{\mathbb A}) \ , \qquad     \mathbb A_{L,R} \equiv \epsilon_{L,R}(\mathbb A) \ , 
\end{equation}
and we are still parametrizing the super gauge field as in eq.~\eqref{mathbb-A-bar-A}. Similarly, we write 
\begin{equation}
A_{L,R} \equiv \epsilon_{L,R}(A) \ , \qquad  B_{L,R} \equiv \epsilon_{L,R}(B) \ , \qquad \rho_{L,R} \equiv \epsilon_{L,R}(\rho) \ , \qquad \bar \rho_{L,R} \equiv \epsilon_{L,R}(\bar \rho) \ ,  
\end{equation}
and so on. Expanding in components the various terms entering the action \eqref{symmetric-ss-action-LR} we find 
\begin{align}
    I[\mathbb G, & \mathbb A, \bar{\mathbb A}] = I[\mathbb G] + \int \frac{\text d^2v}{\pi} \, \text{Tr} \Bigl( \partial_{v}g g^{-1}\bar A_L  - A_R g^{-1} \partial_{\bar v}g  - \mfrac{1}{2} A_L \bar A_L - \mfrac{1}{2} A_R \bar A_R + A_R g^{-1} \bar A_L g \nonumber \\
    & + A_R g^{-1} \bar \psi^2 g - A_R g^{-1}\bar \rho_L \bar \psi g - A_R g^{-1} \bar \psi \bar \rho_L g + g \psi^2 g^{-1} \bar A_L - \rho_R g^{-1} \bar A_L g \psi + \rho_R \psi g^{-1} \bar A_L g \nonumber  \\
    & - \lambda_R g^{-1} \bar \psi g - \mfrac{1}{2} \lambda_L \bar \rho_L - \mfrac{1}{2} \lambda_R \bar \rho_R +\lambda_R g^{-1}\bar \rho_L g  -\mfrac{1}{2} \rho_L \bar \lambda_L - \mfrac{1}{2} \rho_R \bar \lambda_R  + g \psi g^{-1} \bar \lambda_L + \rho_R g^{-1}\bar \lambda_L g  \nonumber \\
    & + \mfrac{1}{2} \bar B_L B_L + \mfrac{1}{2} \bar B_R B_R + \rho_R g^{-1} B_L \bar \psi g - \rho_R g^{-1} \bar \psi B_L g + \bar B_R g^{-1}\bar \rho_L g \psi - \bar B_R g^{-1}B_L g + \bar B_R \psi g^{-1} \bar \rho_L g \nonumber \\
    & + \rho_R \bar \nabla \psi  + \nabla \bar \psi \bar \rho_L - \rho_R g^{-1}\bar \rho_L g \psi g^{-1} \bar \psi g  - \rho_R \psi g^{-1} \bar \rho_L \bar \psi g + \rho_R g^{-1} \bar \psi \bar \rho_L g \psi + \rho_R g^{-1} \bar \psi g \psi g^{-1} \bar \rho_L g  \nonumber    \\
     & - \bar B_R a + \rho_R [a, g^{-1} \bar \psi g] - g a g^{-1} B_L + g [a, \psi] g^{-1} \bar \rho_L + \rho_R g^{-1}\bar \rho_L g a - \rho_R a g^{-1} \bar \rho_L g \Bigr)   \ . 
     \label{compos-ss-wzw-action-LR}%
\end{align}
The bosonic asymmetrically gauged WZW action reads, see eq.~\eqref{asym-act-A},
\begin{equation}
    S^B(g, A, \bar A) = S^\WZW(g) +\int \frac{\text d^2v}{\pi} \, \text{Tr} ( \partial_{v}g g^{-1}\bar A_L  - A_R g^{-1} \partial_{\bar v}g  - \mfrac{1}{2} A_L \bar A_L - \mfrac{1}{2} A_R \bar A_R + A_R g^{-1}\bar A_L g) \ , 
    \label{IBgAA-LR}
\end{equation}
so that the terms in the first line of \eqref{compos-ss-wzw-action-LR} can be rewritten as 
\begin{equation}
    S^B(g, A, \bar A) + \int \frac{\text d^2v}{2\pi} \psi \bar \nabla \psi + \int\frac{\text d^2v}{2\pi} \bar \psi \nabla \bar \psi \ . 
\end{equation}
The terms in the second line of \eqref{compos-ss-wzw-action-LR} can be reorganized as
\begin{equation}
     -\frac{1}{2} \bar \psi \, [g A_R g^{-1}, \bar \psi] - [g A_R g^{-1}, \bar \psi] \bar \rho_L -\frac{1}{2} \psi \, [g^{-1} \bar A_L g, \psi] - \rho_R [g^{-1} \bar A_L g, \psi] \ .  \label{psi-constr-LR}
\end{equation}
Integrating out the auxiliary field $a$ we find the constraint 
\begin{equation}
\bar B_R + g^{-1} \bar \psi g \rho_R + \rho_R  g^{-1} \bar \psi g  + g^{-1} B_L g - \psi g^{-1} \bar \rho_L g - g^{-1} \bar \rho_L g \psi - \rho_R g^{-1}\bar \rho_L g - g^{-1} \bar \rho_L g \rho_R = 0 \ . 
\label{aeom-LR}
\end{equation}
With all these simplifications eq.~\eqref{compos-ss-wzw-action-LR} becomes 
\begin{align}
    I[\mathbb G, \mathbb A, \bar{\mathbb A}] = &\,  S^B(g, A, \bar A) + \int \frac{\text d^2v}{\pi} \, \text{Tr} \Bigl( \, \tfrac{1}{2}\psi \bar \nabla \psi - \tfrac{1}{2} \psi \, [g^{-1} \bar A_L g, \psi] + \tfrac{1}{2}\bar \psi \nabla \bar \psi - \tfrac{1}{2} \bar \psi \, [g A_R g^{-1}, \bar \psi] \nonumber \\
    & + \rho_R \bar \nabla \psi - \rho_R [g^{-1} \bar A_L g, \psi] + \nabla \bar \psi \bar \rho_L - [g A_R g^{-1}, \bar \psi] \bar \rho_L \nonumber \\
    & - \lambda_R \bigl( \bar \rho_R + g^{-1} (\bar \psi - \bar \rho_L)g \bigr) - \bigl(\rho_L - g (\psi + \rho_R)g^{-1} \bigr) \bar \lambda_L \nonumber \\
    & +\mfrac{1}{2} \bar B_L B_L + \mfrac{1}{2} \bar B_R B_R + \rho_R g^{-1} B_L \bar \psi g - \rho_R g^{-1} \bar \psi B_L g \nonumber\nonumber\nonumber\nonumber \\
    & + \bar B_R g^{-1}\bar \rho_L g \psi - \bar B_R g^{-1}B_L g + \bar B_R \psi g^{-1} \bar \rho_L g \nonumber \\
    & -\rho_R g^{-1}\bar \rho_L g \psi g^{-1} \bar \psi g  - \rho_R \psi g^{-1} \bar \rho_L \bar \psi g + \rho_R g^{-1} \bar \psi \bar \rho_L g \psi + \rho_R g^{-1} \bar \psi g \psi g^{-1} \bar \rho_L g \Bigr) \ ,  \label{compos-ss-wzw-action-2-LR} 
\end{align}
where we used that, by the anomaly cancellation condition \eqref{no-anomal}, 
\begin{equation}
    \text{Tr} \left( \lambda_L \bar \rho_L \right) = \text{Tr} \left(  \lambda_R \bar \rho_R \right) \ ,  \qquad \text{Tr} \left( \rho_L \bar \lambda_L \right) = \text{Tr} \left(  \rho_R \bar \lambda_R \right) \ . 
\end{equation}
Adding to the action \eqref{compos-ss-wzw-action-2-LR} the following terms  
\begin{subequations}
\begin{align}
    0 & = \tfrac{1}{2} \psi [g^{-1} \bar \rho^2_L g, \psi] + \psi^2 g^{-1} \bar \rho^2_L g  \ , \\
    0 & = \tfrac{1}{2} \bar \psi [g \rho^2_R g^{-1}, \bar \psi] + \bar \psi^2 g \rho^2_R g^{-1}  \ , \\
    0 & = \rho_R [g^{-1} \bar \rho^2_L g, \psi] - \rho_R g^{-1} \bar \rho^2_L g \psi + \rho_R \psi g^{-1} \bar \rho^2_L g  \ , \\
    0 & =  [g \rho^2_R g^{-1}, \bar \psi]\bar \rho_L -  g \rho^2_R g^{-1} \bar \psi \bar \rho_L + \bar \psi g \rho^2_R g^{-1} \bar \rho_L  \ ,  
\end{align}
\end{subequations}
introducing the covariant derivatives
\begin{equation}
    \nabla_{\mathcal A_R} = \nabla - [g \mathcal A_R g^{-1} , \ ] \ , \qquad  \bar \nabla_{\bar{\mathcal A}_L} = \bar \nabla - [g^{-1} \bar{\mathcal{A}}_L g , \ ]  \ , 
    \label{bar-nabla-bar-A-LR}
\end{equation}
and the gauge fields
\begin{equation}
    \mathcal A_{L,R} = A_{L,R} - \rho^2_{L,R} \ , \qquad    \bar{\mathcal A}_{L,R} = \bar A_{L,R} - \bar \rho^2_{L,R} \ ,
    \label{mathcalAdef-LR}
\end{equation}
we can rewrite the action \eqref{compos-ss-wzw-action-2-LR} as 
\begin{align}
    I[\mathbb G, \mathbb A, \bar{\mathbb A}] & = S^B(g,A, \bar A) + \int \frac{\text d^2v}{\pi} \, \text{Tr} \Bigl( \, \tfrac{1}{2} \psi \bar \nabla_{\bar{\mathcal A}_L} \psi + \tfrac{1}{2} \bar \psi \nabla_{\mathcal{A}_R} \bar \psi + \rho_R \bar \nabla_{\bar{\mathcal A}_L} \psi + \nabla_{\mathcal{A}_R} \bar \psi \bar \rho_L \nonumber \\
     & -\lambda_R \bigl(\bar \rho_R + g^{-1} (\bar \psi - \bar \rho_L)g \bigr) - \bigl(\rho_L - g (\psi + \rho_R)g^{-1} \bigr) \bar \lambda_L \nonumber \\
    &  - \rho_R g^{-1}\bar \rho_L g \psi g^{-1} \bar \psi g  - \rho_R \psi g^{-1} \bar \rho_L \bar \psi g + \rho_R g^{-1} \bar \psi \bar \rho_L g \psi + \rho_R g^{-1} \bar \psi g \psi g^{-1} \bar \rho_L g \nonumber \\
    &  + \psi^2 g^{-1} \bar \rho^2_L g + \bar \psi^2 g \rho^2_R g^{-1} - \rho_R g^{-1} \bar \rho^2_L g \psi + \rho_R \psi g^{-1} \bar \rho^2_L g - g \rho^2_R g^{-1} \bar \psi \bar \rho_L + \bar \psi g \rho^2_R g^{-1} \bar \rho_L  \nonumber  \\
     &   +  \tfrac{1}{2} \bar B_L B_L + \tfrac{1}{2} \bar B_R B_R + \rho_R g^{-1} B_L \bar \psi g - \rho_R g^{-1} \bar \psi B_L g \nonumber \\
    &   + \bar B_R g^{-1}\bar \rho_L g \psi - \bar B_R g^{-1}B_L g + \bar B_R \psi g^{-1} \bar \rho_L g \Bigr) \ . 
\label{compos-ss-wzw-action-4-LR}
\end{align}
Using eq.~\eqref{aeom-LR} we find
\begin{align}
\bar B_R g^{-1} B_L g & = - \bar \psi g \rho_R g^{-1} B_L - \rho_R g^{-1} \bar \psi B_L g - B^2_L + \psi g^{-1} \bar \rho_L B_L g + \bar \rho_L g \psi g^{-1} B_L \nonumber \\
& \quad + \rho_R g^{-1} \bar \rho_L B_L g + \bar \rho_L g \rho_R g^{-1} B_L \\
& = -\bar B^2_R - \bar B_R g^{-1} \bar \psi g \rho_R - \bar B_R \rho_R g^{-1} \bar \psi g + \bar B_R \psi g^{-1} \bar \rho_L g + \bar B_R g^{-1} \bar \rho_L g \psi \nonumber \\
& \quad + \bar B_R \rho_R g^{-1} \bar \rho_L g + \bar B_R g^{-1} \bar \rho_L g \rho_R  \ . 
\end{align}
We can then rewrite the last two lines of \eqref{compos-ss-wzw-action-4-LR} as
\begin{align}
    & \tfrac{1}{2} B^2_L + \tfrac{1}{2} \bar B^2_R + \tfrac{1}{2} B_L \bar B_L + \tfrac{1}{2} B_R \bar B_R + \bar B_R [g^{-1}(\bar \psi - \bar \rho_L)g, \rho_R] - 2 B_L [g(\psi+\rho_R)g^{-1}, \bar \rho_L]  \nonumber \\ 
    & + \bar \rho_L g \psi^2 g^{-1} \bar \rho_L - g^{-1} \bar \rho_L g \psi g^{-1} \bar \psi g \rho_R - \bar \rho_L g \psi \rho_R g^{-1} \bar \psi + \bar \rho_L g \psi \rho_R g^{-1} \bar \rho_L \nonumber \\
    & - \psi g^{-1} \bar \rho_L \bar \psi g \rho_R + \psi g^{-1} \bar \rho^2_L g \rho_R - \psi g^{-1} \bar \rho_L g \rho_R g^{-1} \bar \psi g \nonumber \\
    & + \bar \psi g \rho^2_R g^{-1} \bar \psi - \bar \psi g \rho^2_R g^{-1} \bar \rho_L - \bar \rho_L g \rho^2_R g^{-1} \bar \psi + \bar \rho_L g \rho^2_R g^{-1} \bar \rho_L     \ . 
\end{align}
Many terms cancel against opposite terms in the third and fourth line of \eqref{compos-ss-wzw-action-4-LR} and the supersymmetric action reduces to
\begin{align}
    & I[ \mathbb G, \mathbb A, \bar{\mathbb A}] \nonumber \\
    & = \, S^B (g,A,\bar A) + \int \frac{\text d^2v}{\pi} \, \text{Tr} \Bigl( \, \tfrac{1}{2} \psi \bar \nabla_{\bar{\mathcal A}_L} \psi + \tfrac{1}{2} \bar \psi \nabla_{\mathcal{A}_R} \bar \psi + \rho_R \bar \nabla_{\bar{\mathcal A}_L} \psi + \nabla_{\mathcal A_R} \bar \psi \bar \rho_L - \rho^2_R g^{-1} \bar \rho^2_L g \nonumber \\
    & \qquad - (  \rho_L -  g (\psi + \rho_R) g^{-1} )\bar \lambda_L - \lambda_R (\bar \rho_R + g^{-1} (\bar \psi - \bar \rho_L) g)  \nonumber \\
    & \qquad + \tfrac{1}{2} B_L^2 + \tfrac{1}{2} \bar B_R^2 + \tfrac{1}{2} B_L \bar B_L + \tfrac{1}{2} B_R \bar B_R + \bar B_R [g^{-1}(\bar \psi - \bar \rho_L) g, \rho_R] - B_L[g(\psi + \rho_R )g^{-1},\bar \rho_L] \Bigr)    \ . \label{compos-ss-wzw-action-5-LR}
\end{align}
The terms in the second line of \eqref{compos-ss-wzw-action-5-LR} can be rewritten as  
\begin{subequations}
\begin{align}
   &   S^B (g,A,\bar A) + \int \frac{\text d^2v}{\pi} \, \text{Tr} \Bigl( -\tfrac{1}{2} \rho_R \bar \nabla_{\bar{\mathcal A}_L} \rho_R -\tfrac{1}{2} \bar \rho_L \nabla_{\mathcal{A}_R} \bar \rho_L - \rho^2_R g^{-1} \bar \rho^2_L g \nonumber \\
   &  \hspace{120pt} + \tfrac{1}{2} (\psi + \rho_R) \bar \nabla_{\bar{\mathcal A}_L}(\psi + \rho_R) + \tfrac{1}{2} (\bar \psi - \bar \rho_L) \nabla_{\mathcal{A}_R}(\bar \psi - \bar \rho_L) \Bigr) \\
   = \, &  S^B(g, \mathcal A, \bar{\mathcal A}) +  \int \frac{\text d^2v}{2\pi} \, \text{Tr} \Bigl( -\rho_L \bar \partial_{\bar{\mathcal A}_L} \rho_L - \bar \rho_R \partial_{\mathcal{A}_R} \bar \rho_R - \rho_L^2 \bar \rho^2_L - \rho_R^2 \bar \rho^2_R  \nonumber \\
   &  \hspace{60pt} +g(\psi + \rho_R)g^{-1} \bar \partial_{\bar{\mathcal A}_L} \left( g(\psi + \rho_R)g^{-1} \right) + g^{-1}(\bar \psi - \bar \rho_L)g \partial_{{\mathcal A}_R} \left( g^{-1}(\bar \psi - \bar \rho_L)g \right)  \Bigr) \ , 
\end{align}
\end{subequations}
where we expressed $S^B (g,A,\bar A)$ in terms of $S^B(g, \mathcal A, \bar{\mathcal A})$, see eq.~\eqref{mathcalAdef-LR}, and $\bar \nabla_{\bar{\mathcal A}_L}$, $\nabla_{{\mathcal A}_R}$ in terms of $\bar \partial_{\bar{\mathcal A}_L}$, $\partial_{{\mathcal A}_R}$, 
\begin{equation}
    \partial_{\mathcal A} = \partial_{v}- [\mathcal A, \cdot \ ] \ ,   \qquad  \bar \partial_{\bar{\mathcal A}} = \partial_{\bar v}- [\bar{ \mathcal A}, \cdot \ ] \ . 
    \label{bar-partial-bar-A}  
\end{equation}
We also used that by eq.~\eqref{no-anomal}
\begin{equation}
    \text{Tr} \left( \rho_R \partial_{\bar v}\rho_R \right) = \text{Tr} \left( \rho_L \partial_{\bar v}\rho_L \right) \ , 
\end{equation}
and 
\begin{equation}
      \text{Tr} \left( \rho_L^2 \bar A_L \right) =   \text{Tr} \left( \rho^2_R \bar A_R \right) \ , \qquad   \text{Tr} \left( A_R \bar \rho^2_R \right) =   \text{Tr} \left( A_L \bar \rho^2_L \right) \ , 
\end{equation}
In fact, notice that $\epsilon_L$, $\epsilon_R$ are Lie algebra homomorphisms, \ie~linear maps preserving the Lie braket and hence
\begin{equation}
    \epsilon_L(\rho^2) = \frac{1}{2} \epsilon_L([\rho, \rho]) = \frac{1}{2} [\epsilon_L(\rho), \epsilon_L(\rho)] = \epsilon_L(\rho)^2 = \rho_L^2 \ . 
\end{equation}
The action \eqref{compos-ss-wzw-action-5-LR} thus becomes
\begin{align}
    I[\mathbb G, \mathbb A, \bar{\mathbb A}] = \, & S^B(g, \mathcal A, \bar{\mathcal A}) +\int \frac{\text d^2v}{2\pi} \Bigl( - \rho_L \bar \partial_{\bar{\mathcal A}_L} \rho_L - \bar \rho_R \partial_{\mathcal{A}_R} \bar \rho_R - \rho_L^2 \bar \rho^2_L - \rho_R^2 \bar \rho^2_R  \nonumber \\ 
    & + g(\psi + \rho_R)g^{-1} \bar \partial_{\bar{\mathcal A}_L} \left( g(\psi + \rho_R)g^{-1} \right) + g^{-1}(\bar \psi - \bar \rho_L)g \partial_{{\mathcal A}_R} \left( g^{-1}(\bar \psi - \bar \rho_L)g \right) \nonumber \\
    & -2 (  \rho_L -  g (\psi + \rho_R) g^{-1} )\bar \lambda_L - 2 \lambda_R (\bar \rho_R + g^{-1} (\bar \psi - \bar \rho_L) g) \nonumber  \\
    & +B_L^2 + \bar B_R^2 + B_L \bar B_L + B_R \bar B_R  \nonumber \\
    & +2 \bar B_R [g^{-1}(\bar \psi - \bar \rho_L) g, \rho_R] - 2 B_L[g(\psi + \rho_R )g^{-1},\bar \rho_L] \Bigr) \ . \label{compos-ss-wzw-action-7-LR}%
\end{align}
We now decompose $\mathfrak g$ as 
\begin{equation}
    \mathfrak g = \mathfrak h_L \oplus \mathfrak h_L^\perp = \mathfrak h_R \oplus \mathfrak h_R^\perp \ . 
    \label{g=hL+hLp}
\end{equation}
Notice that this decomposition is not always possible. In fact, when the gauging is null, we have $\mathfrak h_L^\perp \supset \mathfrak h_L$ and eq.~\eqref{g=hL+hLp} cannot possibly hold. A sufficient condition for eq.~\eqref{g=hL+hLp} to hold is requiring that $\mathfrak h_L$ is \emph{anisotropic} \cite[Theorem 4.1]{nacinovich}. A vector subspace $\mathfrak h_L$ of a vector space $\mathfrak g$ is said to be anisotropic with respect to a symmetric bilinear form $\langle \, \cdot \, , \, \cdot \, \rangle $ if for each non-zero vector $v \in \mathfrak h_L$ there exist a vector $w \in \mathfrak{h}_L$ such that $\langle v,w \rangle \neq 0$. In the rest of this subsection, we will assume $\mathfrak h_L$ and $\mathfrak h_R$ to be anisotropic, so that eq.~\eqref{g=hL+hLp} holds. In the next subsection we will then separately consider the case of null gauging.

Integrating out $\lambda$ and $\bar \lambda$, we obtain the equations of motion 
\begin{equation}
    \rho_L =  g(\psi + \rho_R) g^{-1}|_{\mathfrak h_L} \ , \qquad \bar \rho_R  = - g^{-1}(\bar \psi - \bar \rho_L)g|_{\mathfrak h_R} \ . 
\label{lambdaeom-LR}
\end{equation}
Making use of eq.~\eqref{lambdaeom-LR}, the terms involving $B$ and $\bar B$ in the last two lines of \eqref{compos-ss-wzw-action-7-LR} can be rewritten as
\begin{align}
    & 2 \bar B_R [g^{-1}(\bar \psi - \bar \rho_L) g|_{\mathfrak h_R} + g^{-1}(\bar \psi - \bar \rho_L) g|_{\mathfrak h_R^\perp}, \rho_R] -2 B_L[g(\psi + \rho_R )g^{-1}|_{\mathfrak h_L} + g(\psi + \rho_R )g^{-1}|_{\mathfrak h_L^\perp},\bar \rho_L] \nonumber \\
    =\, &  2 \bar B_R [g^{-1}(\bar \psi - \bar \rho_L) g|_{\mathfrak h_R}, \rho_R] -2 B_L[g(\psi + \rho_R )g^{-1}|_{\mathfrak h_L},\bar \rho_L] \\
    = \, & -2 \bar B_R [\bar \rho_R, \rho_R] -2 B_L[\rho_L,\bar \rho_L] \ . 
\end{align}
In the second equality we used that 
\begin{equation}
    \text{Tr} \left( \bar B_R [g^{-1}(\bar \psi + \bar \rho_L) g|_{\mathfrak h_R^\perp}, \rho_R] \right)  = 0
\end{equation}
and a similar condition for the term involving $B_L$. Let us explain how this comes about. Using the properties of the trace, 
\begin{align}
     \text{Tr} \left(  \bar B_R [g^{-1}(\bar \psi - \bar \rho_L) g|_{\mathfrak h_R^\perp}, \rho_R] \right) & =  \text{Tr} \left(  - \rho_R \, \bar B_R \, g^{-1}(\bar \psi - \bar \rho_L) g|_{\mathfrak h_R^\perp} + \bar B_R \, \rho_R \, g^{-1}(\bar \psi - \bar \rho_L) g|_{\mathfrak h_R^\perp} \right) \\
    & =  \text{Tr} \left( [\bar B_R, \rho_R] g^{-1}(\bar \psi - \bar \rho_L) g|_{\mathfrak h_R^\perp} \right) = 0 \ . 
\end{align}
In fact, $[\bar B_R, \rho_R] \in \mathfrak h_R$ since $\epsilon_R$ is a Lie algebra homomorphism. The action \eqref{compos-ss-wzw-action-7-LR} thus becomes
\begin{align}
    I[\mathbb G, \mathbb A, \bar{\mathbb A}] = & \, S^B(g, \mathcal A, \bar{\mathcal A}) + \int \frac{\text d^2v}{2 \pi} \, \text{Tr} \Bigl(- \rho_L \bar \partial_{\bar{\mathcal A}_L} \rho_L - \bar \rho_R \partial_{\mathcal{A}_R} \bar \rho_R - \rho_L^2 \bar \rho^2_L - \rho_R^2 \bar \rho^2_R   \nonumber \\ 
    & + g(\psi + \rho_R)g^{-1} \bar \partial_{\bar{\mathcal A}_L} \left( g(\psi + \rho_R)g^{-1} \right) + g^{-1}(\bar \psi - \bar \rho_L)g \partial_{{\mathcal A}_R} \left( g^{-1}(\bar \psi - \bar \rho_L)g \right) \nonumber \\
     & +B_L^2 + \bar B_R^2 + B_L \bar B_L + B_R \bar B_R -2 \bar B_R [\bar \rho_R, \rho_R] -2 B_L[\rho_L,\bar \rho_L] \Bigr) \ . \label{compos-ss-wzw-action-6-LR}
\end{align}
Finally, using again eq.~\eqref{no-anomal} and invariance of the Lie braket under $\epsilon_{L,R}$ we can rewrite the last line in \eqref{compos-ss-wzw-action-6-LR} as 
\begin{equation}
 (B_L + \bar B_L)(B_L + \bar B_L) - 2(B_L + \bar B_L)[\rho_L, \bar \rho_L] \ . 
\end{equation}
Integrating out $B_L + \bar B_L$ we find 
\begin{equation}
    B_L + \bar B_L = [\rho_L, \bar \rho_L] \ , 
\end{equation}
and replacing it back into the action, we have 
\begin{align}
    I[\mathbb G, \mathbb A, \bar{\mathbb A}] = & \, S^B(g, \mathcal A, \bar{\mathcal A}) +\int \frac{\text d^2v}{2\pi} \, \text{Tr} \Bigl( - \rho_L \bar \partial_{\bar{\mathcal A}_L} \rho_L - \bar \rho_R \partial_{\mathcal{A}_R} \bar \rho_R \nonumber \\
    & + g(\psi + \rho_R)g^{-1} \bar \partial_{\bar{\mathcal A}_L} \left( g(\psi + \rho_R)g^{-1} \right) + g^{-1}(\bar \psi - \bar \rho_L)g \partial_{{\mathcal A}_R} \left( g^{-1}(\bar \psi - \bar \rho_L)g \right) \Bigr)  \ ,  \label{compos-ss-wzw-action-9-LR}
    \end{align}
where we used that
\begin{equation}
 \text{Tr}\left(   \rho^2_L \bar \rho^2_L \right) = \text{Tr}\left( \rho^2_R \bar \rho^2_R \right) \ , 
\end{equation}
again following from \eqref{no-anomal} and invariance of the Lie bracket. 

We now wish to show that the last two terms on the first line of~\eqref{compos-ss-wzw-action-9-LR} cancel the part of the terms in the last line that are in $\mathfrak{h}$.
Decomposing $g(\psi + \rho_R)g^{-1}$ and $g^{-1}(\bar \psi - \bar \rho_L)g$ as 
\begin{align}
    g(\psi + \rho_R)g^{-1} & =  \rho_L + g(\psi + \rho_R)g^{-1}|_{\mathfrak h_L^\perp} \ ,  \\
    g^{-1}(\bar \psi - \bar \rho_L)g & = -\rho_R + g^{-1}(\bar \psi - \bar \rho_L)g|_{\mathfrak h_R^\perp} \ , 
\end{align}
the action takes the final form 
\begin{equation}
    I[\mathbb G, \mathbb A, \bar{\mathbb A}] = S^B(g, \mathcal A, \bar{\mathcal A}) +\int \frac{\text d^2v}{2 \pi} \, \text{Tr} \Biggl(  \Psi|_{\mathfrak h_L^\perp} \bar \partial_{\bar{\mathcal A}_L}  \Psi|_{\mathfrak h_L^\perp} + \bar \Psi|_{\mathfrak h_R^\perp} \partial_{{\mathcal A}_R} \bar \Psi|_{\mathfrak h_R^\perp} \Biggr) \ , 
\label{compos-ss-wzw-action-10-LR}%
\end{equation}
where we introduced 
\begin{equation}
    \Psi = g(\psi + \rho_R)g^{-1}|_{\mathfrak h^\perp} \ , \qquad \bar \Psi|_{\mathfrak h^\perp} = g^{-1}(\bar \psi - \bar \rho_L)g|_{\mathfrak h^\perp} \ , 
\end{equation}
and used that \eg
\begin{equation}
   \text{Tr} \left( \Psi|_{\mathfrak h_L} [\bar{\mathcal A}_L, \Psi|_{\mathfrak h_L^\perp}] \right) = \text{Tr} \left( [\Psi|_{\mathfrak h_L}, {\mathcal A}_L] \, \Psi|_{\mathfrak h_L^\perp} \right) = 0 \ .
\end{equation}

\subsection{Null gauging}

We already noticed that when the gauging is null, it is not possible to decompose $\mathfrak g$ as in \eqref{g=hL+hLp}. We should then restart from eq.~\eqref{compos-ss-wzw-action-7-LR} and use that by eq.~\eqref{null-condition-embeddings} we have 
\begin{equation}
\begin{aligned}
    \text{Tr} \left( \rho_L \bar \partial_{\bar{\mathcal A}_L} \rho_L \right) & = \text{Tr} \left(  \rho_L^2 \bar \rho^2_L \right) = \text{Tr}\left( \rho_L \bar \lambda_L \right) = \text{Tr}\left( B_L^2 \right) =  \text{Tr}\left( B_L \bar B_L \right) =  0 \ , \\
    \text{Tr}\left( \bar \rho_R \partial_{\mathcal{A}_R} \bar \rho_R \right) & = \text{Tr}\left( \rho_R^2 \bar \rho^2_R \right) = \text{Tr}\left( \lambda_R \bar \rho_R \right) = \text{Tr}\left( \bar B_R^2 \right) = \text{Tr}\left( B_R \bar B_R \right) = 0 \ . 
\end{aligned}
\end{equation}
Eq.~\eqref{compos-ss-wzw-action-7-LR} reduces to 
\begin{align}
    I[\mathbb G, \mathbb A, \bar{\mathbb A}] = & \, S^B(g, \mathcal A, \bar{\mathcal A}) +\int \frac{\text d^2v}{2 \pi} \, \text{Tr} \Bigl( 2 g (\psi+\rho_R) g^{-1} \bar \lambda_L - 2 \lambda_R \, g^{-1} (\bar \psi - \bar \rho_L) g  \nonumber  \\ 
    & + g(\psi + \rho_R)g^{-1} \bar \partial_{\bar{\mathcal A}_L} \left( g(\psi + \rho_R)g^{-1} \right) + g^{-1}(\bar \psi - \bar \rho_L)g \partial_{{\mathcal A}_R} \left( g^{-1}(\bar \psi - \bar \rho_L)g \right) \nonumber \\
    & +2 \bar B_R [g^{-1}(\bar \psi - \bar \rho_L) g, \rho_R] - 2 B_L[g(\psi + \rho_R )g^{-1},\bar \rho_L]  \Bigr) \ . \label{compos-ss-wzw-action-1-null}
\end{align}
Let us now decompose $\mathfrak g$ as 
\begin{equation}
\label{gdecomp}
    \mathfrak g = \mathfrak h_L^\perp \oplus \mathfrak h_L^{\notin}  = \mathfrak h_R^\perp \oplus \mathfrak h_R^{\notin} \ , 
\end{equation}
where \eg~$\mathfrak h_L^\perp$ is defined as usual,  
\begin{equation}
 \mathfrak h_L^\perp = \{ v \in \mathfrak g \, | \, \langle v,w \rangle = 0 \ \forall \, w \in \mathfrak h_L \} \ , 
\end{equation}
while $\mathfrak h_L^{\notin} $ is constructed by completing a basis of $\mathfrak h_L^\perp$ to a basis of $\mathfrak g$. 
Similarly for $\mathfrak h_R^\perp$ and $\mathfrak h_R^{\notin} $. Notice that, by construction, this decomposition of $\mathfrak g$ is always possible. When the gauging is not null, we have $\mathfrak h_L^{\notin}  = \mathfrak h_L$, while for null gauging we have $\mathfrak h_L \subseteq \mathfrak h_L^\perp$. 

Let us decompose the fermions as 
\begin{align}
    g (\psi+\rho_R) g^{-1} & = g (\psi + \rho_R) g^{-1}|_{h_L^\perp} + g (\psi + \rho_R) g^{-1}|_{h_L^{\notin} } \ , \\
    g^{-1} (\bar \psi - \bar \rho_L) g & = g^{-1} (\bar \psi - \bar \rho_L) g|_{h_R^\perp} + g^{-1} (\bar \psi - \bar \rho_L) g|_{h_R^{\notin} } \ , 
\end{align}
By definition of $\mathfrak h_L^\perp$, 
\begin{equation}
    \text{Tr}\left( g (\psi + \rho_R) g^{-1}|_{h_L^\perp} \, \bar \lambda_L \right) = 0 \qquad \text{and} \qquad \text{Tr}\left( \lambda_R \, g^{-1} (\bar \psi - \bar \rho_L) g|_{h_R^\perp} \right) = 0 \ ,
\end{equation}
so that the null gauged action \eqref{compos-ss-wzw-action-1-null} becomes 
\begin{align}
    I[\mathbb G, \mathbb A, \bar{\mathbb A}] = & \, S^B(g, \mathcal A, \bar{\mathcal A})  + \int \frac{\text d^2v}{\pi} \, \text{Tr} \Bigl( 2 g (\psi + \rho_R) g^{-1}|_{h_L^{\notin} } \, \bar \lambda_L - 2 \lambda_R \, g^{-1} (\bar \psi - \bar \rho_L) g|_{h_R^{\notin} }   \nonumber \\ 
    & + g(\psi + \rho_R)g^{-1} \bar \partial_{\bar{\mathcal A}_L} \left( g(\psi + \rho_R)g^{-1} \right) + g^{-1}(\bar \psi - \bar \rho_L)g \, \partial_{{\mathcal A}_R} \left( g^{-1}(\bar \psi - \bar \rho_L)g \right) \nonumber \\
    & +2 \bar B_R [g^{-1}(\bar \psi - \bar \rho_L) g, \rho_R] -2 B_L[g(\psi + \rho_R )g^{-1},\bar \rho_L]  \Bigr) \ . \label{compos-ss-wzw-action-2-null}
\end{align}
Integrating out $\lambda$ and $\bar \lambda$ we find 
\begin{equation}
\label{lambda constraint}
    g (\psi + \rho_R) g^{-1}|_{h_L^{\notin} } = 0 \qquad \text{and} \qquad g^{-1} (\bar \psi - \bar \rho_L) g|_{h_R^{\notin} } = 0  \ . 
\end{equation}
The second line of eq.~\eqref{compos-ss-wzw-action-2-null} then simplifies to 
\begin{equation}
   g(\psi + \rho_R)g^{-1}|_{h_L^\perp} \bar \partial_{\bar{\mathcal A}_L} \left( g(\psi + \rho_R)g^{-1}|_{h_L^\perp} \right) + g^{-1}(\bar \psi - \bar \rho_L)g|_{h_R^\perp} \partial_{{\mathcal A}_R} \left( g^{-1}(\bar \psi - \bar \rho_L)g|_{h_R^\perp} \right) \ , 
\end{equation}
while the last line of eq.~\eqref{compos-ss-wzw-action-2-null} vanishes,
\begin{multline}
\text{Tr} \Bigl( 2 \bar B_R [g^{-1}(\bar \psi - \bar \rho_L) g|_{h_R^\perp} , \rho_R] - 2 B_L[g(\psi + \rho_R )g^{-1}|_{h_L^\perp} ,\bar \rho_L] \Bigr) \\
= \text{Tr} \Bigl( 2 \underbrace{[\bar B_R, \rho_R]}_{~~~\in \,\mathfrak h_R} \, g^{-1}(\bar \psi - \bar \rho_L) g|_{h_R^\perp} - 2 \underbrace{[B_L, \bar \rho_L]}_{~~~\in\, \mathfrak h_L} \, g(\psi + \rho_R )g^{-1}|_{h_L^\perp} \Bigr) = 0 \ . 
\end{multline}
The action \eqref{compos-ss-wzw-action-2-null} thus collapses to 
\begin{multline}
    I[\mathbb G, \mathbb A, \bar{\mathbb A}] = S^B(g, \mathcal A, \bar{\mathcal A})  +\int \frac{\text d^2v}{2 \pi} \, \text{Tr} \Bigl( g(\psi + \rho_R)g^{-1}|_{h_L^\perp} \bar \partial_{\bar{\mathcal A}_L} \bigl( g(\psi + \rho_R)g^{-1}|_{h_L^\perp} \bigr) \\
    + g^{-1}(\bar \psi - \bar \rho_L)g|_{h_R^\perp} \partial_{{\mathcal A}_R} \bigl( g^{-1}(\bar \psi - \bar \rho_L)g|_{h_R^\perp} \bigr) \Bigr) \ . 
\label{compos-ss-wzw-action-3-null}
\end{multline}
This is exactly the same result we obtained in the previous Section, \cf~eq.~\eqref{compos-ss-wzw-action-10-LR}. The difference is that when the gauging is null 
\begin{equation}
    \mathfrak h_L \subseteq \mathfrak h_L^\perp \ , \qquad     \mathfrak h_R \subseteq \mathfrak h_R^\perp \ ,
\end{equation}
and hence the worldsheet fermions taking values in $\mathfrak h_L$ also decouple from the action (as they must~-- they are shifted linearly by a fermionic gauge transformation; thus they can be set to zero as a gauge choice).

The Lie algebra of $\mathfrak g$ is a direct sum of $\mathfrak h_L^\perp$ and its complement $\mathfrak h_L^{\notin}$, see eq.~\eqref{gdecomp}; $\mathfrak h_L^\perp$ further decomposes into $\mathfrak h_L$ and its complement in $\mathfrak h_L^\perp$ which we denote ${\mathfrak g}_L^{\ttt}$ and refer to as the ``transverse space''. We thus have 
\begin{align}
\begin{split}
    \mfg &= \mfh_L^\perp \oplus \mfh_L^\nin
    = \mfh_L\oplus \mfg_L^\ttt \oplus \mfh_L^\nin \ , 
\\
    \mfg &= \mfh_R^\perp \oplus \mfh_R^\nin
    = \mfh_R\oplus \mfg_R^\ttt \oplus \mfh_R^\nin \ .
\end{split}
\label{g-null-decomposition}
\end{align}
The contributions of $\mfh_L$ (respectively $\mfh_R$) to the kinetic terms vanish due to the null condition \eqref{null-condition-embeddings} and orthogonality to $\mfg_L^\ttt$ (respectively $\mfg_R^\ttt$); thus we have
\begin{multline}
    I[\mathbb G, \mathbb A, \bar{\mathbb A}] = S^B(g, \mathcal A, \bar{\mathcal A})  + \int \frac{\text d^2v}{2 \pi} \, \text{Tr} \Bigl(  g(\psi + \rho_R)g^{-1}|_{\mfg_L^\ttt} \, \bar \partial_{\bar{\mathcal A}_L} \bigl( g(\psi + \rho_R)g^{-1}|_{\mfg_L^\ttt} \, \bigr) \\
    + g^{-1}(\bar \psi - \bar \rho_L)g|_{\mfg_R^\ttt} \, \partial_{{\mathcal A}_R} \bigl( g^{-1}(\bar \psi - \bar \rho_L)g|_{\mfg_R^\ttt} \, \bigr)  \Bigr) \ . 
\label{compos-ss-wzw-action-4-null}
\end{multline}
Hence the fermion kinetic term reduces to the ``transverse space'' under null gauging. The last step consists of decoupling the free fermions from the gauge group $G$. This has the effect of shifting the level in front of the bosonic action by the dual Coxeter number \cite{DiVecchia:1984nyg, Goddard:1986bp, Giveon:1998ns,  Ferreira:2017pgt}.

The above treatment is suitable for simple cosets such as those discussed in Sections~\ref{sec:su2/u1} and~\ref{sec:sl2/u1}, where the transverse space $\mfg^\ttt$ sits simply in $\mfg$ and one retains the fermions $\psi^\pm$ while removing the rest from the path integral.  However, the situation is a good deal more complicated when one comes to examples such as the supertube coset~\eqref{st-coset}; now the removed fermions are some particular linear combinations of $\psi^3_{sl},\psi^3_{su},\psi_t,\psi_y$, and the retained ones some other linear combinations.  Furthermore, this procedure suffers from a lack of manifest supersymmetry in intermediate steps of the calculation.

Writing down the partition function for these fermions is a trivial matter, since they don't couple to the gauge group and yield simply a free fermion partition function~\eqref{trans fermions}.  However, for any computation of operator correlators it is simpler to retain {\it all} of the fermions and impose a BRST constraint to eliminate redundant degrees of freedom.

One proceeds as follows.  Instead of integrating over the fermionic Lagrange multipliers $\lambda,\bar\lambda$ in the gauge multiplet to obtain the constraints~\eqref{lambdaeom-LR}, one maintains manifest supersymmetry by writing them as
\be
\lambda = -\partial\chi \ , \qquad \bar\lambda = -\bar\partial \chi \ , 
\ee
analogous to the Hodge decomposition of the gauge fields~\eqref{Hodge-decomp-01}.  The Jacobian for this transformation in the functional measure then cancels against the functional determinant of the extra pair of fermions one gets because the constraint~\eqref{lambda constraint} is not imposed as a delta function in the path integral.  In the end, one arrives at the same result obtained from the above procedure of eliminating one fermion via the fermionic gauge constraint and another via gauge fixing.

The Jacobian can be represented via a path integral over spinor ghosts $\beta,\gamma$ which are superpartners of the gauge field ghosts $b,c$, and there is an additional term $\gamma\zeta$ in the BRST current, where $\zeta$ is the superpartner of the gauge current $\cJ$ (and similarly for the right-movers).  The role of this term is to ensure that physical operators only contain fermions living in the transverse space~${\mfg}^\ttt$ (with corresponding constraints on spin fields, as discussed in~\rcite{Bufalini:2021ndn,Martinec:2022okx}).


\section{Periodicity properties of the supertube partition function}
\label{app:period-ST}

In this appendix we derive some periodicity properties of the supertube partition function that are used in the main text. In particular, we are going to show that the integrand $\mathcal I(\tau, u, z)$ introduced in eq.~\eqref{ST-pf-final} is periodic in the integration variable $u$ with periods $\tau$ and $1$, \ie
\begin{equation}
\mathcal I(u + m_1 \tau + m_2, \tau, z) = \mathcal I(u, \tau, z)  \ , \qquad \forall \, m_1, m_2 \in \mathds Z \  .  
\label{ST-integrand-periodicity}
\end{equation}
Equivalently, in terms of the holonomies $s_1$ and $s_2$ introduced in eq.~\eqref{u-def}, 
\begin{equation}
\mathcal I(s_1 + m_1, s_2 + m_2, \tau, z) = \mathcal I(s_1, s_2, \tau, z)  \ , \qquad \forall \, m_1, m_2 \in \mathds Z \  .  
\end{equation}
Let us analyse the behavior of the various factors entering \eqref{ST-pf-final}. It is easy to check that
\begin{equation}
e^{- \pi n_5 \tau_2 (s_1 + m_1)^2} e^{- 4 \pi z_2 (s_1 + m_1)} = e^{-\pi n_5 \tau_2 m_1^2} \, e^{-2 \pi n_5 u_2 m_1} \, e^{- 4 \pi z_2 m_1 } \, e^{- \pi n_5 \tau_2 s_1^2} e^{- 4 \pi z_2 s_1} \ , 
\end{equation}
and that 
\begin{equation}
\left|\frac{\theta_1(\tau, u + \frac{n_5+2}{n_5}z+m_1 \tau + m_2)}{\theta_1(\tau, u + \frac{2}{n_5}z+m_1 \tau + m_2)} \right|^2 = e^{4 \pi m_1 z_2} \left| \frac{\theta_1(\tau, u + \frac{n_5+2}{n_5}z)}{\theta_1(\tau, u + \frac{2}{n_5}z)} \right|^2 \ ,
\end{equation}
where we used the periodicity properties of $\theta_1$, see eq.~\eqref{theta1-period}. Since $\ell_2, r_2 \in \mathds Z$, see Section~\ref{sec:supertube-cosets}, eq.~\eqref{theta1-period} also implies
\begin{align}
& \theta_1(\tau, \ell_2 u + \tfrac{n_5-2}{n_5}z + \ell_2(m_1 \tau + m_2)) \nonumber \\
& \hspace{50pt} = (-1)^{\ell_2(m_1+m_2)} \, q^{-\frac{\ell_2^2 m_1^2 }{2}} \, e^{-2 \pi i \ell_2^2 m_1 u} \, e^{-2 \pi i \ell_2 m_1 \frac{n_5-2}{n_5}z} \, \theta_1(\tau, \ell_2 u + \tfrac{n_5-2}{n_5}z ) \ , \\
& \overline{\theta_1(\tau, r_2 u + \tfrac{n_5-2}{n_5}z + r_2 (m_1 \tau + m_2))} \nonumber \\
& \hspace{50pt} = (-1)^{r_2(m_1+m_2)} \, \bar q^{-\frac{r_2^2 m_1^2}{2}} \, e^{2 \pi i r_2^2 m_1 \bar u} \, e^{2 \pi i r_2 m_1 \frac{n_5-2}{n_5} \bar z} \, \overline{\theta_1(\tau, r_2 u + \tfrac{n_5-2}{n_5}z )} \ . 
\end{align}
The $\mathds R_t$ contribution transforms as 
\begin{align}
\int_{-\infty}^\infty \text d E  \, q^{-\frac{E^2}{4}} \, \bar q^{-\frac{E^2}{4}} \, e^{-2 \pi \ell_3 E (u_2 + m_1 \tau_2)} = q^{\frac{m_1^2 \ell_3^2}{4}} \, \bar q^{\frac{m_1^2 \ell_3^2}{4}} \, e^{-2 \pi \ell_3^2 m_1 u_2} \,  \int_{-\infty}^\infty \text d E  \, q^{-\frac{E^2}{4}} \, \bar q^{-\frac{E^2}{4}} \, e^{-2 \pi \ell_3 E u_2} \ . 
\end{align}
Using the quantization conditions \eqref{li ri vals}, one sees that the $U(1)_y$ partition function obeys
\begin{align}
& \sum_{w, n \in \mathds Z} e^{\frac{\pi i}{2} \tau (\frac{n}{R} + w R)^2} \, e^{-\frac{\pi i}{2} \bar \tau (\frac{n}{R} - w R)^2} \, e^{\pi i \ell_4 (\frac{n}{R} + w R)(u + m_1 \tau + m_2)} \, e^{-\pi i r_4 (\frac{n}{R} - w R)(\bar u + m_1 \bar \tau + m_2)} \nonumber \\
 & \hspace{30pt} = q^{-\frac{\ell_4^2 \, m_1^2}{4}} \, \bar q^{-\frac{r_4^2 \, m_1^2}{4}} \, e^{-\pi i \ell_4^2 m_1 u} \, e^{\pi i r_4^2 m_1 u} \nonumber \\
 & \hspace{100pt} \times \, \sum_{w, n \in \mathds Z} e^{\frac{\pi i}{2} \tau (\frac{n}{R} + w R)^2} \, e^{-\frac{\pi i}{2} \bar \tau (\frac{n}{R} - w R)^2} \, e^{\pi i \ell_4 (\frac{n}{R} + w R)u} \, e^{-\pi i r_4 (\frac{n}{R}- w R)\bar u} \ . 
\end{align}
Finally, making use of the fact that $l_2-r_2=2\sfn\in2\bZ$ together with \eqref{su2-ch-period-even} and \eqref{su2-ch-period-odd}, we obtain
\begin{multline}
\mathcal Z_{B}^{\mathfrak{su}(2)}\left(\tau, \bar \tau, \ell_2 u-\tfrac{2}{n_5  }z + \ell_2 (m_1 \tau + m_2), r_2 \bar u-\tfrac{2}{n_5  } \bar z + r_2 (m_1 \tau + m_2) \right) \\
= q^{-\frac{(n_5  -2)}{4}\ell_2^2 m_1^2} \, \bar q^{-\frac{(n_5  -2)}{4}r_2^2 m_1^2} \, e^{-\pi i (n_5  -2)\ell_2^2 m_1 u} \, e^{\pi i (n_5  -2)r_2^2 m_1 \bar u} \, e^{2 \pi i \frac{n_5  -2}{n_5  } \ell_2 m_1 z} \, e^{-2 \pi i \frac{n_5  -2}{n_5  } r_2 m_1 \bar z} \\
\times \, \mathcal Z_{B}^{\mathfrak{su}(2)}\left(\tau, \bar \tau, \ell_2 u-\tfrac{2}{n_5  }z, r_2 \bar u-\tfrac{2}{n_5  } \bar z \right) \ . 
\end{multline}
Using the null constraints \eqref{ST-null-constraints} it is now easy to derive eq.~\eqref{ST-integrand-periodicity}. 

The periodicity property \eqref{ST-integrand-periodicity} of the integrand \eqref{ST-pf-final} implies the following identity for the integral, 
\begin{equation}
    \int_0^1 \text ds_1 \int_0^1 \text ds_2 \, \mathcal I(\tau, u, z) = \int_0^1 \text ds_1 \int_0^1 \text ds_2 \, \mathcal I(\tau, u - a \tau - b, z) \ , \qquad \forall \, a, b \in \mathds R \ . 
    \label{ST-integral-periodicity}
\end{equation}
In the special case of $a, b \in \mathds Z$, eq.~\eqref{ST-integral-periodicity} directly follows from \eqref{ST-integrand-periodicity}. Let us derive \eqref{ST-integral-periodicity} for $a$, $b$ arbitrary real numbers. We have 
\begin{align}
\int_0^1 & \text d s_1  \, \mathcal I(u)  = \int_a^{1+a} \text d s_1 \, \mathcal I((s_1-a) \tau + s_2)  \nonumber \\
& = \int_0^1 \text d s_1 \, \mathcal I((s_1-a) \tau + s_2)-\int_0^a \text d s_1 \, \mathcal I((s_1-a) \tau + s_2) + \int_0^a \text d s_1 \, \mathcal I((s_1-a+1) \tau + s_2) \nonumber \\
& = \int_0^1 \text d s_1 \, \mathcal I((s_1-a) \tau + s_2) = \int_0^1 \text d s_1 \, \mathcal I(u - a \tau) \ , 
\end{align}
where we used eq.~\eqref{ST-integrand-periodicity} in the third equality. Since a similar derivation holds for $s_2$ in place of $s_1$ and $b$ in place of $a$, we obtain eq.~\eqref{ST-integral-periodicity}.

\bibliographystyle{JHEP}      

\bibliography{fivebranes}


\end{document}
